\documentclass[12pt,english]{article}
\usepackage[T1]{fontenc}
\usepackage{amssymb}
\usepackage{amsmath}
\usepackage{cite}
\usepackage{epsfig}
\usepackage[unicode=true,
 bookmarks=true,bookmarksnumbered=false,bookmarksopen=false,
 breaklinks=false,pdfborder={0 0 1},backref=false,colorlinks=true]
 {hyperref}
\usepackage[hang,flushmargin]{footmisc}
\usepackage{relsize}
\usepackage{color}
\usepackage{amsmath}
\usepackage{amsfonts}
\usepackage{amssymb}
\usepackage{graphicx}
\usepackage[compat=1.1.0]{tikz-feynman}
\usepackage{xparse}
\usepackage{subcaption}
\usetikzlibrary{arrows.meta,decorations.markings}

\tikzset{
	hard/.style={postaction={decorate},
		line width=0.5mm
	},
		soft/.style={postaction={decorate},
		line width=0.2mm, dashed
	},
	momentum/.style={postaction={decorate},
	line width=0.5mm,
	color=gray,
	decoration={
    markings,
    mark=at position 0.8 with {\arrow{stealth}}}
    },
	hardarrow/.style={postaction={decorate},
	line width=0.5mm,
	decoration={
    markings,
    mark=at position 0.6 with {\arrow{stealth}}}},
  graviton/.style={decorate, decoration={snake, amplitude=.4mm, segment length=2mm, pre length=0.5mm, post length=0mm}, double}
}

\usepackage{pifont}

\makeatletter
\def\simgt{\mathrel{\lower2.5pt\vbox{\lineskip=0pt\baselineskip=0pt
           \hbox{$>$}\hbox{$\sim$}}}}
\def\simlt{\mathrel{\lower2.5pt\vbox{\lineskip=0pt\baselineskip=0pt
           \hbox{$<$}\hbox{$\sim$}}}}
\makeatother

\newcommand{\bomega}{\bar{\omega}}

\def\I{\text{I}}
\def\II{\text{II}}

\newcommand{\be}{\begin{equation}}
\newcommand{\ee}{\end{equation}}

\newcommand{\MC}[1]{\footnotesize{MC{#1}}}
\newcommand{\NMC}[2]{\footnotesize{N$^{#1}$MC{#2}}}

\newcommand{\Eq}[1]{Eq.~\eqref{#1}}

\def\deltabar{{\mathchar '26\mkern -10mu\delta}}

\begin{document}

\hypersetup{citecolor=blue,linkcolor=blue,urlcolor=blue}

\interfootnotelinepenalty=10000
\baselineskip=18pt

\hfill

\vspace{2cm}
\thispagestyle{empty}
\begin{center}
{\LARGE \bf
Generalized Unitarity Method for\\ Worldline Field Theory \\
}
\bigskip\vspace{1cm}{
{\large Vincent F. He, Julio Parra-Martinez}
} \\[7mm]
 {\it Institut des Hautes \'{E}tudes Scientifiques, 91440 Bures-sur-Yvette, France }
 \end{center}

\bigskip
\centerline{\large\bf Abstract}
\begin{quote} \small
We present a generalized unitarity method for theories of point-particle worldlines coupled to gravity, analogous to that of scattering amplitudes in quantum field theory. This method allows the computation of perturbative observables from basic principles such as locality and unitarity, thus avoiding gauge redundancies and the use of Feynman diagrams.
We illustrate the method with a variety of examples, including the gravitational waveform for the scattering of two point masses at next-to-leading order (or ${\cal O}(G^{5/2})$), reproducing known results. Our method further streamlines the calculation of the scattering dynamics of compact binary systems and opens the door to further applications and systematic exploration of the structure in this class of observables.

\end{quote}

\setcounter{footnote}{0}

\newpage

\tableofcontents
\newpage

\section{Introduction}\label{sec:intro}
The past decade of advances in observational gravitational wave (GW) physics, led by the LIGO/Virgo/Kagra experiment \cite{LIGOScientific:2016aoc, LIGOScientific:2017vwq, Purrer:2019jcp}, and the promise of upcoming experiments such as LISA, Cosmic Explorer or the Einstein Telescope, have kindled the need for high-precision calculations of gravitational observables.  In recent years, a plethora of methods based on scattering amplitudes have resulted in tremendous progress in the study of the classical interactions and radiation from black hole and neutron star binaries \cite{Damour:2017zjx,Cheung:2018wkq,Bern:2019nnu,KoemansCollado:2019ggb,Bern:2019crd,Damour:2019lcq,Cristofoli:2020uzm,Damour:2020tta,Kalin:2020mvi,AccettulliHuber:2020dal, Bern:2021yeh,Bern:2021dqo,Herrmann:2021lqe,DiVecchia:2021ndb,DiVecchia:2021bdo,Herrmann:2021tct,Bjerrum-Bohr:2021vuf,Bjerrum-Bohr:2021din,Jakobsen:2021smu,Dlapa:2021npj,Brandhuber:2021eyq,Dlapa:2021vgp,Cristofoli:2021vyo,Manohar:2022dea,Bern:2022jvn,Dlapa:2022lmu,Herderschee:2023fxh,Georgoudis:2023eke,Brandhuber:2023hhy,Barack:2023oqp,Dlapa:2023hsl,Damgaard:2023ttc,Cheung:2023lnj,Kosmopoulos:2023bwc,Ivanov:2024sds,Correia:2024jgr,Driesse:2024xad,Bern:2024adl,Cheung:2024byb,Driesse:2024feo, Heissenberg:2025ocy, Caron-Huot:2025tlq, Ivanov:2025ozg, Georgoudis:2025vkk, Mogull:2025cfn, Hoogeveen:2025tew,Bern:2025zno}. These techniques provide an efficient way to study this problem in the weak-field or post-Minkowskian (PM) limit \cite{Bertotti:1956pxu,Kerr:1959zlt,Bertotti:1960wuq,Portilla:1979xx,Westpfahl:1979gu,Portilla:1980uz,Bel:1981be,Westpfahl:1985tsl,Ledvinka:2008tk}, in which the separation of the masses is much larger than their Schwarzschild radii but velocities may remain relativistic, and
have produced significant improvements in analytical computations in gravity by incorporating methods from collider physics and the traditional amplitudes program.

In this context, a remarkable tool to perform weak-field calculations is the modern incarnation of the worldline formalism \cite{Goldberger:2004jt,Porto:2016pyg,Cheung:2018wkq, Kalin:2020mvi,Mogull:2020sak}, in which compact objects are modeled as a point particle with worldline action
\begin{equation}
\label{eq:wl}
  S^{\text{wl}}=-\frac{m}{2}\int d\tau \, g_{\mu\nu}\dot{x}^\mu\dot{x}^\nu + \cdots.
\end{equation}
where $m$ is its mass, dots denote derivatives with respect to proper time $\tau$ and the $\cdots$ can include higher-order operators on the worldline, encoding tidal effects and additional degrees of freedom such as spin. This particle is coupled to Einstein gravity described by the bulk action
\begin{equation}
\label{eq:EH}
    S^{\text{EH}}= \frac{1}{16\pi G} \int d^4x \,\sqrt{-g} R + \text{gauge-fixing terms}.
\end{equation}
Practical calculations in this formalism are performed by expanding the worldline and the gravitons  around their respective background values, corresponding to a straight trajectory and flat spacetime:
\begin{equation}
\label{eq:fluct}
  x^\mu(\tau)=b^\mu+u^\mu \tau+z^\mu(\tau),\qquad g_{\mu\nu}=\eta_{\mu\nu}+\kappa h_{\mu\nu}.
\end{equation}
Here $b$ is the impact parameter, $u$ is the background worldline four-velocity (with $u^2=1$), $z$ is the worldline fluctuation, and $\kappa=\sqrt{32\pi G}$.
Then, one proceeds to solve perturbatively the gravitational dynamics of the system, which can be conveniently encoded in the language of Feynman diagrams. This approach has been dubbed worldline quantum field theory (WQFT) \cite{Mogull:2020sak,Jakobsen:2022psy,Jakobsen:2023oow,Jakobsen:2021zvh,Driesse:2024feo} and is the one we follow in this paper.

The worldline formalism greatly streamlines the computations of classical observables, bypassing the need to take the subtle classical limit of quantum amplitudes. Powerful integration methods from scattering amplitudes, such as integration-by-parts (IBP) reduction \cite{Tkachov:1981wb,Chetyrkin:1981qh},  (canonical) differential equations \cite{Kotikov:1990kg,Bern:1992em,Gehrmann:1999as,Henn:2013pwa,Henn:2014qga} and reverse unitarity \cite{Anastasiou:2002yz,Anastasiou:2002qz,Anastasiou:2003yy,Anastasiou:2015yha} can be straightforwardly applied in the worldline setting, as the integrals that appear are the same as those in the classical limit of QFT amplitudes \cite{Parra-Martinez:2020dzs}. Nevertheless, powerful structures and methods for constructing integrands, such as generalized unitarity \cite{Bern:1994zx,Bern:1994cg,Britto:2004nc,Bern:2007ct,Bern:2008pv,Bern:2010tq,Carrasco:2011hw, Bern:2024vqs} and the double copy \cite{Bern:2008qj,Bern:2010ue,Bern:2019prr}, have not been fully understood in the worldline context (see however \cite{Goldberger:2016iau,Shen:2018ebu,Shi:2021qsb,Comberiati:2022cpm, Edison:2022cdu,Edison:2023qvg,Edison:2024owb} for some initial explorations).

The generalized unitarity method leverages the fact that, as a consequence of locality and unitarity, tree-level scattering amplitudes and loop integrands factorize as a given momentum goes on-shell
\begin{align}
        \mathcal{A} &\xrightarrow{k^2\to 0} i\frac{\mathcal{A}_L\mathcal{A}_R}{k^2}
\end{align}
where $\mathcal{A}_L$ and $\mathcal{A}_R$ are sub-amplitudes separated by the on-shell particle, and a sum over all possible intermediate states with such on-shell momentum is implied. In the gravitational context, the limit $k^2\to 0$ corresponds to imposing the on-shell condition on the graviton momentum, $k$, which stems from its free equation of motion in the usual de Donder gauge:
\begin{equation}\label{eq:eomh}
  \square h_{\mu\nu}(x)=0,\qquad k^2=0.
\end{equation}
More generally, one might compute the residue of any given series of momentum poles as a product of simpler amplitudes. This can be recursed down to the most basic building blocks, which are local (i.e., polynomial) matrix elements, whose coefficients correspond to the irreducible coupling constants of the theory. Such a method thus allows to compute amplitudes from basic principles, sidestepping the need for Feynman diagrams and their associated off-shell and gauge redundancies.

While the current bottleneck for precision calculation of gravitational observables lies not in the construction of integrands but in integration, it is still desirable to extend generalized unitarity methods to the context of worldline theories, as they will streamline future calculations, and might reveal hidden structures. Naively this seems straightforward, as fluctuations of the worldline have an associated free equation of motion and on-shell condition
\begin{equation}\label{eq:eomz}
 m \ddot z^\mu(\tau) = 0\,, \qquad m \omega^2 = 0\, ,
\end{equation}
where $\omega$ is the Fourier conjugate frequency to $\tau$.
Hence one might expect that in the limit $\omega \to 0$ when a worldline fluctuation goes on-shell observables should factorize.
Indeed, the coefficients of double poles in $\omega$ in worldline observables factorize into simpler quantities, but worldline observables also contain simple poles in frequency of the form $1/\omega$ which a priori do not seem to be fixed on shell
\begin{align}
        \mathcal{A}& {\xrightarrow{\omega\to0}}-i\frac{\mathcal{A}_L\mathcal{A}_R}{m\omega^2} +\frac{?}{\omega} \,.\label{eq:cutomegasinglepole}
\end{align}
In the philosophy of effective field theory (EFT), whenever a contact term in an amplitude is unfixed by basic considerations, it corresponds to a new local operator in the theory that needs to be matched. However, a simple pole is not local in any sense, so its interpretation is unclear. The main technical challenge in implementing the unitarity method in the presence of worldline degrees of freedom is then to devise a way to compute the coefficient of such simple poles in the worldline frequencies.

The solution to this problem turns out to be remarkably simple: one might complexify the worldline energy, so that the on-shell condition takes the form
\begin{equation}
m \omega \overline{\omega} = 0\,.
\end{equation}
Then the on-shell limit can be taken by setting either $\omega$ or $\overline{\omega}$ to zero, and considering both possibilities unambiguously fixes the coefficient of the single poles in frequency. The complexification of energies and momenta is familiar from the analysis of three-point amplitudes in the spinor-helicity formalism, which have special collinear kinematics \cite{Benincasa:2007xk}. Worldline momenta are one-dimensional and hence always collinear, so, with the benefit of hindsight, the need to introduce complex kinematics does not come as a surprise.

This paper is organized as follows: In Section~\ref{sec:obs} we describe the class of observables in worldline field theories that we will compute, introduce their extension to complex frequencies, and explain their properties which will enable the unitarity method, including a soft theorem. In Section~\ref{sec:rational} we will explain how basic principles fix a class of rational observables in worldline theories (analogous to tree amplitudes in QFT), without the need to introduce a Lagrangian. In Section~\ref{sec:loop}, we show how to use the rational observables as building blocks for more general observables, analogous to loop amplitudes, and develop a full generalized unitarity method. As an application and check of this method we reproduce the integrands of the conservative on-shell action (or radial action) up to $\mathcal{O}(G^3)$, as well as the $\mathcal{O}(G^2)$ waveform, and we check that they agree with the known results upon integration.

\emph{Conventions:} We use mostly-minus metric signature. The momenta of all external states are taken to be outgoing. We use $k$'s to denote graviton momenta, $u$'s to denote worldline velocities, and $\omega$'s to denote worldline fluctuation energies.
We define the conveniently normalized Dirac delta function and $D$-dimensional integration measure as
\begin{equation}
  \deltabar(x) = 2\pi \delta(x)\,, \qquad \int_k = \int \frac{d^D k}{(2\pi)^D}\,.
\end{equation}

\section{Observables in worldline field theory}
\label{sec:obs}

In this section we introduce the class of observables that we will be concerned with in the rest of the paper, and describe some of their properties which enable the generalized unitarity method.

\subsection{In-out vs. in-in observables and their integrands}

Time-ordered worldline expectation values are defined as the result of a path integral:
\begin{equation}
  \langle \mathcal{TO}(h_i,z_j)\rangle=\int Dh\,Dz\,\mathcal{O}(h_i,z_j)\,e^{i S}\,.
\end{equation}
where $S=S^{\text{wl}}+S^{\text{EH}}+\cdots$ is the sum over the actions in Eqs.~\eqref{eq:wl}, \eqref{eq:EH} supplemented by appropriate gauge-fixing terms, and $\cal O$ is some operator composed of graviton and worldline fluctuations. These expectation values can be computed using familiar Feynman rules with time-ordered, or Feynman, propagators. For gravitons with momentum $k$ and worldlines with frequency $\omega$ these are respectively of the form
\begin{equation}
    G^F_h(k) = \frac{i\Pi^{\mu\nu\alpha\beta}}{k^2+i\epsilon}\,, \qquad  G^F_h(\omega) = \frac{-i\eta^{\mu\nu}}{\omega^2+i\epsilon}\,,
\end{equation}
where $\Pi$ is an appropriate projector, which depends on the chosen gauge. For instance in de Donder gauge we have
\begin{align}
\Pi^{\mu\nu\alpha\beta}=\frac{1}{2}\Big(\eta^{\mu\alpha}\eta^{\nu\beta}+\eta^{\mu\beta}\eta^{\nu\alpha}-\tfrac{1}{D-2}\eta^{\mu\nu}\eta^{\alpha\beta}\Big)\,.
\label{eq:gproj}
\end{align}
Note that as we are expanding the worldline in fluctuations about background trajectories $\bar x_i^\mu(\tau) = b_i^\mu + u_i^\mu \tau$ as in Eq.~\eqref{eq:fluct}, the Feynman rules generically contain insertions of $\bar x_i^\mu(\tau)$, which we will refer to as sources.

The worldline amplitudes which we focus on in this work are obtained by applying LSZ reduction on the graviton and worldline fluctuation states. In momentum space, this reads:
\begin{align}
\begin{aligned}
  \label{LSZ}
  &A(k_1,\ldots k_n, \varepsilon_1,\ldots \varepsilon_n, \omega_1, \ldots,\omega_m, \zeta_1,\ldots, \zeta_m)
  =\text{LSZ}\langle \mathcal{T} h_1\cdots  h_n\, \ z_1\cdots  z_m\rangle \\
  &\hspace{2cm}= \lim_{k_i^2,\omega_j^2\to 0}
  \prod_a (-i k^2_a)\prod_b (im_b \omega^2_b)\,\langle h_1 \cdots h_n\,  z_1\cdots  z_m\rangle\,.
  \end{aligned}
\end{align}
Here $h_i = \varepsilon_i^{\mu\nu}h_{\mu\nu}(k_i)$, $z_i = \zeta_i^\mu z_\mu(\omega_i)$ (or their Fourier transforms), where the polarization tensors of gravitons are denoted by $\varepsilon_i$ and henceforth written as the product of two spin-one polarizations $\varepsilon_i^{\mu\nu}=\varepsilon_i^\mu\varepsilon_i^\nu$, which are transverse, $\varepsilon\cdot k =0$, from which we can always extract the traceless symmetric part by appropriate projection. In order to use index-free notation throughout, we also introduce dummy polarization vectors for worldline fluctuations, denoted by $\zeta_i^\mu$.
In a system with $n$  sources,\footnote{The number of sources corresponds to the number of sub-amplitudes that are connected solely by graviton intermediate states. This is not necessarily the same as the number of the background worldlines, since multiple sources can correspond to the same worldline.} worldline energy conservation and bulk four-momentum conservation implies that these amplitude take the generic form:
\begin{equation}
  \label{eq:WQFTamp}
  \begin{aligned}
    A(k_1,\ldots k_n, \varepsilon_1,\ldots \varepsilon_n, &\omega_1, \ldots,\omega_m, \zeta_1,\ldots, \zeta_m)\\
    &=\int_{q_1,\ldots,q_{m}}\deltabar^{D}(\sum_{i=1}^n k_i-\sum_{j=1}^m q_j)\left(\prod_{j=1}^{m}\deltabar(q_j\cdot u_j)e^{iq_j\cdot b_j}\right) {\cal A}(q_i, k_i,u_i, \varepsilon_i,\zeta_i)\,
  \end{aligned}
\end{equation}
where $k_i$ are the four-momenta of the external gravitons and $q_j$ is the four-momentum exchanged by gravitons and the $j$-th worldline source.
The integrand ${\cal A}$ is a function of Lorentz invariant products of the $k_i,q_i,u_i$.\footnote{Note that at this stage, ${\cal A}$ is independent of external worldline energies $\omega_j$ because they vanish on-shell. However, later we will extend the notion of worldline amplitude to include non-vanishing external worldline energies.} This form is dictated by translational invariance, in the same way that regular QFT amplitudes are proportional to a momentum-conserving delta function. The presence of the worldline sources breaks translation invariance  in the  directions transverse to their four-velocities $u_i$. Thus only energy conservation is satisfied by the worldline interactions, which yields a factor of  $\int d\tau e^{i q_j.\bar x_j(\tau)}=\deltabar(q_j\cdot u_j)e^{iq_j\cdot b_j}$ per source, where the exponential factor is required by the fact that the graviton momentum eigenstates must transform by a phase upon spacetime translations $b^\mu\to b^\mu+\Delta b^\mu$. On the other hand, the graviton interactions are translation invariant, so graviton momentum is conserved up to the total momentum exchanged with the worldline sources, yielding the factor of $\deltabar^{D}(\sum_i^n k_i-\sum_j^m q_j)$. This fixes the form in Eq.~\eqref{eq:WQFTamp}, where the remaining exchanged momenta $q_i$ are to be integrated over.

Since we are only concerned with classical observables, all the integrals in the formalism come from the lack of four-momentum conservation on the worldlines. Closed loops of gravitons or worldline fluctuations do not contribute in the classical limit, which one can check by power counting in the transfer momenta and/or $\hbar$. In other words, classical observables in this setting are simply tree diagrams integrated against sources, as one might expect from the solution to classical equations of motion. In analogy with QFT amplitudes we will still refer to these remaining integrals as ``loops'' and to the corresponding amplitudes as ``loop amplitudes''. The corresponding integrand ${\cal A}$ is then a rational function.

\begin{figure}[t]
    \centering  \includegraphics[width=0.9\linewidth]{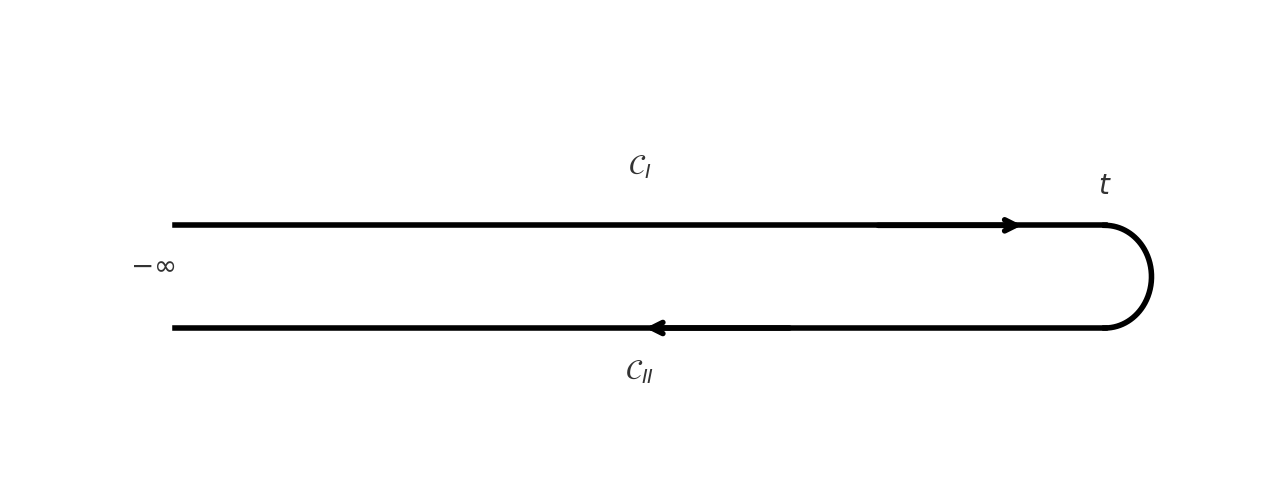}
    \caption{Schwinger-Keldysh contour for in-in correlators.}
    \label{fig:SKcontour}
\end{figure}

Note that the amplitudes we define in Eq.~\eqref{LSZ}, being a result of a single path integral, are \emph{in-out} amplitudes. In contrast, the observables we are usually concerned with in classical scattering are \emph{in-in} observables\cite{Jakobsen:2022psy, Kalin:2022hph}. These in-in amplitudes can be computed by a folded Schwinger-Keldysh contour with shape shown in Fig.~\ref{fig:SKcontour}\cite{Keldysh:1964ud,Schwinger:1960qe}. Equivalently, if we label fields on the first half, $\mathcal{C}_{\I}$, of the contour with $\I$ and those on the second half, $\mathcal{C}_{\II}$, with $\II$, in-in observables are computed by a path integral over two copies of the fields identified in the future
\begin{equation}
\label{eq:SK}
  \langle \mathcal{O}(h_i,z_j)\rangle=\int Dh^{\I}\, Dh^{\II}\,Dz^{\I}\,Dz^{\II}\,\mathcal{O}(h_i,z_j)\,e^{i S^{\I}- i S^{\II}}\,.
\end{equation}
with boundary conditions such that the two copies are identified in the future
\begin{equation}
    h^{\I}(t=+\infty)=h^{\II}(t=+\infty)\,, \quad z^{\I}(t=+\infty)=z^{\II}(t=+\infty)\,.
\end{equation}
The propagator matrix is then given by:
\begin{equation}
   \mathbf{G}= \begin{pmatrix}
        G_{\I,\I} & G_{\I,\II}\\
        G_{\II,\I} & G_{\II,\II}
\end{pmatrix}=
\begin{pmatrix}
        G^F & G^>\\
        G^< & G^{\bar{F}}
    \end{pmatrix}\,.
\end{equation}
$G^F$ and $G^{\bar{F}}$ are the time-ordered (i.e., Feynman) and anti-time-ordered (i.e., anti-Feynman) propagators respectively, and the off-diagonal terms, $G^\lessgtr$, are the Wightman functions.

For gravitons the matrix of propagators is
\begin{equation}
   \mathbf{G}_h(k)=
\begin{pmatrix}
        \frac{i}{k^2+i\epsilon} &  \theta(k^0)\deltabar(k^2) \\
         \theta(-k^0)\deltabar(k^2) & \frac{-i}{k^2-i\epsilon}
    \end{pmatrix} \Pi^{\mu\nu\alpha\beta}\,.
\end{equation}
and for the worldline fluctuations it is given by
\begin{equation}
   \mathbf{G}_z(\omega)=
\begin{pmatrix}
        \frac{-i}{\omega^2+i\epsilon} &  \frac12 \delta'(\omega) \\
         \frac12 \delta'(\omega) & \frac{i}{\omega^2-i\epsilon}
    \end{pmatrix} \eta^{\mu\nu}\,.
\end{equation}
The in-in observables are computed by summing over all Feynman diagrams, which only differ from the in-out ones in the choice of $i\epsilon$ prescription and the complex conjugation of the $\II$ vertices arising from the difference in signs in the exponential $e^{i S^{\I}- i S^{\II}}$.

Note that a different choice of field variables can be made for the path integral in Eq.~\eqref{eq:SK}.  For instance, one might choose the so-called Keldysh basis,\footnote{Given by \begin{equation}
    h^+ = \frac12 (h^\I+h^\II)\,, \qquad h^- = h^\I-h^\II\,, \qquad z^+ = \frac12 (z^\I+z^\II)\,, \qquad z^- = z^\I-z^\II\,.
\end{equation}} which is used in Refs.~\cite{Jakobsen:2022psy,Kalin:2022hph}.
This basis has some advantages: for instance it makes causality manifest, as the propagator matrices involve the retarded and advanced propagators, more familiar from the solution of classical equations of motion. Furthermore the integrals do not contain contour-pinching poles such as those in the worldline Feynman propagator $1/(\omega^2+i\epsilon)$, so their evaluation is straightforward (and unambiguous when one of the poles is pinched by a kinematic numerator).
However, the combinatorics of the Feynman rules in the Keldysh basis is different, which might pose a challenge for applying generalized unitarity methods.\footnote{This is likely mitigated in the classical limit, as only vertices with a single minus-type field are allowed \cite{Caron-Huot:2010fvq,Biswas:2024ept}.}

In this paper, we are only concerned with constructing integrands. And indeed the integrands for any in-in amplitude in the $\I/\II$ basis can be obtained directly from that of the corresponding in-out amplitudes, by cutting each diagram in all possible ways, i.e., replacing the cut Feynman propagators by Wightman functions and complex conjugating all terms on one side of the cut, as done e.g., in Refs.~\cite{Herrmann:2021lqe,Herrmann:2021tct}. Thus, from now on we will ignore the differences between the $i\epsilon$ prescriptions of in-in and in-out amplitudes and focus our attention on the properties of the integrand.

\subsection{Locality and unitarity}

Let us now discuss the general implications of locality and unitarity for the structure of worldline amplitudes.  Locality implies that the integrand in \Eq{eq:WQFTamp}, can be decomposed into a sum of diagrams corresponding to different space-time processes and the only allowed poles correspond to particles or fluctuations propagating according to the edges of such graphs. Traditionally this is represented as a sum over Feynman-like diagrams
\begin{equation}
  {\cal A}=\sum_i\frac{N_i}{\prod_{\alpha_i} D_{\alpha_i}}\,,
\end{equation}
where diagram $i$ has edges corresponding to the propagators $(D_i)_\alpha$ of the various modes, and $N_i$'s are polynomials in Lorentz products of momenta and polarization vectors.
The propagator for graviton edges takes the form $D_k=k^2$ and that of worldline fluctuations is $D_\omega=m\omega^2$.

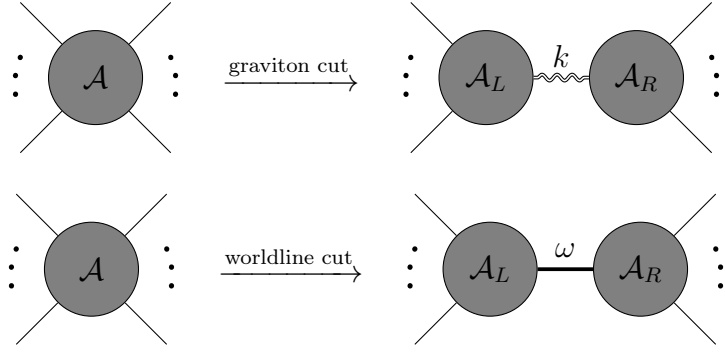
\begin{figure}[t]
    \centering
\raisebox{5pt}{
    \begin{tikzpicture}[baseline=(current bounding box.center)]
    \begin{feynman}
      \vertex (i1);
      \vertex[blob, shape=circle, minimum size=1.25cm, fill=gray] at ($(i1) + (1cm, -1cm)$) (c) {$\mathcal{A}$};
      \vertex at ($(c) + (1cm, 1cm)$)  (i2);
      \vertex at ($(c) + (-1cm,-1cm)$)  (i3);
      \vertex at ($(c) + (1cm, -1cm)$)  (i4);
      \vertex at ($(i1) + (0, -0.73cm)$) (u1);
      \vertex at ($(u1) + (0, -0.73cm)$) (d1);
      \vertex at ($(i2) + (0, -0.73cm)$) (u2);
      \vertex at ($(u2) + (0, -0.73cm)$) (d2);
      \diagram* {
        (i1)--(c)--(i2),(i3)--(c)--(i4), (u1) -- [line cap=round, dash pattern=on 0pt off 7pt, line width=2pt, bend right=20] (d1),  (u2) -- [line cap=round, dash pattern=on 0pt off 7pt, line width=2pt, bend left=20] (d2)
      };
    \end{feynman}
  \end{tikzpicture}}~$\quad\xrightarrow{\text{graviton cut}}\quad$~\raisebox{5pt}{\begin{tikzpicture}[baseline=(current bounding box.center)]
    \begin{feynman}
      \vertex (i1);
      \vertex[blob, shape=ellipse, minimum size=1.25cm, fill=gray] at ($(i1) + (1cm, -1cm)$) (c1) {$\mathcal{A}_L$};
      \vertex[blob, shape=ellipse, minimum size=1.25cm, fill=gray] at ($(c1) + (2cm, 0cm)$) (c2) {$\mathcal{A}_R$};
      \vertex at ($(c2) + (1cm, 1cm)$)  (i2);
      \vertex at ($(c1) + (-1cm,-1cm)$)  (i3);
      \vertex at ($(c2) + (1cm, -1cm)$)  (i4);
      \vertex at ($(i1) + (0, -0.73cm)$) (u1);
      \vertex at ($(u1) + (0, -0.73cm)$) (d1);
      \vertex at ($(i2) + (0, -0.73cm)$) (u2);
      \vertex at ($(u2) + (0, -0.73cm)$) (d2);
      \diagram* {
        (i1)--(c1)--[graviton, edge label=$k$](c2)--(i2),(i3)--(c1),(c2)--(i4), (u1) -- [line cap=round, dash pattern=on 0pt off 7pt, line width=2pt, bend right=20] (d1),  (u2) -- [line cap=round, dash pattern=on 0pt off 7pt, line width=2pt, bend left=20] (d2)
      };
    \end{feynman}
  \end{tikzpicture}}
   \vspace{0.5cm}\\
    \raisebox{5pt}{\begin{tikzpicture}[baseline=(current bounding box.center)]
    \begin{feynman}
      \vertex (i1);
      \vertex[blob, shape=circle, minimum size=1.25cm, fill=gray] at ($(i1) + (1cm, -1cm)$) (c) {$\mathcal{A}$};
      \vertex at ($(c) + (1cm, 1cm)$)  (i2);
      \vertex at ($(c) + (-1cm,-1cm)$)  (i3);
      \vertex at ($(c) + (1cm, -1cm)$)  (i4);
      \vertex at ($(i1) + (0, -0.73cm)$) (u1);
      \vertex at ($(u1) + (0, -0.73cm)$) (d1);
      \vertex at ($(i2) + (0, -0.73cm)$) (u2);
      \vertex at ($(u2) + (0, -0.73cm)$) (d2);
      \diagram* {
        (i1)--(c)--(i2),(i3)--(c)--(i4), (u1) -- [line cap=round, dash pattern=on 0pt off 7pt, line width=2pt, bend right=20] (d1),  (u2) -- [line cap=round, dash pattern=on 0pt off 7pt, line width=2pt, bend left=20] (d2)
      };
    \end{feynman}
  \end{tikzpicture}}~$\quad\xrightarrow{\text{worldline cut}}\quad$~\raisebox{5pt}{\begin{tikzpicture}[baseline=(current bounding box.center)]
    \begin{feynman}
      \vertex (i1);
      \vertex[blob, shape=ellipse, minimum size=1.25cm, fill=gray] at ($(i1) + (1cm, -1cm)$) (c1) {$\mathcal{A}_L$};
      \vertex[blob, shape=ellipse, minimum size=1.25cm, fill=gray] at ($(c1) + (2cm, 0cm)$) (c2) {$\mathcal{A}_R$};
      \vertex at ($(c2) + (1cm, 1cm)$)  (i2);
      \vertex at ($(c1) + (-1cm,-1cm)$)  (i3);
      \vertex at ($(c2) + (1cm, -1cm)$)  (i4);
      \vertex at ($(i1) + (0, -0.73cm)$) (u1);
      \vertex at ($(u1) + (0, -0.73cm)$) (d1);
      \vertex at ($(i2) + (0, -0.73cm)$) (u2);
      \vertex at ($(u2) + (0, -0.73cm)$) (d2);
      \diagram* {
        (i1)--(c1)--[hard, edge label=$\omega$](c2)--(i2),(i3)--(c1),(c2)--(i4), (u1) -- [line cap=round, dash pattern=on 0pt off 7pt, line width=2pt, bend right=20] (d1),  (u2) -- [line cap=round, dash pattern=on 0pt off 7pt, line width=2pt, bend left=20] (d2)
      };
    \end{feynman}
  \end{tikzpicture}}
  \caption{The amplitudes factorize into products of sub-amplitudes upon cutting an internal propagator. Wavy lines represent gravitons and thick solid lines represent worldline fluctuations. }
  \label{fig:factorization}
\end{figure}

By unitarity, the integrand factorizes when an internal propagator goes on shell according to the equations of motion in Eqs.~\eqref{eq:eomh}-\eqref{eq:eomz}.
That is,
\begin{subequations}
\begin{align}
    \lim_{k^2\to 0}   k^2 \mathcal{A} &= i\sum_{\rm pol.} \mathcal{A}_L(k, \varepsilon)\mathcal{A}_R(-k,\varepsilon^*)\,,
    \label{eq:unitaritycutg}\\
    \lim_{\omega^2\to 0}   m\omega^2 \mathcal{A} &=-i\sum_{\rm pol.} \mathcal{A}_L(\omega, \zeta)\mathcal{A}_R(-\omega,\zeta^*)\,,
    \label{eq:unitaritycutz}
\end{align}
\end{subequations}
where the sum over physical polarizations of the graviton is performed by
\begin{equation}
  \begin{aligned}
  \sum_{\text{pol}}\varepsilon^{*\mu}(-k)\varepsilon^{*\nu}(-k)\varepsilon_\alpha(k)\varepsilon_\beta(k)&=\frac{1}{2}\left(P^{\mu\alpha}P^{\nu\beta}+P^{\mu\beta}P^{\nu\alpha}-\frac{1}{D-2}P^{\mu\nu}P^{\alpha\beta}\right)\,,
  \end{aligned}
\end{equation}
with physical state projector
\begin{equation}
     P^{\mu\nu}(k)=\eta^{\mu\nu}-\frac{k^\mu q^\nu+k^\nu q^\mu}{k\cdot q}\,,
\end{equation}
and $q^\mu$ is some reference null momentum.  Similarly, the polarization sum of the worldline fluctuation is given by
\begin{equation}
  \sum_{\text{pol}}\zeta^{*\mu}(-\omega) \zeta^\nu (\omega)=\eta^{\mu\nu}\,.
\end{equation}
Diagrammatically, \Eq{eq:unitaritycutg} and \Eq{eq:unitaritycutz} can be represented by Fig.~\ref{fig:factorization}. In our diagrammatic convention, we use wavy lines to represent gravitons and thick solid lines to represent worldline fluctuations. We take the procedure defined by Eq. (22b) to be the definition for ''cutting'' in our work, which amounts to taking the residue of the integrand at the pole corresponding to the on-shell condition of some internal leg.

The principle of generalized unitarity \cite{Bern:2011qt} allows us to perform multiple cuts at the same time and the factorization can be generalized easily.

\subsection{Complexified kinematics}
Notice that the on-shell condition for a worldline fluctuation $\omega^2=0$ (\Eq{LSZ}) implies that $\omega=0$ because $\omega$ is a scalar. Thus, the cut \Eq{eq:unitaritycutz} only contains information about the double pole in $\omega$ but not the simple pole. Conversely, one can only recover the double pole from sub-amplitudes. 
As we have explained in the introduction, in the spirit of EFT, we would like to use unitarity to fix the amplitudes up to contact terms. To this end, we complexify the worldline fluctuation energy such that the LSZ procedure also extracts the single pole in $\omega$. More precisely, we denote the complexified energies by $\omega$ and $\bomega$, which are independent variables with $\omega^*\neq\bomega$. The on-shell condition now becomes $\bomega\omega=0$, allowing either $\bomega$ or $\omega$ to be non-zero on shell. We will denote amplitudes with complexified external worldline fluctuation energies as\footnote{The regularity of the correlation functions in the limit is guaranteed by a soft theorem that we will prove in Sec.~\ref{sec:softth}.}
\begin{subequations}
\begin{align}
A(\mathfrak{z}_i,\cdots)&=\lim_{\bomega\to 0}im_i\omega\bomega\,\langle z_i(\omega,\bomega)\cdots\rangle\,,\\
A(\bar{\mathfrak{z}}_i,\cdots)&=\lim_{\omega\to 0}im_i\omega\bomega\,\langle z_i(\omega,\bomega) \cdots\rangle\,,
\end{align}
\label{eq: complexamp}
\end{subequations}
and similarly for amplitudes with more external states.  The operator
$z_i(\omega,\bar\omega)$ denotes a complexified off-shell
worldline fluctuation whose quadratic kernel is
$m_i\omega\bomega$.
The correlators on the right-hand side of \Eq{eq: complexamp} are the associated off-shell currents with one unamputated worldline leg\footnote{These off-shell currents may equivalently be viewed as complexified
Berends--Giele currents built from complexfied local operators and propagators.}. Their local vertices are defined by symmetrizing the dependence on the two
worldline energies under
$\omega\leftrightarrow\bar\omega$.
The two amplitudes in \Eq{eq: complexamp} are the residues of the same off-shell
current on the two branches $\bar\omega=0$ and $\omega=0$ of the
complexified on-shell condition $\omega\bar\omega=0$. Since one of the two complex energies is non-vanishing, these amplitudes live on the support of modified energy conservation. For example, $A(\mathfrak{z}_j,\cdots) \propto \deltabar\left(\sum_i u\cdot k_i+\omega_j\right)$, while $A(\bar{\mathfrak{z}}_j,\cdots) \propto \deltabar\left(\sum_i u\cdot k_i+\bomega_j\right)$, where $k_i$ are the momenta of gravitons. This is to be contrasted with the usual definition of the worldline amplitudes
\begin{equation}
A(z_i,\cdots)=\lim_{\omega\to0}im_i\omega^2\,\langle z_i(\omega) \cdots\rangle\,,
\end{equation}
where the real-kinematics relation $\bomega=\omega$ has been imposed before taking the on-shell limit. These usual amplitudes are fully independent of $\omega$.
In other words, $z$ represents an external state whose energy is real and on-shell, whereas $\mathfrak{z}$ and $\bar{\mathfrak{z}}$ represent external states whose energies are complex, with the on-shell conditions evaluated on the conjugate energy.

Such complexification is not unusual in amplitude-based methods. For instance, while the three-point amplitudes of massless particles in four dimensions have no support for real kinematics, they have one-dimensional support in complex kinematics in spinor-helicity variables, where one can choose either the angle or square brackets to vanish so that the amplitude is non-vanishing \cite{Benincasa:2007xk}.  Note that the usual spinor-helicity complexification scheme does not modify momentum conservation because the conserved real momenta are quadratic in the spinors. In contrast, our worldline amplitudes can be defined for real $\omega$ (with $\bomega=0$, or vice versa), because energy conservation is modified.

\begin{figure}[t!]
    \centering
\raisebox{5pt}{\begin{tikzpicture}[baseline=(current bounding box.center)]
    \begin{feynman}
      \vertex (i1);
      \vertex[blob, shape=circle, minimum size=1.25cm, fill=gray] at ($(i1) + (1cm, -1cm)$) (c) {$\mathcal{A}$};
      \vertex at ($(c) + (1cm, 1cm)$)  (i2);
      \vertex at ($(c) + (-1cm,-1cm)$)  (i3);
      \vertex at ($(c) + (1cm, -1cm)$)  (i4);
      \vertex at ($(i1) + (0, -0.73cm)$) (u1);
      \vertex at ($(u1) + (0, -0.73cm)$) (d1);
      \vertex at ($(i2) + (0, -0.73cm)$) (u2);
      \vertex at ($(u2) + (0, -0.73cm)$) (d2);
      \diagram* {
        (i1)--(c)--(i2),(i3)--(c)--(i4), (u1) -- [line cap=round, dash pattern=on 0pt off 7pt, line width=2pt, bend right=20] (d1),  (u2) -- [line cap=round, dash pattern=on 0pt off 7pt, line width=2pt, bend left=20] (d2)
      };
    \end{feynman}
  \end{tikzpicture}}~$\quad\xrightarrow{\mathfrak{z}\text{ cut}}\quad$~\raisebox{5pt}{\begin{tikzpicture}[baseline=(current bounding box.center)]
    \begin{feynman}
      \vertex (i1);
      \vertex[blob, shape=ellipse, minimum size=1.25cm, fill=gray] at ($(i1) + (1cm, -1cm)$) (c1) {$\mathcal{A}_L$};
      \vertex at ($(c1) + (1.1cm, 0cm)$) (mid);
      \vertex[blob, shape=ellipse, minimum size=1.25cm, fill=gray] at ($(c1) + (2.5cm, 0cm)$) (c2) {$\mathcal{A}_R$};
      \vertex at ($(c2) + (1cm, 1cm)$)  (i2);
      \vertex at ($(c1) + (-1cm,-1cm)$)  (i3);
      \vertex at ($(c2) + (1cm, -1cm)$)  (i4);
      \vertex at ($(i1) + (0, -0.73cm)$) (u1);
      \vertex at ($(u1) + (0, -0.73cm)$) (d1);
      \vertex at ($(i2) + (0, -0.73cm)$) (u2);
      \vertex at ($(u2) + (0, -0.73cm)$) (d2);
      \diagram* {
        (i1)--(c1)--[hard, edge label=$\omega$](mid)--[hard, dashed, edge label=$\bomega$](c2)--(i2),(i3)--(c1),(c2)--(i4), (u1) -- [line cap=round, dash pattern=on 0pt off 7pt, line width=2pt, bend right=20] (d1),  (u2) -- [line cap=round, dash pattern=on 0pt off 7pt, line width=2pt, bend left=20] (d2)
      };
    \end{feynman}
  \end{tikzpicture}}\vspace{0.5cm}\\
  \raisebox{5pt}{\begin{tikzpicture}[baseline=(current bounding box.center)]
    \begin{feynman}
      \vertex (i1);
      \vertex[blob, shape=circle, minimum size=1.25cm, fill=gray] at ($(i1) + (1cm, -1cm)$) (c) {$\mathcal{A}$};
      \vertex at ($(c) + (1cm, 1cm)$)  (i2);
      \vertex at ($(c) + (-1cm,-1cm)$)  (i3);
      \vertex at ($(c) + (1cm, -1cm)$)  (i4);
      \vertex at ($(i1) + (0, -0.73cm)$) (u1);
      \vertex at ($(u1) + (0, -0.73cm)$) (d1);
      \vertex at ($(i2) + (0, -0.73cm)$) (u2);
      \vertex at ($(u2) + (0, -0.73cm)$) (d2);
      \diagram* {
        (i1)--(c)--(i2),(i3)--(c)--(i4), (u1) -- [line cap=round, dash pattern=on 0pt off 7pt, line width=2pt, bend right=20] (d1),  (u2) -- [line cap=round, dash pattern=on 0pt off 7pt, line width=2pt, bend left=20] (d2)
      };
    \end{feynman}
  \end{tikzpicture}}~$\quad\xrightarrow{\bar{\mathfrak{z}}\text{ cut}}\quad$~\raisebox{5pt}{\begin{tikzpicture}[baseline=(current bounding box.center)]
    \begin{feynman}
      \vertex (i1);
      \vertex[blob, shape=ellipse, minimum size=1.2cm, fill=gray] at ($(i1) + (1cm, -1cm)$) (c1) {$\mathcal{A}_L$};
      \vertex at ($(c1) + (1.3cm, 0cm)$) (mid);
      \vertex[blob, shape=ellipse, minimum size=1.2cm, fill=gray] at ($(c1) + (2.5cm, 0cm)$) (c2) {$\mathcal{A}_R$};
      \vertex at ($(c2) + (1cm, 1cm)$)  (i2);
      \vertex at ($(c1) + (-1cm,-1cm)$)  (i3);
      \vertex at ($(c2) + (1cm, -1cm)$)  (i4);
      \vertex at ($(i1) + (0, -0.73cm)$) (u1);
      \vertex at ($(u1) + (0, -0.73cm)$) (d1);
      \vertex at ($(i2) + (0, -0.73cm)$) (u2);
      \vertex at ($(u2) + (0, -0.73cm)$) (d2);
      \diagram* {
        (i1)--(c1)--[hard,dashed, edge label=$\omega$](mid)--[hard, edge label=$\bomega$](c2)--(i2),(i3)--(c1),(c2)--(i4), (u1) -- [line cap=round, dash pattern=on 0pt off 7pt, line width=2pt, bend right=20] (d1),  (u2) -- [line cap=round, dash pattern=on 0pt off 7pt, line width=2pt, bend left=20] (d2)
      };
    \end{feynman}
  \end{tikzpicture}}
    \caption{Factorization of worldline amplitudes upon cutting the complex energies. The thick solid lines represent worldline fluctuations with real (thus vanishing) energy and the thick dashed lines represent worldline fluctuations with complexified energy. In all our diagrams, we assign the conjugate energy to the sub-amplitude on the right. }
    \label{fig:complexfactor}
\end{figure}
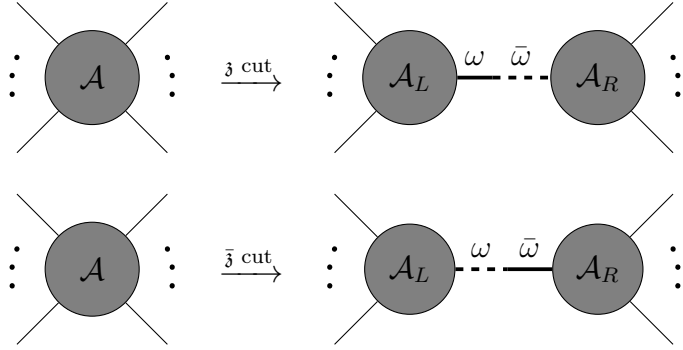

With these definitions, the complexified worldline cuts are imposed by requiring the analytic continuation of the integrand to factorize as:
\begin{subequations}
\begin{align}
  \lim_{\omega\to0}m\omega\bomega\mathcal{A}&=-i\sum_\text{pol.}\mathcal{A}_L(\mathfrak{z})\mathcal{A}_R(\bar{\mathfrak{z}})\mid_{\omega=0}=-i\sum_\text{pol.}\mathcal{A}_L(z)\mathcal{A}_R(\bar{\mathfrak{z}})\,,\label{eq:factorizationa}\\
\lim_{\bomega\to0}m\omega\bomega\mathcal{A}&=-i\sum_\text{pol.}\mathcal{A}_L(\mathfrak{z})\mathcal{A}_R(\bar{\mathfrak{z}})\mid_{\bomega=0}=-i\sum_\text{pol.}\mathcal{A}_L(\mathfrak{z})\mathcal{A}_R(z)\,,\label{eq:factorizationb}
\end{align}
\label{eq:factorization}
\end{subequations}
which is depicted in Fig.~\ref{fig:complexfactor}. This is the worldline analogue of imposing factorization on complex poles in ordinary generalized unitarity. The RHS of \Eq{eq:factorizationa} contains information about the residue in $\omega$ while that of \Eq{eq:factorizationb} the residue in $\bomega$; they both contain an overlapping piece which corresponds to the coefficient of the double pole in $\mathcal{A}$, namely, the terms that are $\mathcal{O}(\omega^0)$ and $\mathcal{O}(\bomega^0)$. The factorization of the double pole follows from the usual assumptions of generalized unitarity, namely, the analytic structure of the integrands. The factorization of the simple pole uses the same analytic continuation away from the physical cut, together with the choice of complexified energy flow in \Eq{eq:factorization}.
Naively, given an expression for $\mathcal{A}$ with real energy variables, one cannot arbitrarily choose to assign them to either $\omega$ or $\bomega$ and expect the factorization to hold. Instead, the complexification is chosen so that the analytic continuation of the integrand has the factorization property in \Eq{eq:factorization}. From the perspective of building higher order amplitudes from lower order ones (i.e., going from the RHS to the LHS of \Eq{eq:factorization}), this assignment is natural. Thus, we can use sub-amplitudes with complexified external worldline energies to reconstruct $\mathcal{A}$ up to contact terms.

 These generalized amplitudes with complexified kinematics are gauge-invariant up to linear order in $\omega$ or $\bomega$, because the replacement of $\varepsilon$ with $k$ generates terms of order $\omega^2$. That is, when imposing the Ward identity in any graviton leg $\varepsilon_i^{(\mu}\varepsilon_i^{\nu)}\to k_i^{(\mu}\varepsilon_i^{\nu)}$, we have
\begin{equation}
  \mathcal{A}\mid_{\varepsilon_i^{(\mu}\varepsilon_i^{\nu)}\to k_i^{(\mu}\varepsilon_i^{\nu)}}=\mathcal{O}(\omega^2)\,.
\end{equation}
This means that these generalized amplitudes contain gauge invariant physical information up to linear order in the complexified energies.

Ultimately, we consider these objects $\mathcal{A}$ as generalized amplitudes rather than off-shell currents because of their quasi-gauge-invariance. We will use this condition to constrain any contact terms which are not fixed by imposing the on-shell factorization of amplitudes.


\subsection{A soft theorem}
\label{sec:softth}
An additional useful property of these in-out complexified worldline amplitudes is that they satisfy the soft theorem:
\begin{align}
    \label{eq:softheoremb}
 A(h_1,\ldots,h_m, \mathfrak{z}_1,\ldots, \mathfrak{z}_{n+1}(\omega)) &= \zeta^\mu\frac{\partial }{\partial b^\mu}A(h_1,\ldots,h_m, \mathfrak{z}_1,\ldots, \mathfrak{z}_{n})  \\
&+ i \omega\, \zeta^\mu\frac{\partial }{\partial u^\mu}A(h_1,\ldots,h_m, \mathfrak{z}_1,\ldots, \mathfrak{z}_{n}) + {\cal O}(\omega^2) \,,\nonumber
\end{align}
where $\zeta$ is the polarization of the soft fluctuation and $\omega$ its frequency. This soft theorem captures both the leading (i.e., ${\cal O}(\omega^0)$) and subleading behavior (i.e., ${\cal O}(\omega)$) of the amplitude as $\omega\to 0$.

  The proof is analogous to that of the geometric soft theorems in Refs.~\cite{Cheung:2021yog,Derda:2024jvo}. We can think of the worldline fluctuation as the would-be Goldstone boson of the spontaneously broken symmetry of spacetime translations of the worldline as the result of choosing some background trajectory $\bar x^\mu(\tau) = b^\mu + u^\mu \tau$. More concretely, to prove the leading soft theorem, we note that the action $S$ is invariant under the spurionic transformation
\begin{equation}
    z^\mu\to z^\mu+a^\mu\,, \qquad b^\mu\to b^\mu-a^\mu\,.
\end{equation}
This implies the operator equation:
\begin{equation}
\label{eq:shiftward}
  \frac{\partial S}{\partial b^\mu}=  \frac{\delta S}{\delta z^\mu} = \partial_\tau\frac{\delta S}{\delta \dot{z}^\mu} \,,
\end{equation}
where in the second equality we have used the equation of motion. Evaluating the LHS of this relation in the Fourier domain on an on-shell state $|\alpha \rangle$ we find
\begin{align}
\label{eq:db}
  \zeta^\mu\langle0|\frac{\partial S}{\partial b^\mu}|\alpha\rangle= -i \zeta^\mu\frac{\partial }{\partial b^\mu} \langle0|\alpha\rangle=-i\zeta^\mu\frac{\partial}{\partial b^\mu} A(\alpha)\,.
\end{align}
On the RHS we have
\begin{align}
 \zeta^\mu \langle0|\partial_\tau\frac{\delta S}{\delta \dot{z}^\mu}(\omega)|\alpha\rangle= -\zeta^\mu\langle0|m \ddot{z}_\mu(\omega)+{\cal O}(z^2)|\alpha\rangle = \zeta^\mu m\omega \bomega\langle0| z_\mu(\omega)|\alpha\rangle + {\cal O}(z^2) \,.
\end{align}
The soft limit automatically performs the LSZ reduction\footnote{The ${\cal O}(z^2)$ are disconnected terms which do not contribute in the on-shell limit due to the amputation.}
\begin{equation}
 \lim_{\omega\to 0}\lim_{\bomega\to 0}   \zeta^\mu \langle0|\partial_\tau\frac{\delta S}{\delta \dot{z}^\mu}(\omega)|\alpha\rangle = -i \lim_{\omega\to 0}   A(\alpha, \mathfrak{z})\,,
\end{equation}
thus yielding the soft theorem above by equating with Eq.~\eqref{eq:db}.

Note that the soft limit is in fact the on-shell limit of the $z_{n+1}$ state with real kinematics $\lim_{\omega\to 0}   A(\alpha, \mathfrak{z}_{n+1})=A(\alpha,z)$, so the leading soft limit computes an impulse-like quantity. This theorem is therefore a generalization of the formula relating the impulse to the on-shell action.

The proof of the subleading soft theorem is analogous, and follows by noticing that the action also enjoys the spurionic symmetry
\begin{equation}
    z^\mu\to z^\mu+a^\mu \tau\,, \qquad u^\mu\to u^\mu - a^\mu\,.
\end{equation}
which in turn implies
\begin{equation}
  \frac{\partial S}{\partial u^\mu}=  \tau \frac{\delta S}{\delta z^\mu} = \tau \partial_\tau\frac{\delta S}{\delta \dot{z}^\mu} \,.
\end{equation}
After integrating along the worldline, the LHS is treated as above and simply gives the $u$ derivative of the lower-point amplitude; and the RHS in the Fourier domain and for soft frequency yields
\begin{align}
 \zeta^\mu \langle 0|\frac{\partial S}{\partial u^\mu}|\alpha\rangle=\zeta^\mu i\frac{\partial}{\partial\omega} \langle0|\partial_\tau\frac{\delta S}{\delta \dot{z}^\mu}(\omega)|\alpha\rangle= - \lim_{\omega\to 0} \frac{\partial}{\partial\omega}  A(\alpha, \mathfrak{z}) \,.
 \label{eq:subleadingsoft}
\end{align}

We will use the leading soft theorem to constrain local amplitudes in the next section. Indeed, in Refs.~\cite{Mogull:2020sak,Jakobsen:2021zvh} it was observed that the Feynman vertices in WQFT satisfy these relationships. This has been extended to the subleading order in Ref.~\cite{Haddad:2025cmw}. Note, however, that the soft theorem in Eq.~\eqref{eq:softheoremb} is more general, as it applies not only to local amplitudes but to the full complexified amplitudes. We  check this explicitly with  examples in the next section.

The soft theorem in Eq.~\eqref{eq:softheoremb} guarantees that  the complexified amplitudes are unambiguous and contain gauge-invariant physical information up to ${\cal O}(\omega)$, which indeed will be sufficient to fix the aforementioned single poles in integrands via unitarity. Thus we take Eq.~\eqref{eq:softheoremb} as a defining feature of this set of amplitudes.

\section{Bootstrapping rational worldline amplitudes}
\label{sec:rational}

In this section we construct the rational building blocks, which are analogous to tree-level amplitudes in QFT. In worldline field theory, rational amplitudes are those with one source. In this case, the integral over the momentum exchange with the worldline in Eq.~\eqref{eq:WQFTamp} can be trivially performed in terms of the total graviton momentum $k=\sum_i k_i$ yielding the following form
\begin{equation}
  \label{eq:WQFTamprat}
  A=e^{ik\cdot b} \mathcal{A}\,.
\end{equation}
The final amplitude is then rational up to the exponential factor $e^{ik\cdot b}$ dictated by translation symmetry.

We will now illustrate that the rational amplitudes are completely fixed by the properties of locality, unitarity, gauge invariance, and the \textit{leading} soft theorem. In practice we will show this by writing a local ansatz for the amplitudes as a sum over diagrams, which guarantees locality. The numerators of each diagram in the ansatz are polynomials in Lorentz products which satisfy the following properties

\begin{enumerate}

  \item \emph{Diagram symmetry:} Each numerator must be invariant under the symmetries (i.e., automorphisms) of the corresponding diagram. For contact terms this is simply Bose symmetry.

  \item \emph{Little group scaling:} Each term is bilinear in each graviton polarization $\varepsilon_i$, and linear in each worldline polarization $\zeta_i$.

  \item \emph{Power counting:} Each numerator term must contain exactly two factors of $u$ and/or $\omega$ per worldline vertex and two factors of graviton momenta $k_i$ per bulk graviton vertex. This ensures that the amplitudes we construct correspond to the minimal coupling of worldlines to gravity, and it could be relaxed to allow for non-minimal couplings.

\end{enumerate}

In addition, we will also impose the leading soft theorem in Eq.~\eqref{eq:softheoremb}, which for this class of amplitudes, due to \eqref{eq:WQFTamprat}, takes the simpler form
\begin{equation}
    \label{eq:softheorem}
 \lim_{\omega_{n+1}\to0}\mathcal{A}(h_1,\ldots,h_m, \mathfrak{z}_1,\ldots, \mathfrak{z}_{n+1})= i (\zeta_{n+1}\cdot k) \mathcal{A}(h_1,\ldots,h_m, \mathfrak{z}_1,\ldots, \mathfrak{z}_{n})\,,
  \end{equation}
  This encodes the intuitive statement that by conservation of momentum, the impulse exerted on a single worldline must be equal to the total momentum of the gravitational waves scattering against it.

  We will only use this relation as a bootstrap condition to fix local amplitudes.
  We will not make use of the subleading soft theorem in Eq.~\eqref{eq:softheoremb} here, but rather use it as a check\footnote{Naively, one may expect we can construct any amplitude by recursively applying the full soft theorem Eq.~\eqref{eq:softheoremb} to lower-point amplitudes. However, the derivative operator in the subleading soft theorem will in general obscure the locality structure (i.e. propagator structure) of the higher-point amplitudes thus constructed (see the end of Sec.~3.2.2). Thus, to construct manifestly local amplitudes, we only impose the leading soft theorem in Eq.~\eqref{eq:softheorem} as a bootstrap condition.}.

\subsection{Local building blocks}

Let us first explain how the local amplitudes (i.e. amplitudes with no poles) are fixed by the principles above. These will serve as building blocks to construct more complicated amplitudes.

\subsubsection*{Graviton three-point amplitude}
Given that graviton amplitudes enjoy the full Lorentz symmetry of the bulk and that physical polarizations are transverse traceless, the minimal basis of non-vanishing Lorentz invariants are $(k_i\cdot \varepsilon_j)$, and $(\varepsilon_i\cdot \varepsilon_j)$ where $i\neq j$. It is a classic exercise to check that the most general amplitude that satisfies the properties outlined above is
\begin{align}
    {\cal A}_{\rm bulk}(h_1,h_2,h_3)&=
\raisebox{2pt}{\begin{tikzpicture}[baseline=(current bounding box.center)]
    \begin{feynman}
      \vertex (i1);
      \vertex[blob, shape=circle, minimum size=0.4cm, fill=gray] at ($(i1) + (1cm, 0cm)$) (i2) {} ;
      \vertex at ($(i2) + (0.71cm, 0.71cm)$)  (i3);
      \vertex at ($(i2) + (0.71cm, -0.71cm)$)  (i4) ;
      \diagram* {
      (i1)--[graviton](i2)--[graviton](i3), (i2)--[graviton](i4)
      };
      \draw node[left] at (i1) {$h_1(k_1)$};
      \draw node[right] at (i3) {$h_2(k_2)$};
      \draw node[right] at (i4) {$h_3(k_3)$};
    \end{feynman}
    \end{tikzpicture}}\\
    &=a_1 \left((k_1\cdot\varepsilon_3)^2(\varepsilon_1\cdot\varepsilon_2)^2+(k_1\cdot\varepsilon_2)^2(\varepsilon_1\cdot\varepsilon_3)^2+(k_2\cdot\varepsilon_1)^2(\varepsilon_2\cdot\varepsilon_3)^2\right)\nonumber\\
    &\hspace{0.5cm}+ a_2 \big((k_1\cdot\varepsilon_2)(k_1\cdot\varepsilon_3)(\varepsilon_1\cdot\varepsilon_2)(\varepsilon_1\cdot\varepsilon_3) -(k_1\cdot\varepsilon_3)(k_2\cdot\varepsilon_1)(\varepsilon_1\cdot\varepsilon_2)(\varepsilon_1\cdot\varepsilon_3)\nonumber\\
    &\hspace{0.5cm}+(k_1\cdot\varepsilon_2)(k_2\cdot\varepsilon_1)(\varepsilon_1\cdot\varepsilon_3)(\varepsilon_2\cdot\varepsilon_3)\big)\,.\nonumber
\end{align}
Enforcing the Ward identity for any one of the gravitons, we get
\begin{equation}
    a_2=-2a_1.
\end{equation}
The entire amplitude is then fixed up to a single coupling constant, which we will identify with $-i\kappa$ for agreement with Einstein gravity:
\begin{equation}
    A(h_1,h_2,h_3)=-i\kappa\left((k_1\cdot\varepsilon_3)(\varepsilon_1\cdot\varepsilon_2)-(k_1\cdot\varepsilon_2)(\varepsilon_1\cdot\varepsilon_3)+(k_2\cdot\varepsilon_1)(\varepsilon_2\cdot\varepsilon_3)\right)^2.
\end{equation}

Given the graviton 3-point amplitude, higher order tree-level bulk graviton amplitudes are completely determined by locality, unitarity, and gauge invariance \cite{Arkani-Hamed:2016rak, Rodina:2016jyz} or by familiar QFT on-shell recursion relations \cite{Cachazo:2005ca}. In what follows, we take these tree-level pure-graviton amplitudes as known inputs to the worldline recursion relations.

\subsubsection*{Graviton one-point amplitude in the presence of worldline}

 On shell, the only non-trivial Lorentz invariant is $(u\cdot\varepsilon)$, so the only possibility is
\begin{equation}
\label{eq:1pt}
  \mathcal{A}(h_1)=
  \begin{tikzpicture}[baseline=(current bounding box.center)]
    \begin{feynman}
      \vertex (i1);
      \vertex[blob, shape=circle, minimum size=0.4cm, fill=gray] at ($(i1) + (1cm, 0cm)$) (i2) {} ;
      \vertex at ($(i2) + (1cm, 0cm)$)  (i3);
      \vertex at ($(i2) + (0cm, -1cm)$)  (i4) ;
      \diagram* {
      (i1)--[soft](i2)--[soft](i3), (i2)--[graviton](i4)
      };
      \draw node[left] at (i1) {$u$};
      \draw node[below] at (i4) {$h_1(k)$};
    \end{feynman}
    \end{tikzpicture}
  =\kappa'(u\cdot\varepsilon)^2.
\end{equation}
where $\kappa'$ is some coupling constant. We will later see that gauge invariance relates it to the gravitational constant by $\kappa'=\frac{m\kappa}{2}$.

\subsubsection*{One-graviton one-fluctuation amplitude}

  For an amplitude with external worldline fluctuation energy $\omega$, the minimal basis of Lorentz invariants are $(u\cdot\varepsilon)$, $\omega$,  $(u\cdot \zeta)$, $(k\cdot \zeta)$, and $(\varepsilon\cdot \zeta)$. Since its mass dimension is that of $\mathcal{A}(h_1)$ plus one, we can only write terms up to first order in $\omega$:
  \begin{equation}
    \mathcal{A}(h_1,\mathfrak{z}_1)=
    \begin{tikzpicture}[baseline=(current bounding box.center)]
    \begin{feynman}
      \vertex (i1);
      \vertex[blob, shape=circle, minimum size=0.4cm, fill=gray] at ($(i1) + (1cm, 0cm)$) (i2) {};
      \vertex at ($(i2) + (1cm, 0cm)$)  (i3);
      \vertex at ($(i2) + (0cm, -1cm)$)  (i4) ;
      \diagram* {
      (i1)--[soft](i2)--[ hard, dashed ](i3), (i2)--[graviton](i4)
      };
      \draw node[left] at (i1) {$u$};
      \draw node[below] at (i4) {$h_1(k)$};
      \draw node[right] at (i3) {$\mathfrak{z}_1(\omega)$};
    \end{feynman}
    \end{tikzpicture}
    =  a_1 (u\cdot \varepsilon)^2(k\cdot \zeta) + a_2\omega (u\cdot \varepsilon)(\varepsilon\cdot \zeta)
  \end{equation}
  where $a_1$ and $a_2$ are constants to be fixed. Enforcing the Ward identity (up to linear order in $\omega$) gives $a_2=2a_1$. Using the leading soft theorem relating this amplitude to $\mathcal{A}(h_1)$, we have
  \begin{equation}
    \mathcal{A}(h_1,\mathfrak{z}_1)= i\kappa' ((u\cdot \varepsilon)^2(k\cdot \zeta) + 2\omega (u\cdot \varepsilon)(\varepsilon\cdot \zeta)).
  \end{equation}
  It is easy to check that this amplitude also satisfies the subleading soft theorem.

\subsubsection*{One-graviton $n$-fluctuation amplitude}

The possible Lorentz invariants are $(u\cdot \varepsilon), \omega_i, (k\cdot \zeta_i), (\varepsilon\cdot \zeta_i)$, $(u\cdot \zeta_i)$, and $(\zeta_i\cdot \zeta_j)$ for $i\neq j$. By power counting, this amplitude has mass dimension $n$. By the leading soft theorem, any term that is zeroth order in $\omega_i$ must contain the factor $(k\cdot\zeta_i)$. Thus, one can check that the most general form satisfying this power counting and the Bose symmetry between worldline fluctuations is:
 \begin{align}
  \label{eq:hnz}
  &\mathcal{A}(h_1,\mathfrak{z}_1,\ldots,\mathfrak{z}_n) = \begin{tikzpicture}[baseline=(current bounding box.center)]
    \begin{feynman}
      \vertex (i1);
      \vertex[blob, shape=circle, minimum size=0.4cm, fill=gray] at ($(i1) + (1cm, 0cm)$) (i2) {} ;
      \vertex at ($(i2) + (1cm, 0cm)$)  (i3);
      \vertex at ($(i2) + (0cm, -1cm)$)  (i4) ;
      \vertex at ($(i2) + (1.2cm,1.2cm)$)  (i5);
      \diagram* {
      (i1)--[soft](i2)--[dashed, hard](i3), (i2)--[graviton](i4), (i2)--[ hard, bend left=40, dashed ](i5)
      };
      \draw node[left] at (i1) {$u$};
      \draw node[below] at (i4) {$h_1(k)$};
      \draw node[right] at (i5) {$\mathfrak{z}_1(\omega_1)$};
      \draw node[right] at (i3) {$\mathfrak{z}_n(\omega_n)$};
      \draw node[below] at (i5) {$\vdots$};
    \end{feynman}
    \end{tikzpicture}\\
  =&(i)^n\kappa' (u\cdot \varepsilon)^2 \prod_i (k\cdot \zeta_i)+a_1 \sum_i \left(\omega_i (\varepsilon\cdot \zeta_i)(u\cdot \varepsilon) \prod_{j\neq i}(k\cdot \zeta_j)\right) \nonumber\\
  &\hspace{3.95cm}+ a_2 \sum_{i< j}\left(\omega_i \omega_j (\varepsilon\cdot \zeta_i)(\varepsilon\cdot \zeta_j)\prod_{l\neq i,j}(k\cdot \zeta_l)\right). \nonumber
 \end{align}
The amplitude is automatically linear in each $\omega_i$ because terms proportional to $\omega_i^2$ cannot satisfy power counting and the soft theorem at the same time \footnote{ Without loss of generality, suppose there is some $\omega_1^2$ term in the amplitude in \Eq{eq:hnz}. This term must contain a factor of $(\varepsilon\cdot \zeta_i)$ for some $i\neq 1$. Consider then taking the soft limit of $\omega_i$, then this term will remain in the soft limit, since it is independent of $\omega_i$. However, this term is not proprotional to $(k\cdot \zeta_i)$, contradicting the leading soft theorem.}. Furthermore, enforcing the Ward identity imposes
\begin{equation}
  a_2=a_1=2 (i)^n \kappa'.
\end{equation}
The entire amplitude is then fixed up to an overall coupling constant $\kappa'$, and indeed it also satisfies the subleading soft theorem. This general result coincides with the gauge-invariant part of the WQFT Feynman vertices \cite{Mogull:2020sak}.

\subsection{Examples}
Having fixed the local amplitudes, we can now construct any rational amplitude by factorization.
We illustrate this with some examples. Although the procedure can be carried out in $D$ dimensions, in the following examples we work in $D=4$, for simplicity of the resulting formulae.
\subsubsection{Linear Compton scattering}
\label{sec:comptoncomp}
First we use the computation of leading-order linear Compton amplitude as an example to illustrate how to use the complexified worldline energies.

The Compton amplitude has two channels at leading order:
\begin{align}
  \label{eq:comptoncomplexexpand}
  \mathcal{A}_{\kappa^2}(h_1,h_2)&=\begin{tikzpicture}[baseline=(current bounding box.center)]
    \begin{feynman}
      \vertex (i1);
      \vertex at ($(i1) + (0.8cm, 0cm)$) (i2) ;
      \vertex at ($(i2) + (0.8cm, 0cm)$)  (i3);
      \vertex at ($(i2) + (0cm, -0.5cm)$)  (i4) ;
      \vertex at ($(i4) + (-0.5cm, -0.5cm)$)  (i5);
      \vertex at ($(i4) + (0.5cm, -0.5cm)$)  (i6);
      \diagram* {
      (i1)--[soft](i2)--[soft](i3), (i2)--[graviton](i4)--[graviton](i5),(i4)--[graviton](i6)
      };
      \draw node[left] at (i1) {$u$};
      \draw node[left] at (i5) {$1$};
      \draw node[right] at (i6) {$2$};
    \end{feynman}
    \end{tikzpicture}+\begin{tikzpicture}[baseline=(current bounding box.center)]
    \begin{feynman}
      \vertex (i1);
      \vertex at ($(i1) + (0.5cm, 0cm)$) (i2);
      \vertex at ($(i2) + (1.5cm, 0cm)$)  (i3) ;
      \vertex at ($(i3) + (0.5cm, 0cm)$)  (i4);
      \vertex at ($(i2) + (0cm, -1cm)$)  (i5);
      \vertex at ($(i3) + (0cm, -1cm)$)  (i6);
      \diagram* {
        (i1)--[soft](i2)--[hard, edge label=\(\omega\), dashed](i3)--[soft](i4), (i2)--[graviton](i5), (i3)--[graviton](i6)
      };
      \draw node[left] at (i1) {$u$};
      \draw node[left] at (i5) {$1$};
      \draw node[right] at (i6) {$2$};
    \end{feynman}
  \end{tikzpicture}\nonumber\\
   &=\frac{N_1}{2(k_1\cdot k_2)}+\frac{N_2}{m\bomega\omega}
\end{align}
where $N_1$ and $N_2$ correspond to the kinematic numerators to be determined.  We work with two different basis of Lorentz invariants in the two channels:
\begin{align}
  \text{LI}_1&=\{(u\cdot k_1), (k_1\cdot k_2), (k_1\cdot\varepsilon_2),(k_2\cdot\varepsilon_1),(u\cdot\varepsilon_1), (u\cdot\varepsilon_2),(\varepsilon_1\cdot\varepsilon_2)\},\allowdisplaybreaks\\
  \text{LI}_2&=\{ \omega, \bomega, (k_1\cdot k_2), (k_1\cdot\varepsilon_2),(k_2\cdot\varepsilon_1),(u\cdot\varepsilon_1), (u\cdot\varepsilon_2),(\varepsilon_1\cdot\varepsilon_2)\},
\end{align}
 In general, whenever there is an internal worldline fluctuation, we work in a basis involving $\omega$ and $\bomega$'s instead of $(u\cdot k_i)$'s to make the complexification of worldline energies manifest.\footnote{One needs to be careful about energy conservation when reducing $(u\cdot k_i)$ to $\omega$ or $\bomega$. Since $\omega$ and $\bomega$ are not identified, we cannot use overall energy conservation to convert $(u\cdot k_2)$ to $-(u\cdot k_1)$ and in turn $\omega$; the energy conservation in cut${}_2$ reduces $(u\cdot k_1)$ to $-\omega$ and $(u\cdot k_2)$ to $\bomega$.} However, when merging the cuts, we will convert $\omega$ and $\bomega$'s to $(u\cdot k_i)$'s to express everything in terms of external kinematics.

When constructing the ansatze for $N_1$ and $N_2$, we follow the rules described before. After taking into account the symmetries of the topologies (i.e. symmetric under $1\leftrightarrow2$ and $\omega\leftrightarrow -\bomega)$, the result is:
\begin{subequations}
\begin{align}
  N_1&=a_{1,1} (u\cdot k_1)^2 (\varepsilon_1 \cdot \varepsilon_2)^2 + a_{1,2} (u\cdot k_1)\left[  (u \cdot \varepsilon_1)(k_1 \cdot \varepsilon_2)(\varepsilon_1 \cdot \varepsilon_2)
  - (u \cdot \varepsilon_2)(k_2 \cdot \varepsilon_1)(\varepsilon_1 \cdot \varepsilon_2) \right] \nonumber \\
  & + a_{1,3} \left[ (u \cdot \varepsilon_1)^2 (k_1 \cdot \varepsilon_2)^2
  + (u \cdot \varepsilon_2)^2 (k_2 \cdot \varepsilon_1)^2 \right] + a_{1,4} (u \cdot \varepsilon_1)(u \cdot \varepsilon_2)(k_1 \cdot k_2)(\varepsilon_1 \cdot \varepsilon_2) \nonumber \\
  & + a_{1,5} (u \cdot \varepsilon_1)(u \cdot \varepsilon_2)(k_1 \cdot \varepsilon_2)(k_2 \cdot \varepsilon_1) + a_{1,6} (k_1 \cdot k_2)(\varepsilon_1 \cdot \varepsilon_2)^2 \nonumber\\
  &+ a_{1,7} (k_1 \cdot \varepsilon_2)(k_2 \cdot \varepsilon_1)(\varepsilon_1 \cdot \varepsilon_2)\allowdisplaybreaks \\
   N_2&= a_{2,1} \omega \bomega (u \cdot \varepsilon_1)(u \cdot \varepsilon_2)(\varepsilon_1 \cdot \varepsilon_2) + a_{2,2} \, \omega \bomega (\varepsilon_1 \cdot \varepsilon_2)^2 \nonumber \\
   & + a_{2,3} \left[ \omega (u \cdot \varepsilon_1)^2 (u \cdot \varepsilon_2)(k_1 \cdot \varepsilon_2)
   - \bomega (u \cdot \varepsilon_1)(u \cdot \varepsilon_2)^2 (k_2 \cdot \varepsilon_1) \right] \nonumber \\
   & + a_{2,4} \left[ - \bomega (u \cdot \varepsilon_1)^2 (u \cdot \varepsilon_2)(k_1 \cdot \varepsilon_2)
   + \omega (u \cdot \varepsilon_1)(u \cdot \varepsilon_2)^2 (k_2 \cdot \varepsilon_1) \right] \nonumber \\
   & + a_{2,5} \left[ \omega (u \cdot \varepsilon_1)(k_1 \cdot \varepsilon_2)(\varepsilon_1 \cdot \varepsilon_2)
   - \bomega (u \cdot \varepsilon_2)(k_2 \cdot \varepsilon_1)(\varepsilon_1 \cdot \varepsilon_2) \right] \nonumber \\
   & + a_{2,6} \left[ - \bomega (u \cdot \varepsilon_1)(k_1 \cdot \varepsilon_2)(\varepsilon_1 \cdot \varepsilon_2)
   + \omega (u \cdot \varepsilon_2)(k_2 \cdot \varepsilon_1)(\varepsilon_1 \cdot \varepsilon_2) \right] \nonumber\\
   &+ a_{2,7} (u \cdot \varepsilon_1)^2 (u \cdot \varepsilon_2)^2 (k_1 \cdot k_2) + a_{2,8} \left[ (u \cdot \varepsilon_1)^2 (k_1 \cdot \varepsilon_2)^2
   + (u \cdot \varepsilon_2)^2 (k_2 \cdot \varepsilon_1)^2 \right] \nonumber \\
   & + a_{2,9} (u \cdot \varepsilon_1)(u \cdot \varepsilon_2)(k_1 \cdot k_2)(\varepsilon_1 \cdot \varepsilon_2) + a_{2,10} (u \cdot \varepsilon_1)(u \cdot \varepsilon_2)(k_1 \cdot \varepsilon_2)(k_2 \cdot \varepsilon_1) \nonumber \\
   & + a_{2,11}  (k_1 \cdot k_2)(\varepsilon_1 \cdot \varepsilon_2)^2  + a_{2,12}  (k_1 \cdot \varepsilon_2)(k_2 \cdot \varepsilon_1)(\varepsilon_1 \cdot \varepsilon_2)
\end{align}
\end{subequations}

Next, we determine the free parameters in the ansatze for the numerators by requiring factorization, i.e., by imposing various cuts. Fixing $N_1$ is straightforward: we require that the ansatz agrees with the gluing of one- and three-graviton amplitudes in the limit where the intermediate graviton goes on-shell $(k_1+k_2)^2 = 2 k_1\cdot k_2 = 0$,
\begin{equation}
  \label{eq:cut1}
    N_1|_{(k_1\cdot k_2)=0} = \begin{tikzpicture}[baseline=(current bounding box.center)]
    \begin{feynman}
      \vertex (i1) ;
      \vertex[blob, shape=circle, minimum size=0.3cm, fill=gray] at ($(i1) + (0.8cm, 0cm)$) (i2) {};
      \vertex at ($(i2) + (0.8cm, 0cm)$)  (i3);
      \vertex[blob, shape=circle, minimum size=0.3cm, fill=gray] at ($(i2) + (0cm, -0.8cm)$)  (i4)  {};
      \vertex at ($(i4) + (-0.5cm, -0.5cm)$)  (i5);
      \vertex at ($(i4) + (0.5cm, -0.5cm)$)  (i6);
      \diagram* {
      (i1)--[soft](i2)--[soft](i3), (i2)--[graviton](i4)--[graviton](i5),(i4)--[graviton](i6)
      };
      \draw node[left] at (i1) {$u$};
      \draw node[left] at (i5) {$1$};
      \draw node[right] at (i6) {$2$};
    \end{feynman}
    \end{tikzpicture}.
\end{equation}
More explicitly, the cut equation \Eq{eq:cut1} reads:
\begin{equation}
  \label{eq:cut1explicit}
  \begin{aligned}
    &a_{1,1} (u\cdot k_1)^2 (\varepsilon_1 \cdot \varepsilon_2)^2 + a_{1,2} (u\cdot k_1)\left[(u \cdot \varepsilon_1)(k_1 \cdot \varepsilon_2)(\varepsilon_1 \cdot \varepsilon_2)
    - (u \cdot \varepsilon_2)(k_2 \cdot \varepsilon_1)(\varepsilon_1 \cdot \varepsilon_2) \right]\\
    &+ a_{1,3} \left[ (u \cdot \varepsilon_1)^2 (k_1 \cdot \varepsilon_2)^2
    + (u \cdot \varepsilon_2)^2 (k_2 \cdot \varepsilon_1)^2 \right]+ a_{1,5} (u \cdot \varepsilon_1)(u \cdot \varepsilon_2)(k_1 \cdot \varepsilon_2)(k_2 \cdot \varepsilon_1)\\
    &+ a_{1,7} (k_1 \cdot \varepsilon_2)(k_2 \cdot \varepsilon_1)(\varepsilon_1 \cdot \varepsilon_2)\\
  &= -\frac{i \kappa \kappa'}{2}\left[-(u \cdot \varepsilon_1)(k_1 \cdot \varepsilon_2) + (u \cdot \varepsilon_2)(k_2 \cdot \varepsilon_1) + (u\cdot k_1)(\varepsilon_1 \cdot \varepsilon_2)\right]^2.
  \end{aligned}
\end{equation}
After solving \Eq{eq:cut1explicit}, the ansatz reduces to:
\begin{equation}
\begin{aligned}
  N_1&=-\frac{i \kappa \kappa' }{2}  \left[ -(u \cdot \varepsilon_1)(k_1 \cdot \varepsilon_2) + (u \cdot \varepsilon_2)(k_2 \cdot \varepsilon_1) + \omega(\varepsilon_1 \cdot \varepsilon_2) \right]^2 \\
  & +a_{1,4}(k_1 \cdot k_2)(\varepsilon_1 \cdot \varepsilon_2)(u \cdot \varepsilon_1)(u \cdot \varepsilon_2)+a_{1,6}(k_1 \cdot k_2)(\varepsilon_1 \cdot \varepsilon_2)^2
\end{aligned}
\end{equation}
The only undetermined terms are those that vanish on the cut. To fix $N_2$, we need to reconstruct the worldline cut by summing over physical states of the internal worldline fluctuation. We first consider the cut in $\omega$ :
\begin{equation}
  \label{eq:eqsimplepolecom}
  N_2|_{\omega=0}=\begin{tikzpicture}[baseline=(current bounding box.center)]
    \begin{feynman}
      \vertex (i1);
      \vertex[blob, shape=circle, minimum size=0.3cm, fill=gray] at ($(i1) + (0.5cm, 0cm)$) (i2) {};
      \vertex at ($(i2) + (0.75cm, 0cm)$)  (i3a) ;
      \vertex[blob, shape=circle, minimum size=0.3cm, fill=gray] at ($(i3a) + (0.75cm, 0cm)$)  (i3) {};
      \vertex at ($(i3) + (0.5cm, 0cm)$)  (i4);
      \vertex at ($(i2) + (0cm, -1cm)$)  (i5);
      \vertex at ($(i3) + (0cm, -1cm)$)  (i6);
      \diagram* {
        (i1)--[soft](i2)--[hard](i3a)--[hard, dashed](i3)--[soft](i4), (i2)--[graviton](i5), (i3)--[graviton](i6)
      };
      \draw node[left] at (i1) {$u$};
      \draw node[left] at (i5) {$1$};
      \draw node[right] at (i6) {$2$};
      \draw node[above] at (i3a) {$\omega$};
    \end{feynman}
  \end{tikzpicture}.
\end{equation}
In our diagrammatic convention, the energy always flows from $\omega$ to $\bomega$, i.e., from left to right. The solid black line on the left represents a cut on the worldline with energy $\omega$; had it been on the right, it would represent a cut on the worldline with energy $\bomega$ (see later examples).
Explicitly, the cut equation \Eq{eq:eqsimplepolecom} reads:
\begin{equation}
  \label{eq:eqsimplepolecomexplicit}
  \begin{aligned}
   & - a_{2,3}\bomega (u \cdot \varepsilon_1)(u \cdot \varepsilon_2)^2 (k_2 \cdot \varepsilon_1)- a_{2,4}  \bomega (u \cdot \varepsilon_1)^2 (u \cdot \varepsilon_2)(k_1 \cdot \varepsilon_2) \\
   & - a_{2,5} \bomega (u \cdot \varepsilon_2)(k_2 \cdot \varepsilon_1)(\varepsilon_1 \cdot \varepsilon_2) - a_{2,6} \bomega (u \cdot \varepsilon_1)(k_1 \cdot \varepsilon_2)(\varepsilon_1 \cdot \varepsilon_2)\\
   &+ a_{2,7} (u \cdot \varepsilon_1)^2 (u \cdot \varepsilon_2)^2 (k_1 \cdot k_2) + a_{2,8} \left[ (u \cdot \varepsilon_1)^2 (k_1 \cdot \varepsilon_2)^2
   + (u \cdot \varepsilon_2)^2 (k_2 \cdot \varepsilon_1)^2 \right] \\
   & + a_{2,9} (u \cdot \varepsilon_1)(u \cdot \varepsilon_2)(k_1 \cdot k_2)(\varepsilon_1 \cdot \varepsilon_2) + a_{2,10} (u \cdot \varepsilon_1)(u \cdot \varepsilon_2)(k_1 \cdot \varepsilon_2)(k_2 \cdot \varepsilon_1)  \\
   & + a_{2,11}  (k_1 \cdot k_2)(\varepsilon_1 \cdot \varepsilon_2)^2  + a_{2,12}  (k_1 \cdot \varepsilon_2)(k_2 \cdot \varepsilon_1)(\varepsilon_1 \cdot \varepsilon_2)\\
   &=\frac{i \kappa'^2}{4}(u\cdot \varepsilon_1)^2 (u \cdot \varepsilon_2)[-(u\cdot \varepsilon_2)(k_1\cdot k_2)+2\bomega (k_1\cdot \varepsilon_2)]
  \end{aligned}
\end{equation}
Solving \Eq{eq:eqsimplepolecomexplicit} gives
\begin{equation}
  \begin{aligned}
  N_2=&-\frac{i  \kappa'^2}{4}  (u \cdot \varepsilon_1)^2 (u \cdot \varepsilon_2)^2 (k_1 \cdot k_2) +\frac{i \kappa'^2}{2}\bomega (u\cdot \varepsilon_1)^2(u\cdot \varepsilon_2)(k_1\cdot\varepsilon_2)-\frac{i \kappa'^2}{2}\omega (u\cdot \varepsilon_1)(u\cdot \varepsilon_2)^2(k_2\cdot\varepsilon_1) \\
  &+a_{2,1} \, \omega \bomega (u \cdot \varepsilon_1)(u \cdot \varepsilon_2)(\varepsilon_1 \cdot \varepsilon_2)+ a_{2,2} \, \omega \bomega (\varepsilon_1 \cdot \varepsilon_2)^2
  \end{aligned}
\end{equation}
Due to the symmetry between $\omega$ and $\bomega$, fixing the cut in $\omega$ also fixes the cut in $\bomega$, so the remaining undetermined terms correspond to the purely contact terms. These can be fixed by the gauge invariance of the amplitude, enforced by the Ward identity.\footnote{When merging the cuts, we need to use energy conservation to convert $\omega$ and $\bomega$ to $-(u\cdot k_1)$ and $(u\cdot k_2)$, respectively, in order to work in the basis LI$_1$.} In fact, we must also have $\kappa'=\frac{m\kappa}{2} $ in order to satisfy the Ward identity. Finally, setting $\omega=\bomega$ gives us the final result for the amplitude
\begin{align}
\label{eq:comptonres}
  \mathcal{A}(h_1,h_2)|_{\kappa^2}=
  &\tfrac{-i m\kappa^2}{4}\left(\tfrac{\left((k_1\cdot u)(\varepsilon_1\cdot\varepsilon_2)+(k_2\cdot \varepsilon_1)(u\cdot\varepsilon_2)-(k_1\cdot\varepsilon_2)(u\cdot\varepsilon_1)\right)^2}{(k_1\cdot k_2)}+\tfrac{(k_1\cdot k_2)(u\cdot \varepsilon_1)^2(u\cdot\varepsilon_2)^2}{(u\cdot k_1)^2}\right.\nonumber\\
  &\left.+\tfrac{2(u\cdot \varepsilon_1)(u\cdot \varepsilon_2)((k_1\cdot\varepsilon_2)(u\cdot\varepsilon_1)-(k_2\cdot \varepsilon_1)(u\cdot\varepsilon_2))}{(u\cdot k_1)}-2(u\cdot\varepsilon_1)(u\cdot \varepsilon_2)(\varepsilon_1\cdot\varepsilon_2)\right).
  \end{align}
The fact that all parameters in the ansatz are fixed or cancel in the sum over diagrams is a non-trivial consistency check of our procedure. Furthermore, this agrees with the existing result for this amplitude \cite{Ivanov:2026icp, Bjerrum-Bohr:2025bqg}.

  \subsubsection{Impulse exerted by a gravitational wave}
  Next we wish to use the tree-level amplitude $\mathcal{A}(h_1,h_2,\mathfrak{z})$ to illustrate fixing amplitudes with external worldline fluctuations.
  This amplitude is composed of the following diagrams:
  \begin{align}
    \mathcal{A}(h_1,h_2,\mathfrak{z})|_{\kappa^2}=&\begin{tikzpicture}[baseline=(current bounding box.center)]
    \begin{feynman}
      \vertex (i1);
      \vertex at ($(i1) + (0.8cm, 0cm)$) (i2) ;
      \vertex at ($(i2) + (0.8cm, 0cm)$)  (i3);
      \vertex at ($(i2) + (0cm, -0.5cm)$)  (i4) ;
      \vertex at ($(i4) + (-0.5cm, -0.5cm)$)  (i5);
      \vertex at ($(i4) + (0.5cm, -0.5cm)$)  (i6);
      \diagram* {
      (i1)--[soft](i2)--[hard, dashed](i3), (i2)--[graviton](i4)--[graviton](i5),(i4)--[graviton](i6)
      };
      \draw node[left] at (i1) {$u$};
      \draw node[left] at (i5) {$1$};
      \draw node[right] at (i6) {$2$};
      \draw node[right] at (i3) {$\mathfrak{z}(\omega_1)$};
    \end{feynman}
 \end{tikzpicture}+\begin{tikzpicture}[baseline=(current bounding box.center)]
    \begin{feynman}
      \vertex (i1);
      \vertex at ($(i1) + (0.5cm, 0cm)$) (i2);
      \vertex at ($(i2) + (1.2cm, 0cm)$)  (i3);
      \vertex at ($(i3) + (0.7cm, 0cm)$)  (i4);
      \vertex at ($(i2) + (0cm, -1cm)$)  (i5);
      \vertex at ($(i3) + (0cm, -1cm)$)  (i6);
      \diagram* {
        (i1)--[soft](i2)--[edge label=$\omega_2$, hard, dashed](i3)--[hard, dashed](i4), (i2)--[graviton](i5), (i3)--[graviton](i6)
      };
      \draw node[left] at (i1) {$u$};
      \draw node[left] at (i5) {$1$};
      \draw node[right] at (i6) {$2$};
      \draw node[right] at (i4) {$\mathfrak{z}(\omega_1)$};
    \end{feynman}
\end{tikzpicture}+\begin{tikzpicture}[baseline=(current bounding box.center)]
    \begin{feynman}
      \vertex (i1);
      \vertex at ($(i1) + (0.5cm, 0cm)$) (i2);
      \vertex at ($(i2) + (1.2cm, 0cm)$)  (i3);
      \vertex at ($(i3) + (0.7cm, 0cm)$)  (i4);
      \vertex at ($(i2) + (0cm, -1cm)$)  (i5);
      \vertex at ($(i3) + (0cm, -1cm)$)  (i6);
      \vertex at ($(i2) + (0.6cm, -0.5cm)$)  (int) {};
      \diagram* {
        (i1)--[soft](i2)--[edge label=$\omega_2$, hard, dashed](i3)--[hard, dashed](i4), (i2)--[graviton](int)--[graviton](i6), (i3)--[graviton](i5)
      };
      \draw node[left] at (i1) {$u$};
      \draw node[left] at (i5) {$1$};
      \draw node[right] at (i6) {$2$};
    \draw node[right] at (i4) {$\mathfrak{z}(\omega_1)$};
    \end{feynman}
  \end{tikzpicture}\nonumber\\
  =&\frac{N_1}{2(k_1\cdot k_2)}+ \frac{N_2}{m\bomega_2\omega_2}+\frac{N_3}{m\bomega_2\omega_2}\,,
    \label{eq:A21}
  \end{align}
where the last two channels are simply related by relabeling of the gravitons
\begin{equation}
    N_3= N_2\mid_{k_1\leftrightarrow k_2, \varepsilon_1\leftrightarrow\varepsilon_2}\,.
\end{equation}
The internal fluctuation energy flows from $\omega_2$ to $\bomega_2$. If we are only interested in the physical observable related to this amplitude, then we can set $\omega_1=0$ and the amplitude can be obtained conveniently by the soft theorem:
\begin{equation}
  \mathcal{A}(h_1,h_2,z)\mid_{\omega_1=0}=i((k_1+k_2)\cdot \zeta)\mathcal{A}(h_1,h_2).
\end{equation}
However, if we want to use this amplitude in intermediate steps for higher order calculations, we also need the $\mathcal{O}(\omega_1)$ part of the amplitude.
Similar to before, we work with two sets of Lorentz invariants to construct the ansatze for the numerators of the diagrams
\begin{subequations}
  \begin{align}
    \text{LI}_1&=\{ \omega_1, (u\cdot k_1),(u\cdot \zeta),  (u\cdot\varepsilon_1),(u \cdot \varepsilon_2), (k_1\cdot \varepsilon_2), (k_2\cdot \varepsilon_1), (k_1\cdot \zeta), (k_2\cdot \zeta),(\varepsilon_1\cdot \zeta),(\varepsilon_2\cdot \zeta), \nonumber\\ &\hspace{12cm}(k_1\cdot k_2),(\varepsilon_1\cdot \varepsilon_2)\}\\
    \text{LI}_2&=\{ \omega_1, \omega_2, \bomega_2,  (u\cdot \zeta), (u\cdot\varepsilon_1),(u \cdot \varepsilon_2), (k_1\cdot \varepsilon_2), (k_2\cdot \varepsilon_1), (k_1\cdot \zeta), (k_2\cdot \zeta),(\varepsilon_1\cdot \zeta),(\varepsilon_2\cdot \zeta),\nonumber\\ &\hspace{12cm} (k_1\cdot k_2), (\varepsilon_1\cdot \varepsilon_2)\}
  \end{align}
\end{subequations}
We make ansatze for $N_1$ and $N_2$, taking into account the symmetries of $N_1$\footnote{When implementing the symmetries, we need to keep working at linear order in $\omega_1$.}
\begin{subequations}
\begin{align}
&\begin{aligned}
N_1=&\,a_{1,1} \omega_1 (k_1\cdot k_2) (\varepsilon_1\cdot\varepsilon_2)\left[(u\cdot\varepsilon_2)(\varepsilon_1\cdot \zeta)+(u\cdot\varepsilon_1)(\varepsilon_2\cdot \zeta)\right]\\
+& a_{1,2} \omega_1 (\varepsilon_1\cdot\varepsilon_2)\left[(u\cdot\varepsilon_1)(k_1\cdot \varepsilon_2)(k_1\cdot \zeta)+(u\cdot\varepsilon_2)(k_2\cdot \varepsilon_1)(k_2\cdot \zeta)\right]+\cdots\\
+& a_{1,53} (u\cdot k_1) (\varepsilon_1\cdot\varepsilon_2)\left[(u\cdot k_1)(k_1\cdot\varepsilon_2)(\varepsilon_1\cdot\zeta)+(\varepsilon_2\cdot\zeta)(k_2\cdot \varepsilon_1)\left(2\omega_1+(u\cdot k_1)\right)\right]\\
+& a_{1,55}(u\cdot u)(k_1\cdot\varepsilon_2)(k_2\cdot\varepsilon_1)\left[(k_1\cdot \varepsilon_2)(\varepsilon_1\cdot \zeta)+(k_2\cdot \varepsilon_1)(\varepsilon_2\cdot \zeta)\right]
\end{aligned}\allowdisplaybreaks\\
    &\begin{aligned}
      N_2=& \,a_{2,1} \omega_1 (u\cdot\varepsilon_1)^2 (u\cdot\varepsilon_2) (k_1\cdot k_2) (z_1\cdot\varepsilon_2) +a_{2,2} \omega_1 (u\cdot\varepsilon_1)^2 (u\cdot\varepsilon_2) (k_1\cdot z_1) (k_1\cdot\varepsilon_2) \allowdisplaybreaks\\+& a_{2,3} \omega_1 (u\cdot\varepsilon_1)^2 (u\cdot\varepsilon_2) (k_1\cdot\varepsilon_2) (k_2\cdot z_1) +
a_{2,4} \omega_1 \omega_2 (u\cdot\varepsilon_1)^2 (k_1\cdot\varepsilon_2) (z_1\cdot\varepsilon_2) +\cdots\\ +& a_{2,99} \omega_2 \bomega_2 (u\cdot u) (k_1\cdot\varepsilon_2) (z_1\cdot\varepsilon_1) (\varepsilon_1\cdot\varepsilon_2)+
a_{2,100} \omega_2 \bomega_2 (u\cdot u) (k_2\cdot\varepsilon_1) (z_1\cdot\varepsilon_2) (\varepsilon_1\cdot\varepsilon_2) \allowdisplaybreaks \\ +&
a_{2,101} (u\cdot u)^2 (k_1\cdot\varepsilon_2)^2 (k_2\cdot\varepsilon_1) (z_1\cdot\varepsilon_1)+
a_{2,102} (u\cdot u)^2 (k_1\cdot\varepsilon_2) (k_2\cdot\varepsilon_1)^2 (z_1\cdot\varepsilon_2)
    \end{aligned}
    \end{align}
  \end{subequations}
    We only present explicitly the first and last few terms of the ansatze, as they are rather lengthy.

We first constrain $N_1$ on the cut by requiring
\begin{equation}
  \label{eq:cutN1}
    N_1|_{(k_1\cdot k_2)=0} = \begin{tikzpicture}[baseline=(current bounding box.center)]
    \begin{feynman}
      \vertex (i1);
      \vertex[blob, shape=circle, minimum size=0.3cm, fill=gray] at ($(i1) + (0.8cm, 0cm)$) (i2) {};
      \vertex at ($(i2) + (0.8cm, 0cm)$)  (i3);
      \vertex[blob, shape=circle, minimum size=0.3cm, fill=gray] at ($(i2) + (0cm, -0.8cm)$)  (i4) {};
      \vertex at ($(i4) + (-0.5cm, -0.5cm)$)  (i5);
      \vertex at ($(i4) + (0.5cm, -0.5cm)$)  (i6);
      \diagram* {
      (i1)--[soft](i2)--[hard, dashed](i3), (i2)--[graviton](i4)--[graviton](i5),(i4)--[graviton](i6)
      };
      \draw node[left] at (i1) {$u$};
      \draw node[left] at (i5) {$1$};
      \draw node[right] at (i6) {$2$};
      \draw node[right] at (i3) {$\mathfrak{z}(\omega_1)$};
    \end{feynman}
    \end{tikzpicture}+\mathcal{O}(\omega_1^2).
\end{equation}
In practice, the RHS is computed by summing over the graviton polarizations then taking the series expansion up to linear order in $\omega_1$. The RHS is a gauge invariant expression independent of the reference null momentum $q$. Explicitly,
\begin{align}
  \begin{aligned}
    \text{RHS}=&\frac{m \kappa^2}{2} \Big[
 (u \cdot \varepsilon_1)(k_1 \cdot \varepsilon_2)
- (u \cdot \varepsilon_2)(k_2 \cdot \varepsilon_1)
- (u \cdot k_1)(\varepsilon_1 \cdot \varepsilon_2)
\Big] \times \\
&\quad\Big[
  (u \cdot \varepsilon_1)(k_1 \cdot \varepsilon_2) \big( (k_1 \cdot \zeta) + (k_2 \cdot \zeta) \big)
  + 2\omega_1 (k_1 \cdot \varepsilon_2)(\zeta \cdot \varepsilon_1) \\
  &\quad - (k_2 \cdot \varepsilon_1)\Big(
      (u \cdot \varepsilon_2)\big((k_1 \cdot \zeta) + (k_2 \cdot \zeta)\big)
      + 2\omega_1 (\zeta \cdot \varepsilon_2)
    \Big) \\
&\quad - \Big(2\omega_1 (k_1 \cdot \zeta) + (u \cdot k_1)\big((k_1 \cdot \zeta) + (k_2 \cdot \zeta)\big) \Big)
         (\varepsilon_1 \cdot \varepsilon_2)
\Big]+\mathcal{O}(\omega_1^2).
  \end{aligned}
\end{align}
As in the previous example this fixes the result for this numerator up to contact terms.

For $N_2$, we first consider the cut in $\omega_2$:
\begin{equation}
  \label{eq:cutw2}
  N_2|_{\omega_2=0} = \begin{tikzpicture}[baseline=(current bounding box.center)]
    \begin{feynman}
      \vertex (i1);
      \vertex[blob, shape=circle, minimum size=0.3cm, fill=gray] at ($(i1) + (0.5cm, 0cm)$) (i2) {};
      \vertex at ($(i2) + (1cm, 0cm)$)  (i3a);
      \vertex at ($(i3a) + (0cm, 0cm)$)  (i3b);
      \vertex[blob, shape=circle, minimum size=0.3cm, fill=gray] at ($(i3b) + (1cm, 0cm)$)  (i3) {};
      \vertex at ($(i3) + (0.7cm, 0cm)$)  (i4);
      \vertex at ($(i2) + (0cm, -1cm)$)  (i5);
      \vertex at ($(i3) + (0cm, -1cm)$)  (i6);
      \diagram* {
        (i1)--[soft](i2)--[hard](i3a), (i3b)--[hard, dashed](i3)--[hard, dashed](i4), (i2)--[graviton](i5), (i3)--[graviton](i6)
      };
      \draw node[left] at (i1) {$u$};
      \draw node[left] at (i5) {$1$};
      \draw node[right] at (i6) {$2$};
    \draw node[above] at (i3a) {$\omega_2$};
    \draw node[right] at (i4) {$\mathfrak{z}(\omega_1)$};
    \end{feynman}
  \end{tikzpicture}. .
\end{equation}
Here, there is no need to keep track of powers of $\omega_1$, because the RHS is automatically $\mathcal{O}(\omega_1)$. Similarly, we can constrain $N_2$ by imposing the $\bomega_2$-cut:
\begin{equation}
  \label{eq:cutbw2}
  N_2|_{\bomega_2=0} = \begin{tikzpicture}[baseline=(current bounding box.center)]
    \begin{feynman}
      \vertex (i1);
      \vertex[blob, shape=circle, minimum size=0.3cm, fill=gray] at ($(i1) + (0.5cm, 0cm)$) (i2) {};
      \vertex at ($(i2) + (1cm, 0cm)$)  (i3a);
      \vertex at ($(i3a) + (0cm, 0cm)$)  (i3b);
      \vertex[blob, shape=circle, minimum size=0.3cm, fill=gray] at ($(i3b) + (1cm, 0cm)$)  (i3) {};
      \vertex at ($(i3) + (0.7cm, 0cm)$)  (i4);
      \vertex at ($(i2) + (0cm, -1cm)$)  (i5);
      \vertex at ($(i3) + (0cm, -1cm)$)  (i6);
      \diagram* {
        (i1)--[soft](i2)--[hard, dashed](i3a), (i3b)--[hard](i3)--[hard, dashed](i4), (i2)--[graviton](i5), (i3)--[graviton](i6)
      };
      \draw node[left] at (i1) {$u$};
      \draw node[left] at (i5) {$1$};
      \draw node[right] at (i6) {$2$};
    \draw node[above] at (i3a) {$\omega_2$};
    \draw node[right] at (i4) {$\mathfrak{z}(\omega_1)$};
    \end{feynman}
  \end{tikzpicture}.
\end{equation}

For the contact terms, we substitute the ansatze $N_1$ and $N_2$ into the expression for the full amplitude \Eq{eq:A21}. Due to the symmetry between two gravitons, enforcing Ward identity on graviton 1 up to linear order in $\omega_1$ we can fix the amplitude to be
\begin{equation}
  \label{eq:imp}
\begin{aligned}
&\mathcal{A}(h_1(k_1),h_2(k_2), z(\omega_1))=
  m \kappa^2\Big[\Big(\tfrac{(u\cdot\varepsilon_1)^2(u\cdot\varepsilon_2)^2(k_1\cdot k_2)(k_2\cdot \zeta)+2\omega_1(u\cdot \varepsilon_1)^2(u\cdot\varepsilon_2)(\zeta\cdot\varepsilon_2)(k_1\cdot k_2)}{4(u\cdot k_1)^2}\\
  &+\tfrac{(u\cdot\varepsilon_1)(u\cdot\varepsilon_2)(k_2\cdot\zeta)[(u\cdot \varepsilon_1)(k_1\cdot\varepsilon_2)\!-\!(u\cdot\varepsilon_2)(k_2\cdot\varepsilon_1)]\!+\!\omega_1 (u\cdot\varepsilon_1)(\zeta\cdot\varepsilon_2)[(u\cdot\varepsilon_1)(k_1\cdot\varepsilon_2)\!-\!2(u\cdot\varepsilon_2)(k_2\cdot\varepsilon_1)]}{2(u\cdot k_1)}
  +(k_1\leftrightarrow k_2,\varepsilon_1\leftrightarrow\varepsilon_2)\Big)\\  &+\tfrac{((k_1+k_2)\cdot\zeta)[(u\cdot k_1)(\varepsilon_1\cdot\varepsilon_2)+(u\cdot\varepsilon_2)(k_2\cdot \varepsilon_1)-(u\cdot\varepsilon_1)(k_1\cdot\varepsilon_2)]^2}{4(k_1\cdot k_2)}\\
&+\tfrac{2\omega_1[(u\cdot\varepsilon_1)(k_1\cdot\varepsilon_2)-(u\cdot\varepsilon_2)(k_2\cdot\varepsilon_1)-(u\cdot k_1)(\varepsilon_1\cdot\varepsilon_2)][(\zeta\cdot\varepsilon_1)(k_1\cdot \varepsilon_2)-(\zeta\cdot\varepsilon_2)(k_2\cdot\varepsilon_1)-(k_1\cdot\zeta)(\varepsilon_1\cdot\varepsilon_2)]}{4(k_1\cdot k_2)}
  \\
  &-\frac{1}{2}\big(((k_1+k_2)\cdot\zeta)(u\cdot\varepsilon_1)(u\cdot\varepsilon_2)(\varepsilon_1\cdot\varepsilon_2)+\omega_1(\varepsilon_1\cdot\varepsilon_2)[(u\cdot\varepsilon_2)(\zeta\cdot\varepsilon_1)+(u\cdot\varepsilon_1)(\zeta\cdot\varepsilon_2)]\big)\Big].
\end{aligned}
\end{equation}
One can check that this amplitude obeys the soft theorem in Eq.~\eqref{eq:softheoremb} in the $\omega_1\to 0$ limit.  Nevertheless, note that this amplitude is not itself linear in $\omega_1$ due to the presence of $(u\cdot k_2)$ in the denominator, which generates $\mathcal{O}(\omega_1^2)$ terms upon using momentum conservation.

  \subsubsection{Non-linear Compton scattering}
  \label{sec:3gravexpect}
  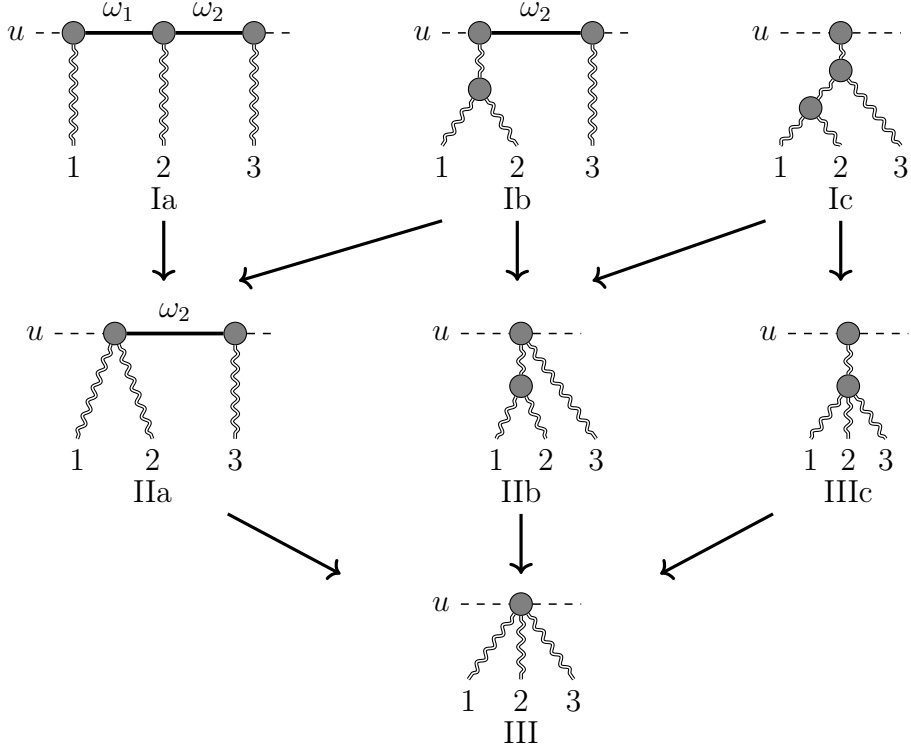
\begin{figure}[t!]
  \centering
    \begin{tikzpicture}[baseline=(current bounding box.center)]
      \begin{feynman}
        \vertex (i1a);
        \vertex[blob, shape=circle, minimum size=0.3cm, fill=gray] at ($(i1a) + (0.5cm, 0cm)$) (i2a) {};
        \vertex[blob, shape=circle, minimum size=0.3cm, fill=gray] at ($(i2a) + (1.2cm, 0cm)$)  (i3a) {};
        \vertex[blob, shape=circle, minimum size=0.3cm, fill=gray] at ($(i3a) + (1.2cm, 0cm)$)  (i4a) {};
        \vertex at ($(i4a) + (0.5cm, 0cm)$)  (i5a);
        \vertex at ($(i2a) + (0cm, -1.5cm)$)  (i6a);
        \vertex at ($(i3a) + (0cm, -1.5cm)$)  (i7a);
        \vertex at ($(i4a) + (0cm, -1.5cm)$)  (i8a);
        \diagram* {
        (i1a)--[soft](i2a)--[hard,  edge label=$\omega_1$](i3a)--[hard,  edge label=$\omega_2$](i4a)--[soft](i5a), (i2a)--[graviton](i6a),(i3a)--[graviton](i7a), (i4a)--[graviton](i8a)
        };
        \draw node[left] at (i1a) {$u$};
        \draw node[below] at (i6a) {$1$};
        \draw node[below] at (i7a) {$2$};
        \draw node[below] at (i8a) {$3$};
      \end{feynman}
      \node at ($(i7a) + (0cm, -0.7cm)$) {Ia};
        \begin{feynman}
          \vertex at  ($(i5a) + (2cm, 0cm)$) (i1b);
          \vertex[blob, shape=circle, minimum size=0.3cm, fill=gray] at ($(i1b) + (0.5cm, 0cm)$) (i2b) {};
          \vertex[blob, shape=circle, minimum size=0.3cm, fill=gray] at ($(i2b) + (1.5cm, 0cm)$)  (i3b) {};
          \vertex at ($(i3b) + (0.5cm, 0cm)$)  (i4b);
          \vertex[blob, shape=circle, minimum size=0.3cm, fill=gray] at ($(i2b) + (0cm, -0.75cm)$)  (i5b) {};
          \vertex at ($(i5b) + (-0.5cm, -0.75cm)$)  (i6b);
          \vertex at ($(i5b) + (0.5cm, -0.75cm)$)  (i7b);
          \vertex at ($(i3b) + (0cm, -1.5cm)$)  (i8b);
          \diagram* {
          (i1b)--[soft](i2b)--[hard, edge label=$\omega_2$](i3b)--[soft](i4b), (i2b)--[graviton](i5b)--[graviton](i6b),(i5b)--[graviton](i7b), (i3b)--[graviton](i8b)
          };
          \draw node[left] at (i1b) {$u$};
          \draw node[below] at (i6b) {$1$};
          \draw node[below] at (i7b) {$2$};
          \draw node[below] at (i8b) {$3$};
        \end{feynman}
        \node at ($(i7b) + (0cm, -0.7cm)$) {Ib};
          \begin{feynman}
            \vertex at  ($(i4b) + (2cm, 0cm)$) (i1c);
            \vertex[blob, shape=circle, minimum size=0.3cm, fill=gray] at ($(i1c) + (0.8cm, 0cm)$) (i2c) {};
            \vertex at ($(i2c) + (0.8cm, 0cm)$)  (i3c);
            \vertex[blob, shape=circle, minimum size=0.3cm, fill=gray] at ($(i2c) + (0cm, -0.5cm)$)  (i4c) {};
            \vertex at ($(i4c) + (0.8cm, -1cm)$)  (i5c);
            \vertex[blob, shape=circle, minimum size=0.3cm, fill=gray] at ($(i4c) + (-0.4cm, -0.5cm)$)  (i6c) {};
            \vertex at ($(i6c) + (0.4cm, -0.5cm)$)  (i7c);
            \vertex at ($(i6c) + (-0.4cm, -0.5cm)$)  (i8c);
            \diagram* {
            (i1c)--[soft](i2c)--[soft](i3c), (i2c)--[graviton](i4c)--[graviton](i5c), (i4c)--[graviton](i6c)--[graviton](i7c),(i6c)--[graviton](i8c)
            };
            \draw node[left] at (i1c) {$u$};
            \draw node[below] at (i5c) {$3$};
            \draw node[below] at (i7c) {$2$};
            \draw node[below] at (i8c) {$1$};
          \end{feynman}
          \node at ($(i7c) + (0cm, -0.7cm)$) {Ic};
          \draw[->, very thick] ($(i7a) + (0cm, -1cm)$) -- ($(i7a) + (0cm, -1.8cm)$);
          \draw[->, very thick] ($(i7b) + (-1cm, -1cm)$) -- ($(i7a) + (1cm, -1.8cm)$);
          \draw[->, very thick] ($(i7b) + (0cm, -1cm)$) -- ($(i7b) + (0cm, -1.8cm)$);
          \draw[->, very thick] ($(i7c) + (0cm, -1cm)$) -- ($(i7c) + (0cm, -1.8cm)$);
          \draw[->, very thick] ($(i7c) + (-1cm, -1cm)$) -- ($(i7b) + (1cm, -1.8cm)$);
    \begin{feynman}
      \vertex at  ($(i6a) + (-0.25cm, -2.5cm)$) (i1d);
      \vertex[blob, shape=circle, minimum size=0.3cm, fill=gray] at ($(i1d) + (0.8cm, 0cm)$) (i2d) {};
      \vertex[blob, shape=circle, minimum size=0.3cm, fill=gray] at ($(i2d) + (1.6cm, 0cm)$)  (i3d) {};
      \vertex at ($(i3d) + (0.5cm, 0cm)$)  (i4d);
      \vertex at ($(i2d) + (-0.5cm, -1.4cm)$)  (i5d);
      \vertex at ($(i2d) + (0.5cm, -1.4cm)$)  (i6d);
      \vertex at ($(i3d) + (0cm, -1.4cm)$)  (i7d);
      \diagram* {
      (i1d)--[soft](i2d)--[hard,   edge label=$\omega_2$](i3d)--[soft](i4d), (i2d)--[graviton](i5d), (i2d)--[graviton](i6d),(i3d)--[graviton](i7d)
      };
      \draw node[left] at (i1d) {$u$};
      \draw node[below] at (i5d) {$1$};
      \draw node[below] at (i6d) {$2$};
      \draw node[below] at (i7d) {$3$};
    \end{feynman}
    \node at ($(i6d) + (0cm, -0.7cm)$) {IIa};
    \begin{feynman}
      \vertex at  ($(i4d) + (2.5cm, 0cm)$) (i1g);
      \vertex[blob, shape=circle, minimum size=0.3cm, fill=gray] at ($(i1g) + (0.8cm, 0cm)$) (i2g) {};
      \vertex at ($(i2g) + (0.8cm, 0cm)$)  (i3g);
      \vertex[blob, shape=circle, minimum size=0.3cm, fill=gray] at ($(i2g) + (0cm, -0.7cm)$)  (i4g) {};
      \vertex at ($(i2g) + (1cm, -1.4cm)$)  (i5g);
      \vertex at ($(i4g) + (-0.33cm, -0.7cm)$)  (i6g);
      \vertex at ($(i4g) + (0.33cm, -0.7cm)$)  (i7g);
      \diagram* {
      (i1g)--[soft](i2g)--[soft](i3g),(i2g)--[graviton](i4g)--[graviton](i6g), (i2g)--[graviton](i5g),(i4g)--[graviton](i7g)
      };
      \draw node[left] at (i1g) {$u$};
      \draw node[below] at (i5g) {$3$};
      \draw node[below] at (i6g) {$1$};
      \draw node[below] at (i7g) {$2$};
    \end{feynman}
    \node at ($(i7g) + (-0.33cm, -0.7cm)$) {IIb};
    \begin{feynman}
      \vertex at  ($(i3g) + (2.75cm, 0cm)$) (i1e);
      \vertex[blob, shape=circle, minimum size=0.3cm, fill=gray] at ($(i1e) + (0.8cm, 0cm)$) (i2e) {};
      \vertex at ($(i2e) + (0.8cm, 0cm)$)  (i3e);
      \vertex[blob, shape=circle, minimum size=0.3cm, fill=gray] at ($(i2e) + (0cm, -0.7cm)$)  (i4e) {};
      \vertex at ($(i4e) + (-0.5cm, -0.7cm)$)  (i5e);
      \vertex at ($(i4e) + (0cm, -0.7cm)$)  (i6e);
      \vertex at ($(i4e) + (0.5cm, -0.7cm)$)  (i7e);
      \diagram* {
      (i1e)--[soft](i2e)--[soft](i3e), (i2e)--[graviton](i4e)--[graviton](i5e), (i4e)--[graviton](i6e),(i4e)--[graviton](i7e)
      };
      \draw node[left] at (i1e) {$u$};
      \draw node[below] at (i5e) {$1$};
      \draw node[below] at (i6e) {$2$};
      \draw node[below] at (i7e) {$3$};
    \end{feynman}
    \node at ($(i6e) + (0cm, -0.7cm)$) {IIIc};
    \draw[->, very thick] ($(i6d) + (1cm, -1cm)$) -- ($(i6d) + (2.5cm, -1.8cm)$);
    \draw[->, very thick] ($(i6e) + (-1cm, -1cm)$) -- ($(i6e) + (-2.5cm, -1.8cm)$);
    \draw[->, very thick] ($(i7g) + (-0.33cm, -1cm)$) -- ($(i7g) + (-0.33cm, -1.8cm)$);
    \begin{feynman}
      \vertex[blob, shape=circle, minimum size=0.3cm, fill=gray]at  ($(i7g) + (-0.33cm, -2.2cm)$) (i1f) {};
      \vertex at ($(i1f) + (-0.8cm, 0cm)$) (i2f);
      \vertex at ($(i1f) + (0.8cm, 0cm)$)  (i3f);
      \vertex at ($(i1f) + (0cm, -1cm)$)  (i5f);
      \vertex at ($(i1f) + (-0.7cm, -1cm)$)  (i6f);
      \vertex at ($(i1f) + (0.7cm, -1cm)$)  (i7f);
      \diagram* {
      (i2f)--[soft](i1f)--[soft](i3f), (i1f)--[graviton](i5f), (i1f)--[graviton](i6f),(i1f)--[graviton](i7f)
      };
      \draw node[left] at (i2f) {$u$};
      \draw node[below] at (i5f) {$2$};
      \draw node[below] at (i6f) {$1$};
      \draw node[below] at (i7f) {$3$};
    \end{feynman}
    \node at ($(i5f) + (0cm, -0.7cm)$) {III};
          \end{tikzpicture}
    \caption{The cut topologies for $\mathcal{A}(h_1,h_2,h_3)$ with the cut-collapsing procedure. In our terminology, the arrows point from the parent topology to the child topology.}
    \label{fig:h3cuts}
 \end{figure}

 Finally, we would like to study the non-linear analog of the Compton scattering amplitude, which involves multiple gravitons. Fig.~\ref{fig:h3cuts} shows the cut-relaxing procedure which allows us to fix the entire amplitude. In practice, we can recycle our previous results for all the sub-amplitudes appearing in this computation.\footnote{It is known that the four graviton amplitude can be reconstructed using a variety of methods, such as by factorization and gauge invariance \cite{Arkani-Hamed:2016rak}. We will not elaborate here.} Thus, the only remaining contact term to be fixed is that of cut III. However, here we will pretend that we have no information about any of the sub-amplitudes so that we can illustrate the procedure to cut multiple worldline propagators at the same time. Note that this figure only contains one representative diagram for each cut topology, whereas in actual computation, multiple relabelings of the same child topology can contribute to a single parent topology. For example, the cuts contributing to IIa are Ia, Ib, and Ia with gravitons $1$ and $2$ exchanged. Moreover, for each cut worldline, there are two possibilities of cutting the energy or conjugate energy, which we do not display explicitly here.

The cut Ia contains two different worldline energies, $\omega_1$ and $\omega_2$, and we use it as an example to illustrate cutting multiple worldline propagators. Since for each worldline propagator, we can choose to set to zero either of the complex frequencies, we need to consider four different combinations of cuts to completely fix the ansatz on Ia:
\begin{enumerate}
  \item setting $\omega_1= \omega_2 = 0$
  \item setting $\bomega_1=\bomega_2=0$
  \item setting $\omega_1 =\bomega_2=0$
  \item setting $\bomega_1=\omega_2=0$
\end{enumerate}
These cuts have overlaps which will serve as consistency checks. For example, terms in the ansatz that contain a single factor of $\omega_1$ with no other worldline energy will be fixed by both 2 and 4 in the list above.

After constraining the ansatze on the maximal cuts, i.e., those with the highest number of cut propagators, we can then move on to the II and III levels to fix the remaining terms. As before, we can use the Ward identity to fix the contact terms and use products of sub-amplitudes to fix the cuts. For example, in IIa, we use the Ward identity for graviton 1 and 2 to fix the contact term in $\omega_1$. Then the cut is fixed by gluing the two sub-amplitudes. The final amplitude is too long to display here and is included in an ancillary file.


\section{Worldline integrands from generalized unitarity}
\label{sec:loop}

Having fixed the rational amplitudes, we are now ready to fix the loop integrands through generalized unitarity. The principle is no different from that for rational amplitudes.  We will implement unitarity in the form of the method of maximal cuts \cite{Bern:2007ct,Bern:2008pv,Bern:2010tq,Carrasco:2011hw} to fix the integrand. Let us briefly review how this works:
\begin{enumerate}
  \item Identify all the allowed propagator structures that contribute to the integrand of a given amplitude. These correspond to the topologies of Feynman-like diagrams with trivalent vertices (or equivalently to all maximal cut topologies).
  \item Using on-shell conditions and momentum conservation, reduce the set of all Lorentz products of the kinematic data to a minimal basis. For each topology, write down an ansatz for the numerator of the integrand in terms of the minimal basis. The ansatz should be the most general expression obeying the following constraints listed at the beginning of Section 3, namely, diagram symmetry, little-group scaling, and power counting.

  \item Generalized unitarity requires that in the appropriate on-shell limit the ansatz should agree with a spanning set of unitarity cuts. In the method of maximal cuts, we start from the maximal number of cuts where the amplitude factorizes into products of local amplitudes. We then relax the cut conditions one by one and fix the ansatze progressively until we reach the level of interest.\footnote{As in the ordinary unitarity method, one needs to account for the combinatorics and mappings of the ansatze to different labelings of each topology contributing to a cut.} In classical black hole scattering, this means we stop this process when we reach the cut corresponding to the product of Compton-like amplitudes.
\end{enumerate}
Let us now apply this method to some examples.

\subsection{On-shell action}
The simplest example for applying the maximal cut method is the on-shell (or radial) action, $\bar S$, which is simply the sum over all ``vacuum'' diagrams in the worldline theory, or rather all diagrams with no external gravitons or worldline fluctuations but any number of static sources. At order $\kappa^4$ this has the following spanning cut topologies\footnote{We do not consider any topology where the worldlines are in contact with each other, as well as diagrams related to the cut
\begin{center}
  \begin{tikzpicture}[baseline=(current bounding box.center)]
    \begin{feynman}
      \vertex (i1);
      \vertex[blob, shape=circle, minimum size=0.3cm, fill=gray] at ($(i1) + (0.5cm, 0cm)$) (i2) {};
      \vertex at ($(i2) + (1.5cm, 0cm)$)  (i3);
      \vertex[blob, shape=circle, minimum size=0.3cm, fill=gray] at ($(i2) + (0cm, -2cm)$)  (i5) {};
      \vertex[blob, shape=circle, minimum size=0.3cm, fill=gray] at ($(i5) + (1cm, 0cm)$)  (i6) {};
      \vertex at ($(i5) + (-0.5cm, 0cm)$) (i7);
      \vertex at ($(i6) + (0.5cm, 0cm)$) (i8);
      \diagram* {
      (i1)--[soft](i2)--[soft](i3), (i2)--[graviton, edge label=3](i5)--[soft](i7),(i5)--[graviton, edge label=2, inner sep=2pt, bend left=60 ](i6)--[soft](i8), (i5)--[soft](i6)
      };
      \draw node[left] at (i1) {$u_1$};
      \draw node[left] at (i7) {$u_2$};
    \end{feynman}
    \end{tikzpicture}
\end{center}
because they only contain scaleless integrals that vanish in dimensional regularization.}
\begin{center}
  \begin{minipage}{\linewidth}
\centering
  \begin{tikzpicture}[baseline=(current bounding box.center)]
    \begin{feynman}
      \vertex (i1);
      \vertex[blob, shape=circle, minimum size=0.3cm, fill=gray] at ($(i1) + (1cm, 0cm)$) (i2) {};
      \vertex at ($(i2) + (1cm, 0cm)$)  (i3);
      \vertex[blob, shape=circle, minimum size=0.3cm, fill=gray] at ($(i2) + (0cm, -1cm)$)  (i4){};
      \vertex[blob, shape=circle, minimum size=0.3cm, fill=gray] at ($(i4) + (-0.5cm, -1cm)$)  (i5) {};
      \vertex[blob, shape=circle, minimum size=0.3cm, fill=gray] at ($(i4) + (0.5cm, -1cm)$)  (i6) {};
      \vertex at ($(i5) + (-0.5cm, 0cm)$) (i7);
      \vertex at ($(i6) + (0.5cm, 0cm)$) (i8);
      \diagram* {
      (i1)--[soft](i2)--[soft](i3), (i2)--[graviton, edge label=3](i4)--[graviton, edge label'=1, inner sep=2pt](i5)--[soft](i7),(i4)--[graviton, edge label=2, inner sep=2pt](i6)--[soft](i8), (i5)--[soft](i6)
      };
      \draw node[left] at (i1) {$u_1$};
      \draw node[left] at (i7) {$u_2$};
    \end{feynman}
    \node at ($(i5) + (0.5cm, -0.5cm)$) {\MC{1}};
    \end{tikzpicture}\qquad \qquad \begin{tikzpicture}[baseline=(current bounding box.center)]
    \begin{feynman}
      \vertex (i1);
      \vertex[blob, shape=circle, minimum size=0.3cm, fill=gray] at ($(i1) + (0.5cm, 0cm)$) (i2) {};
      \vertex[blob, shape=circle, minimum size=0.3cm, fill=gray] at ($(i2) + (1cm, 0cm)$)  (i3) {};
      \vertex at ($(i3) + (0.5cm, 0cm)$)  (i4);
      \vertex[blob, shape=circle, minimum size=0.3cm, fill=gray] at ($(i2) + (0cm, -1.5cm)$)  (i5) {};
      \vertex[blob, shape=circle, minimum size=0.3cm, fill=gray] at ($(i3) + (0cm, -1.5cm)$)  (i6) {};
      \vertex at ($(i5) + (-0.5cm, 0cm)$)  (i7);
      \vertex at ($(i6) + (0.5cm, 0cm)$)  (i8);
      \diagram* {
        (i1)--[soft](i2)--[hard,   edge label'=4](i3)--[soft](i4), (i2)--[graviton, edge label'=1](i5)--[soft](i7), (i3)--[graviton, edge label=2](i6)--[soft](i8), (i5)--[soft](i6)
      };
      \draw node[left] at (i1) {$u_1$};
      \draw node[left] at (i7) {$u_2$};
    \end{feynman}
    \node at ($(i5) + (0.5cm, -0.85cm)$) {\MC{2}};
  \end{tikzpicture}\\
  \begin{tikzpicture}[baseline=(current bounding box.center)]
    \draw[<-, very thick] (-1.3,-0.8) -- (-1.8,0)  ;
    \draw[<-, very thick] (1.5,-0.8) -- (2,0) ;
  \end{tikzpicture}\\
  \begin{tikzpicture}[baseline=(current bounding box.center)]
    \begin{feynman}
        \vertex (i1);
          \vertex[blob, shape=circle, minimum size=0.3cm, fill=gray] at ($(i1) + (0.8cm, 0cm)$) (i2) {};
          \vertex[blob, shape=circle, minimum size=0.3cm, fill=gray] at ($(i2) + (-0.5cm, -1.5cm)$)  (i3) {};
          \vertex[blob, shape=circle, minimum size=0.3cm, fill=gray] at ($(i2) + (0.5cm, -1.5cm)$)  (i5) {};
          \vertex at ($(i2) + (0.8cm, 0cm)$)  (i4);
          \vertex at ($(i3) + (-0.5cm, 0cm)$)  (i6);
          \vertex at ($(i5) + (0.5cm, 0cm)$)  (i7);
          \diagram* {
            (i1)--[soft](i2)--[soft](i4), (i2)--[graviton, edge label'=1](i3)--[soft](i5),(i5)--[graviton, edge label'=2](i2), (i3)--[soft](i6),(i5)--[soft](i7), (i3)--[soft](i5)
          };
          \draw node[left] at (i1) {$u_1$};
          \draw node[left] at (i6) {$u_2$};
      \end{feynman}
      \node at ($(i3) + (0.5cm, -0.4cm)$) {\NMC{1}{1}};
  \end{tikzpicture}
\end{minipage}
\end{center}
The ansatz is composed only of two diagrams with the topology of the maximal cuts MC1, MC2.

All the sub-amplitudes appearing in the above cut topologies have been constructed in the previous section. Carrying out the method of maximal cuts, we find that the NLO on-shell action in $D$ dimensions has the integrand
\begin{equation}
  \begin{aligned}
\bar S|_{\kappa^2}=&\frac{
    i m_1^2m_2\kappa^4
}{
    16\, (-2 + D)^2 k_1^2 k_2^2
}
\Bigg[
\frac{
    (k_1 \cdot k_2) \left( (-2 + D) \gamma^2 - 1 \right)^2
}{
    (k_1 \cdot u_1)^2
} \\
&+
\frac{
    2  \left(
        (-3 + D)(-2 + D) (k_1 \cdot u_1)^2
        + 2\, (k_1 \cdot k_2) \left( -(-2 + D)^2 \gamma^2 + 1 \right)
    \right)
}{
    (k_1+k_2)^2
}
\Bigg]\\
&+(u_1\leftrightarrow u_2, m_1\leftrightarrow m_2)
\end{aligned}
\end{equation}
and $\gamma=(u_1\cdot u_2)$, $b$ is the relative impact parameter, and we set $u_i^2=1$. Due to the simplicity of this example, there is no contact term that can be moved freely between the diagrams.
Indeed this example is almost trivial, as the gluing of the next-to-maximal-cut N$^1$MC1 gives the full answer and is given by the Compton amplitude in Eq.~\eqref{eq:comptonres} with the replacement $u \to u_1$ and
$
   \varepsilon_i^\mu\varepsilon_i^\nu \to m_2\kappa (u_2^\mu u_2^\nu - \tfrac{1}{D-2} \eta^{\mu\nu})
$,
resulting from the sum over polarizations against the one-point function in Eq.~\eqref{eq:1pt},
plus its $(1\leftrightarrow 2)$ image.

Let us also compute the ${\cal O}(\kappa^6)$ conservative on-shell action at second order in the mass ratio $m_2/m_1$ (or first order in self-force) using the unitarity method. This has the following set of spanning cuts:
\begin{center}
  \begin{tikzpicture}[baseline=(current bounding box.center)]
     \begin{feynman}
      \vertex (i1a);
      \vertex[blob, shape=circle, minimum size=0.3cm, fill=gray] at ($(i1a) + (0.5cm, 0cm)$) (i2a) {};
      \vertex[blob, shape=circle, minimum size=0.3cm, fill=gray] at ($(i2a) + (1cm, 0cm)$)  (i3a) {};
      \vertex[blob, shape=circle, minimum size=0.3cm, fill=gray] at ($(i3a) + (1cm, 0cm)$)  (i4a) {};
      \vertex at ($(i4a) + (0.5cm, 0cm)$)  (i5a);
      \vertex at ($(i1a) + (0cm, -1.2cm)$)  (i6a);
      \vertex[blob, shape=circle, minimum size=0.3cm, fill=gray] at ($(i6a) + (0.5cm, 0cm)$)  (i7a){};
      \vertex[blob, shape=circle, minimum size=0.3cm, fill=gray] at ($(i7a) + (1cm, 0cm)$)  (i8a) {};
      \vertex[blob, shape=circle, minimum size=0.3cm, fill=gray] at ($(i8a) + (1cm, 0cm)$)  (i9a){};
      \vertex at ($(i9a) + (0.5cm, 0cm)$)  (i10a);
      \diagram* {
        (i1a)--[soft](i2a)--[hard,   edge label=6](i3a)--[soft](i4a)--[soft](i5a), (i6a)--[soft](i7a)--[soft](i8a)--[hard,   edge label'=7](i9a)--[soft](i10a), (i2a)--[graviton, edge label'=2](i7a), (i3a)--[graviton, edge label=3](i8a), (i4a)--[graviton, edge label=4](i9a)
      };
      \draw node[left] at (i1a) {$u_1$};
      \draw node[left] at (i6a) {$u_2$};
        \end{feynman}
        \node at ($(i8a) + (0cm, -0.6cm)$) {\MC{1}};
      \begin{feynman}
      \vertex at ($(i5a) + (2cm, 0cm)$) (i1b);
      \vertex[blob, shape=circle, minimum size=0.3cm, fill=gray] at ($(i1b) + (0.5cm, 0cm)$) (i2b) {};
      \vertex[blob, shape=circle, minimum size=0.3cm, fill=gray] at ($(i2b) + (1cm, 0cm)$)  (i3b) {};
      \vertex[blob, shape=circle, minimum size=0.3cm, fill=gray] at ($(i3b) + (1cm, 0cm)$)  (i4b) {};
      \vertex at ($(i4b) + (0.5cm, 0cm)$)  (i5b);
      \vertex at ($(i1b) + (0cm, -1.2cm)$)  (i6b);
      \vertex[blob, shape=circle, minimum size=0.3cm, fill=gray] at ($(i6b) + (0.5cm, 0cm)$)  (i7b) {};
      \vertex at ($(i7b) + (1cm, 0cm)$)  (i8b);
      \vertex[blob, shape=circle, minimum size=0.3cm, fill=gray] at ($(i8b) + (1cm, 0cm)$)  (i9b) {};
      \vertex at ($(i9b) + (0.5cm, 0cm)$)  (i10b);
      \vertex[blob, shape=circle, minimum size=0.3cm, fill=gray] at ($(i4b) + (0cm, -0.6cm)$)  (i11b) {};
      \diagram* {
        (i1b)--[soft](i2b)--[hard,   edge label=6](i3b)--[soft](i4b)--[soft](i5b), (i6b)--[soft](i7b)--[soft](i8b)--[soft](i9b)--[soft](i10b), (i2b)--[graviton, edge label'=2](i7b), (i3b)--[graviton, edge label'=3](i11b), (i4b)--[graviton, edge label=4](i11b)--[graviton, edge label=5](i9b)
      };
      \draw node[left] at (i1b) {$u_1$};
      \draw node[left] at (i6b) {$u_2$};
        \end{feynman}
        \node at ($(i8b) + (0cm, -0.6cm)$) {\MC{2}};
      \begin{feynman}
      \vertex at ($(i5b) + (2cm, 0cm)$) (i1c);
      \vertex[blob, shape=circle, minimum size=0.3cm, fill=gray] at ($(i1c) + (0.5cm, 0cm)$) (i2c) {};
      \vertex at ($(i2c) + (1cm, 0cm)$)  (i3c);
      \vertex[blob, shape=circle, minimum size=0.3cm, fill=gray] at ($(i3c) + (1cm, 0cm)$)  (i4c) {};
      \vertex at ($(i4c) + (0.5cm, 0cm)$)  (i5c);
      \vertex at ($(i1c) + (0cm, -1.2cm)$)  (i6c);
      \vertex[blob, shape=circle, minimum size=0.3cm, fill=gray] at ($(i6c) + (0.5cm, 0cm)$)  (i7c){};
      \vertex[blob, shape=circle, minimum size=0.3cm, fill=gray] at ($(i7c) + (1cm, 0cm)$)  (i8c) {};
      \vertex[blob, shape=circle, minimum size=0.3cm, fill=gray] at ($(i8c) + (1cm, 0cm)$)  (i9c) {};
      \vertex at ($(i9c) + (0.5cm, 0cm)$)  (i10c);
      \vertex at ($(i4c) + (0cm, -0.6cm)$)  (i11c);
      \vertex[blob, shape=circle, minimum size=0.3cm, fill=gray] at ($(i2c) + (0cm, -0.6cm)$)  (i12c) {};
      \diagram* {
        (i1c)--[soft](i2c)--[soft](i3c)--[soft](i4c)--[soft](i5c), (i6c)--[soft](i7c)--[soft](i8c)--[hard,   edge label'=7](i9c)--[soft](i10c), (i2c)--[graviton, edge label'=1](i12c)--[graviton, edge label'=2](i7c), (i12c)--[graviton, edge label=3](i8c), (i4c)--[graviton, edge label=4](i9c)
      };
      \draw node[left] at (i1c) {$u_1$};
      \draw node[left] at (i6c) {$u_2$};
        \end{feynman}
        \node at ($(i8c) + (0cm, -0.6cm)$) {\MC{3}};
    \begin{feynman}
      \vertex at ($(i5c) + (2cm, 0cm)$) (i1d);
      \vertex[blob, shape=circle, minimum size=0.3cm, fill=gray] at ($(i1d) + (0.5cm, 0cm)$) (i2d) {};
      \vertex[blob, shape=circle, minimum size=0.3cm, fill=gray] at ($(i2d) + (1cm, 0cm)$)  (i3d) {};
      \vertex at ($(i3d) + (0.5cm, 0cm)$)  (i4d);
      \vertex[blob, shape=circle, minimum size=0.3cm, fill=gray] at ($(i2d) + (0cm, -0.6cm)$)  (i5d) {};
      \vertex[blob, shape=circle, minimum size=0.3cm, fill=gray] at ($(i5d) + (1cm, 0cm)$)  (i6d){};
      \vertex at ($(i1d) + (0cm, -1.2cm)$)  (i7d);
      \vertex[blob, shape=circle, minimum size=0.3cm, fill=gray] at ($(i7d) + (0.5cm, 0cm)$)  (i8d) {};
      \vertex[blob, shape=circle, minimum size=0.3cm, fill=gray] at ($(i8d) + (1cm, 0cm)$)  (i9d) {};
      \vertex at ($(i9d) + (0.5cm, 0cm)$)  (i10d);
      \diagram* {
        (i1d)--[soft](i2d)--[graviton, edge label'=1](i5d)--[graviton, edge label'=2](i8d)--[soft](i7d), (i4d)--[soft](i3d)--[graviton, edge label=4](i6d)--[graviton, edge label=5](i9d)--[soft](i10d), (i2d)--[soft](i3d), (i5d)--[graviton, edge label=3](i6d), (i9d)--[soft](i8d)
      };
      \draw node[left] at (i1d) {$u_1$};
      \draw node[left] at (i7d) {$u_2$};
        \end{feynman}
        \node at ($(i8d) + (0.5cm, -0.6cm)$) {\MC{4}};
        \draw[->, very thick] ($(i8a) + (1cm, -0.8cm)$) -- ($(i8a) + (4cm, -2cm)$);
        \draw[->, very thick] ($(i8a) + (0cm, -0.8cm)$) -- ($(i8a) + (0cm, -2cm)$);
        \draw[->, very thick] ($(i8b) + (-1cm, -0.8cm)$) -- ($(i8b) + (-4cm, -2cm)$);
        \draw[->, very thick] ($(i8b) + (1cm, -0.8cm)$) -- ($(i8b) + (4cm, -2cm)$);
        \draw[->, very thick] ($(i8c) + (-1cm, -0.8cm)$) -- ($(i8c) + (-4cm, -2cm)$);
        \draw[->, very thick] ($(i8c) + (1cm, -0.8cm)$) -- ($(i8c) + (4cm, -2cm)$);
        \draw[->, very thick] ($(i8d) + (0cm, -0.8cm)$) -- ($(i8d) + (-3cm, -2cm)$);
        \draw[->, very thick] ($(i8d) + (0.5cm, -0.8cm)$) -- ($(i8d) + (0.5cm, -2cm)$);
     \begin{feynman}
      \vertex at ($(i6b) + (0cm, -2.5cm)$) (i1e);
      \vertex[blob, shape=circle, minimum size=0.3cm, fill=gray] at ($(i1e) + (0.5cm, 0cm)$) (i2e) {};
      \vertex[blob, shape=circle, minimum size=0.3cm, fill=gray] at ($(i2e) + (2cm, 0cm)$)  (i4e) {};
      \vertex at ($(i4e) + (0.5cm, 0cm)$)  (i5e);
      \vertex at ($(i1e) + (0cm, -1.2cm)$)  (i6e);
      \vertex[blob, shape=circle, minimum size=0.3cm, fill=gray] at ($(i6e) + (0.5cm, 0cm)$)  (i7e) {};
      \vertex[blob, shape=circle, minimum size=0.3cm, fill=gray] at ($(i7e) + (1cm, 0cm)$)  (i8e) {};
      \vertex[blob, shape=circle, minimum size=0.3cm, fill=gray] at ($(i8e) + (1cm, 0cm)$)  (i9e) {};
      \vertex at ($(i9e) + (0.5cm, 0cm)$)  (i10e);
      \diagram* {
        (i1e)--[soft](i2e)--[soft](i4e)--[soft](i5e), (i6e)--[soft](i7e)--[soft](i8e)--[hard,   edge label'=7](i9e)--[soft](i10e), (i2e)--[graviton, edge label'=2](i7e), (i2e)--[graviton, edge label=3](i8e), (i4e)--[graviton, edge label=4](i9e)
      };
      \draw node[left] at (i1e) {$u_1$};
      \draw node[left] at (i6e) {$u_2$};
        \end{feynman}
        \node at ($(i8e) + (0cm, -0.6cm)$) {\NMC{1}{2}};

    \begin{feynman}
      \vertex at ($(i6a) + (0cm, -2.5cm)$) (i1f);
      \vertex[blob, shape=circle, minimum size=0.3cm, fill=gray] at ($(i1f) + (0.5cm, 0cm)$) (i2f) {};
      \vertex[blob, shape=circle, minimum size=0.3cm, fill=gray] at ($(i2f) + (1cm, 0cm)$)  (i3f) {};
      \vertex[blob, shape=circle, minimum size=0.3cm, fill=gray] at ($(i3f) + (1cm, 0cm)$)  (i4f) {};
      \vertex at ($(i4f) + (0.5cm, 0cm)$)  (i5f);
      \vertex at ($(i1f) + (0cm, -1.2cm)$)  (i6f);
      \vertex[blob, shape=circle, minimum size=0.3cm, fill=gray] at ($(i6f) + (0.5cm, 0cm)$)  (i7f) {};
      \vertex at ($(i7f) + (1cm, 0cm)$)  (i8f);
      \vertex[blob, shape=circle, minimum size=0.3cm, fill=gray] at ($(i8f) + (1cm, 0cm)$)  (i9f) {};
      \vertex at ($(i9f) + (0.5cm, 0cm)$)  (i10f);
      \diagram* {
      (i1f)--[soft](i2f)--[hard,   edge label=6](i3f)--[soft](i4f)--[soft](i5f), (i6f)--[soft](i7f)--[soft](i9f)--[soft](i10f), (i2f)--[graviton, edge label'=2](i7f), (i3f)--[graviton, edge label'=3](i9f), (i4f)--[graviton, edge label=4](i9f)
      };
      \draw node[left] at (i1f) {$u_1$};
      \draw node[left] at (i6f) {$u_2$};
      \end{feynman}
      \node at ($(i8f) + (0cm, -0.6cm)$) {\NMC{1}{1}};
        \begin{feynman}
          \vertex at ($(i6c) + (0.5cm, -2.5cm)$) (i1g);
          \vertex[blob, shape=circle, minimum size=0.3cm, fill=gray] at ($(i1g) + (0.5cm, 0cm)$) (i2g) {};
          \vertex[blob, shape=circle, minimum size=0.3cm, fill=gray] at ($(i2g) + (1cm, 0cm)$)  (i3g) {};
          \vertex at ($(i3g) + (0.5cm, 0cm)$)  (i4g);
          \vertex at ($(i2g) + (0cm, -0.6cm)$)  (i5g);
          \vertex[blob, shape=circle, minimum size=0.3cm, fill=gray] at ($(i5g) + (1cm, 0cm)$)  (i6g) {};
          \vertex at ($(i1g) + (0cm, -1.2cm)$)  (i7g);
          \vertex[blob, shape=circle, minimum size=0.3cm, fill=gray] at ($(i7g) + (0.5cm, 0cm)$)  (i8g) {};
          \vertex[blob, shape=circle, minimum size=0.3cm, fill=gray] at ($(i8g) + (1cm, 0cm)$)  (i9g) {};
          \vertex at ($(i9g) + (0.5cm, 0cm)$)  (i10g);
          \diagram* {
            (i1g)--[soft](i2g)--[graviton, edge label'=2](i8g)--[soft](i7g), (i4g)--[soft](i3g)--[graviton, edge label=4](i6g)--[graviton, edge label=5](i9g)--[soft](i10g), (i2g)--[soft](i3g), (i2g)--[graviton, edge label'=3](i6g), (i9g)--[soft](i8g)
          };
          \draw node[left] at (i1g) {$u_1$};
          \draw node[left] at (i7g) {$u_2$};
            \end{feynman}
            \node at ($(i8g) + (0.5cm, -0.6cm)$) {\NMC{1}{3}};
  \begin{feynman}
    \vertex at ($(i7d) + (0cm, -2.5cm)$) (i1h);
    \vertex[blob, shape=circle, minimum size=0.3cm, fill=gray] at ($(i1h) + (0.5cm, 0cm)$) (i2h) {};
    \vertex[blob, shape=circle, minimum size=0.3cm, fill=gray] at ($(i2h) + (1cm, 0cm)$)  (i3h) {};
    \vertex at ($(i3h) + (0.5cm, 0cm)$)  (i4h);
    \vertex[blob, shape=circle, minimum size=0.3cm, fill=gray] at ($(i2h) + (0cm, -0.6cm)$)  (i5h){};
    \vertex at ($(i5h) + (1cm, 0cm)$)  (i6h);
    \vertex at ($(i1h) + (0cm, -1.2cm)$)  (i7h);
    \vertex[blob, shape=circle, minimum size=0.3cm, fill=gray] at ($(i7h) + (0.5cm, 0cm)$)  (i8h){};
    \vertex[blob, shape=circle, minimum size=0.3cm, fill=gray] at ($(i8h) + (1cm, 0cm)$)  (i9h) {};
    \vertex at ($(i9h) + (0.5cm, 0cm)$)  (i10h);
    \diagram* {
      (i1h)--[soft](i2h)--[graviton, edge label'=1](i5h)--[graviton, edge label'=2](i8h)--[soft](i7h), (i4h)--[soft](i3h)--[graviton, edge label=3](i9h)--[soft](i10h), (i2h)--[soft](i3h), (i5h)--[graviton, edge label=4](i9h), (i9h)--[soft](i8h)
    };
    \draw node[left] at (i1h) {$u_1$};
    \draw node[left] at (i7h) {$u_2$};
      \end{feynman}
      \node at ($(i8h) + (0.5cm, -0.6cm)$) {\NMC{1}{4}};
      \draw[->, very thick] ($(i8f) + (1cm, -0.8cm)$) -- ($(i10e) + (0cm, -2cm)$);
      \draw[->, very thick] ($(i8e) + (1cm, -0.8cm)$) -- ($(i10e) + (0.75cm, -2cm)$);
      \draw[->, very thick] ($(i7g) + (0cm, -0.8cm)$) -- ($(i10e) + (1.25cm, -2cm)$);
      \draw[->, very thick] ($(i7h) + (0cm, -0.8cm)$) -- ($(i10e) + (2cm, -2cm)$);
    \begin{feynman}
    \vertex at ($(i10e) + (0cm, -2.5cm)$) (i1i);
      \vertex[blob, shape=circle, minimum size=0.3cm, fill=gray] at ($(i1i) + (0.5cm, 0cm)$) (i2i) {};
      \vertex[blob, shape=circle, minimum size=0.3cm, fill=gray] at ($(i2i) + (1cm, 0cm)$)  (i3i) {};
      \vertex at ($(i3i) + (0.5cm, 0cm)$)  (i4i);
      \vertex at ($(i2i) + (0cm, -0.6cm)$)  (i5i);
      \vertex at ($(i5i) + (1cm, 0cm)$)  (i6i);
      \vertex at ($(i1i) + (0cm, -1.2cm)$)  (i7i);
      \vertex[blob, shape=circle, minimum size=0.3cm, fill=gray] at ($(i7i) + (0.5cm, 0cm)$)  (i8i) {};
      \vertex[blob, shape=circle, minimum size=0.3cm, fill=gray] at ($(i8i) + (1cm, 0cm)$)  (i9i) {};
      \vertex at ($(i9i) + (0.5cm, 0cm)$)  (i10i);
      \diagram* {
        (i1i)--[soft](i2i)--[graviton, edge label'=2](i8i)--[soft](i7i), (i4i)--[soft](i3i)--[graviton, edge label=4](i9i)--[soft](i10i), (i2i)--[soft](i3i), (i2i)--[graviton, edge label'=3](i9i), (i9i)--[soft](i8i)
      };
      \draw node[left] at (i1i) {$u_1$};
      \draw node[left] at (i7i) {$u_2$};
        \end{feynman}
        \node at ($(i8i) + (0.5cm, -0.4cm)$) {\NMC{2}{1}};
  \end{tikzpicture}
\end{center}
We ignore the cut
\begin{equation}
  \begin{tikzpicture}[baseline=(current bounding box.center)]
    \begin{feynman}
      \vertex (i1i);
        \vertex[blob, shape=circle, minimum size=0.3cm, fill=gray] at ($(i1i) + (0.8cm, 0cm)$) (i2i) {};
        \vertex[blob, shape=circle, minimum size=0.3cm, fill=gray] at ($(i2i) + (1.5cm, 0cm)$)  (i3i) {};
        \vertex at ($(i3i) + (0.8cm, 0cm)$)  (i4i);
        \vertex at ($(i1i) + (0cm, -1.8cm)$)  (i7i);
        \vertex[blob, shape=circle, minimum size=0.3cm, fill=gray] at ($(i7i) + (0.8cm, 0cm)$)  (i8i) {};
        \vertex[blob, shape=circle, minimum size=0.3cm, fill=gray] at ($(i8i) + (1.5cm, 0cm)$)  (i9i) {};
        \vertex at ($(i9i) + (0.8cm, 0cm)$)  (i10i);
        \diagram* {
          (i1i)--[soft](i2i)--[graviton, edge label'=2](i8i)--[soft](i7i), (i4i)--[soft](i3i)--[graviton, edge label=5](i9i)--[soft](i10i), (i2i)--[soft](i3i), (i2i)--[graviton, bend right=60, edge label'=3](i3i), (i9i)--[soft](i8i)
        };
        \draw node[left] at (i1i) {$u_1$};
        \draw node[left] at (i7i) {$u_2$};
          \end{feynman}
    \end{tikzpicture}
\end{equation}
which contributes only to radiative effects.
We also discard the cuts
\begin{equation}
  \begin{tikzpicture}
    \begin{feynman}
      \vertex (i1);
      \vertex[blob, shape=circle, minimum size=0.3cm, fill=gray] at ($(i1) + (0.8cm, 0cm)$) (i2) {};
      \vertex[blob, shape=circle, minimum size=0.3cm, fill=gray] at ($(i2) + (1.5cm, 0cm)$)  (i3) {};
      \vertex at ($(i3) + (0.8cm, 0cm)$)  (i4);
      \vertex[blob, shape=circle, minimum size=0.3cm, fill=gray] at ($(i2) + (0.75cm, -0.6cm)$)  (i5) {};
      \vertex[blob, shape=circle, minimum size=0.3cm, fill=gray] at ($(i5) + (0cm, -0.6cm)$)  (i6) {};
      \vertex[blob, shape=circle, minimum size=0.3cm, fill=gray] at ($(i6) + (-0.75cm, -0.6cm)$)  (i7) {};
      \vertex[blob, shape=circle, minimum size=0.3cm, fill=gray] at ($(i6) + (0.75cm, -0.6cm)$)  (i8) {};
      \vertex at ($(i7) + (-0.8cm, 0cm)$)  (i9);
      \vertex at ($(i8) + (0.8cm, 0cm)$)  (i10);
      \diagram* {
        (i1)--[soft](i2)--[soft](i3)--[soft](i4), (i2)--[graviton](i5)--[graviton](i6)--[graviton](i7)--[soft](i8), (i5)--[graviton](i3), (i6)--[graviton](i8), (i7)--[soft](i9), (i8)--[soft](i10)
      };
    \end{feynman}
  \end{tikzpicture}
  \qquad \begin{tikzpicture}
    \begin{feynman}
      \vertex (i1);
      \vertex[blob, shape=circle, minimum size=0.3cm, fill=gray] at ($(i1) + (0.8cm, 0cm)$) (i2) {};
      \vertex[blob, shape=circle, minimum size=0.3cm, fill=gray] at ($(i2) + (1.5cm, 0cm)$)  (i3) {};
      \vertex at ($(i3) + (0.8cm, 0cm)$)  (i4);
      \vertex at ($(i2) + (0.75cm, -0.9cm)$)  (i5);
      \vertex at ($(i2) + (0.75cm, -0.9cm)$)  (i6);
      \vertex[blob, shape=circle, minimum size=0.3cm, fill=gray] at ($(i6) + (-0.75cm, -0.9cm)$)  (i7) {};
      \vertex[blob, shape=circle, minimum size=0.3cm, fill=gray] at ($(i6) + (0.75cm, -0.9cm)$)  (i8) {};
      \vertex at ($(i7) + (-0.8cm, 0cm)$)  (i9);
      \vertex at ($(i8) + (0.8cm, 0cm)$)  (i10);
      \diagram* {
        (i1)--[soft](i2)--[soft](i3)--[soft](i4), (i2)--[graviton](i5)--[graviton](i7)--[soft](i8), (i5)--[graviton](i3), (i6)--[graviton](i8), (i7)--[soft](i9), (i8)--[soft](i10)
      };
    \end{feynman}
  \end{tikzpicture}
  \label{eq:discard1}
\end{equation}
as they are analytic in the momentum transfer and thus do not contribute to the classical on-shell action in $D=4$ dimensions.

Constructing the various unitarity cuts and imposing them on the ansatze is now a straightforward task that can be easily automated. We include the final result of this process (i.e. numerators of integrands) for each maximal cut topology in an ancillary file.\footnote{Note that the labeling of momenta is different between the diagrams presented above and the expressions in the ancillary file.} Note that there remain unfixed coefficients corresponding either to the freedom of moving contact terms between the N$^2$MC1 topology and its flipped counterpart, or to terms that correspond to the four-graviton contact term. A non-trivial check of a successful merging process is that the final result after integration should be free of these unknown constants.
To verify our integrand, we have thus performed two-loop IBP reduction and integration in the potential region and find that it is in agreement with the known result \cite{Damgaard:2021ipf, Brandhuber:2021eyq, Bjerrum-Bohr:2021din}.

\subsection{Gravitational waveform}
\label{sec:2PMwaveform}
We also compute the ${\cal O}(\kappa^5)$ (or next-to-leading-order) waveform using the method of maximal cuts. We consider the spanning cut topologies  presented in Figs.~\ref{fig:2PMwaveformmc}-\ref{fig:2PMwaveformnnmc}.
\begin{center}
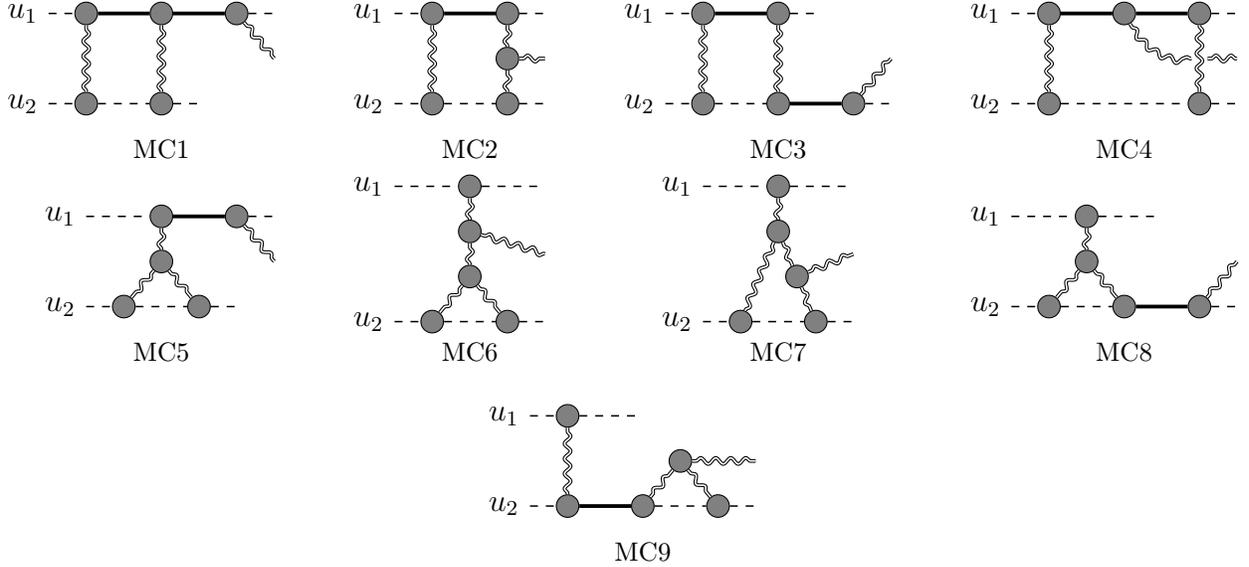
\begin{figure}[t]
  \centering
  \begin{tikzpicture}[baseline=(current bounding box.center)]
     \begin{feynman}
      \vertex (i1a);
      \vertex[blob, shape=circle, minimum size=0.3cm, fill=gray] at ($(i1a) + (0.5cm, 0cm)$) (i2a) {};
      \vertex[blob, shape=circle, minimum size=0.3cm, fill=gray] at ($(i2a) + (1cm, 0cm)$)  (i3a) {};
      \vertex[blob, shape=circle, minimum size=0.3cm, fill=gray] at ($(i3a) + (1cm, 0cm)$)  (i4a) {};
      \vertex at ($(i4a) + (0.5cm, 0cm)$)  (i5a);
      \vertex at ($(i1a) + (0cm, -1.2cm)$)  (i6a);
      \vertex[blob, shape=circle, minimum size=0.3cm, fill=gray] at ($(i6a) + (0.5cm, 0cm)$)  (i7a){};
      \vertex[blob, shape=circle, minimum size=0.3cm, fill=gray] at ($(i7a) + (1cm, 0cm)$)  (i8a) {};
      \vertex at ($(i4a) + (0.5cm, -0.6cm)$)  (i9a);
      \vertex at ($(i8a) + (0.5cm, 0cm)$)  (i10a);
      \diagram* {
        (i1a)--[soft](i2a)--[hard  ](i3a)--[hard  ](i4a)--[soft](i5a), (i6a)--[soft](i7a)--[soft](i8a)--[soft](i10a), (i2a)--[graviton](i7a), (i3a)--[graviton](i8a), (i4a)--[graviton](i9a)
      };
      \draw node[left] at (i1a) {$u_1$};
      \draw node[left] at (i6a) {$u_2$};
        \end{feynman}
        \node at ($(i8a) + (0cm, -0.6cm)$) {\MC{1}};
  \begin{feynman}
      \vertex at ($(i5a) + (1.6cm, 0cm)$) (i1b);
      \vertex[blob, shape=circle, minimum size=0.3cm, fill=gray] at ($(i1b) + (0.5cm, 0cm)$) (i2b) {};
      \vertex[blob, shape=circle, minimum size=0.3cm, fill=gray] at ($(i2b) + (1cm, 0cm)$)  (i3b) {};
      \vertex at ($(i3b) + (0.5cm, 0cm)$)  (i4b);
      \vertex at ($(i1b) + (0cm, -1.2cm)$)  (i6b);
      \vertex[blob, shape=circle, minimum size=0.3cm, fill=gray] at ($(i6b) + (0.5cm, 0cm)$)  (i7b){};
      \vertex[blob, shape=circle, minimum size=0.3cm, fill=gray] at ($(i7b) + (1cm, 0cm)$)  (i8b) {};
      \vertex[blob, shape=circle, minimum size=0.3cm, fill=gray] at ($(i3b) + (0cm, -0.6cm)$)  (i9b) {};
      \vertex at ($(i9b) + (0.5cm, 0cm)$)  (i10b);
      \vertex at ($(i8b) + (0.5cm, 0cm)$)  (i11b);
      \diagram* {
        (i1b)--[soft](i2b)--[  hard](i3b)--[soft](i4b), (i6b)--[soft](i7b)--[soft](i8b)--[soft](i11b), (i2b)--[graviton](i7b),  (i3b)--[graviton](i9b)--[graviton](i8b), (i9b)--[graviton](i10b)
      };
      \draw node[left] at (i1b) {$u_1$};
      \draw node[left] at (i6b) {$u_2$};
        \end{feynman}
        \node at ($(i8b) + (-0.5cm, -0.6cm)$) {\MC{2}};
 \begin{feynman}
      \vertex at ($(i4b) + (1.6cm, 0cm)$) (i1c);
      \vertex[blob, shape=circle, minimum size=0.3cm, fill=gray] at ($(i1c) + (0.5cm, 0cm)$) (i2c) {};
      \vertex[blob, shape=circle, minimum size=0.3cm, fill=gray] at ($(i2c) + (1cm, 0cm)$)  (i3c) {};
      \vertex at ($(i3c) + (0.5cm, 0cm)$)  (i4c);
      \vertex at ($(i1c) + (0cm, -1.2cm)$)  (i6c);
      \vertex[blob, shape=circle, minimum size=0.3cm, fill=gray] at ($(i6c) + (0.5cm, 0cm)$)  (i7c){};
      \vertex[blob, shape=circle, minimum size=0.3cm, fill=gray] at ($(i7c) + (1cm, 0cm)$)  (i8c) {};
      \vertex[blob, shape=circle, minimum size=0.3cm, fill=gray] at ($(i8c) + (1cm, 0cm)$)  (i9c) {};
      \vertex at ($(i9c) + (0.5cm, 0cm)$)  (i10c);
      \vertex at ($(i9c) + (0.5cm, 0.6cm)$)  (i11c);
      \diagram* {
        (i1c)--[soft](i2c)--[hard  ](i3c)--[soft](i4c), (i6c)--[soft](i7c)--[soft](i8c)--[hard  ](i9c)--[soft](i10c), (i2c)--[graviton](i7c),  (i3c)--[graviton](i8c), (i9c)--[graviton](i11c)
      };
      \draw node[left] at (i1c) {$u_1$};
      \draw node[left] at (i6c) {$u_2$};
        \end{feynman}
        \node at ($(i8c) + (0cm, -0.6cm)$) {\MC{3}};
       \begin{feynman}
      \vertex at ($(i11c) + (1.6cm, 0.6cm)$) (i1d);
      \vertex[blob, shape=circle, minimum size=0.3cm, fill=gray] at ($(i1d) + (0.5cm, 0cm)$) (i2d) {};
      \vertex[blob, shape=circle, minimum size=0.3cm, fill=gray] at ($(i2d) + (1cm, 0cm)$)  (i3d) {};
      \vertex[blob, shape=circle, minimum size=0.3cm, fill=gray] at ($(i3d) + (1cm, 0cm)$)  (i4d) {};
      \vertex at ($(i4d) + (0.5cm, 0cm)$)  (i5d);
      \vertex at ($(i1d) + (0cm, -1.2cm)$)  (i6d);
      \vertex[blob, shape=circle, minimum size=0.3cm, fill=gray] at ($(i6d) + (0.5cm, 0cm)$)  (i7d){};
      \vertex[blob, shape=circle, minimum size=0.3cm, fill=gray] at ($(i7d) + (1cm, 0cm)$)  (i8d);
      \vertex[blob, shape=circle, minimum size=0.3cm, fill=gray] at ($(i8d) + (1cm, 0cm)$)  (i9d) {};
      \vertex at ($(i9d) + (0.5cm, 0cm)$)  (i10d);
      \vertex at ($(i9d) + (0.5cm, 0.6cm)$)  (i11d);
      \vertex[blob, shape=circle, minimum size=0.2cm, color=white] at ($(i9d) + (0cm, 0.6cm)$)  (i12d) {};
      \diagram* {
        (i1d)--[soft](i2d)--[hard  ](i3d)--[hard  ](i4d)--[soft](i5d), (i6d)--[soft](i7d)--[soft](i8d)--[soft](i9d)--[soft](i10d), (i2d)--[graviton](i7d),  (i3d)--[graviton, bend right=30](i12d)--[graviton](i11d), (i9d)--[graviton](i4d)
      };
      \draw node[left] at (i1d) {$u_1$};
      \draw node[left] at (i6d) {$u_2$};
        \end{feynman}
        \node at ($(i8d) + (0cm, -0.6cm)$) {\MC{4}};
  \begin{feynman}
      \vertex at ($(i6a) + (0.5cm, -1.5cm)$) (i1e);
      \vertex[blob, shape=circle, minimum size=0.3cm, fill=gray] at ($(i1e) + (1cm, 0cm)$) (i2e) {};
      \vertex[blob, shape=circle, minimum size=0.3cm, fill=gray] at ($(i2e) + (1cm, 0cm)$)  (i3e) {};
      \vertex at ($(i3e) + (0.5cm, 0cm)$)  (i4e);
      \vertex at ($(i4e) + (0cm, -0.6cm)$)  (i11e);
      \vertex[blob, shape=circle, minimum size=0.3cm, fill=gray] at ($(i2e) + (0cm, -0.6cm)$)  (i5e) {};
      \vertex at ($(i1e) + (0cm, -1.2cm)$)  (i6e);
      \vertex[blob, shape=circle, minimum size=0.3cm, fill=gray] at ($(i6e) + (0.5cm, 0cm)$)  (i7e){};
      \vertex[blob, shape=circle, minimum size=0.3cm, fill=gray] at ($(i7e) + (1cm, 0cm)$)  (i8e) {};
      \vertex at ($(i8e) + (0.5cm, 0cm)$)  (i10e);
      \diagram* {
        (i1e)--[soft](i2e)--[hard  ](i3e)--[soft](i4e), (i6e)--[soft](i7e)--[soft](i8e)--[soft](i10e), (i2e)--[graviton](i5e)--[graviton](i7e), (i3e)--[graviton](i11e), (i5e)--[graviton](i8e)
      };
      \draw node[left] at (i1e) {$u_1$};
      \draw node[left] at (i6e) {$u_2$};
        \end{feynman}
        \node at ($(i8e) + (-0.5cm, -0.6cm)$) {\MC{5}};
  \begin{feynman}
      \vertex at ($(i4e) + (1.6cm, 0.4cm)$) (i1f);
      \vertex[blob, shape=circle, minimum size=0.3cm, fill=gray] at ($(i1f) + (1cm, 0cm)$) (i2f) {};
      \vertex[blob, shape=circle, minimum size=0.3cm, fill=gray] at ($(i2f) + (1cm, 0cm)$)  (i3f);
      \vertex[blob, shape=circle, minimum size=0.3cm, fill=gray] at ($(i2f) + (0cm, -0.6cm)$)  (i4f) {};
      \vertex[blob, shape=circle, minimum size=0.3cm, fill=gray] at ($(i4f) + (0cm, -0.6cm)$)  (i9f) {};
      \vertex at ($(i1f) + (0cm, -1.8cm)$)  (i6f);
      \vertex[blob, shape=circle, minimum size=0.3cm, fill=gray] at ($(i6f) + (0.5cm, 0cm)$)  (i7f){};
      \vertex[blob, shape=circle, minimum size=0.3cm, fill=gray] at ($(i7f) + (1cm, 0cm)$)  (i8f) {};
      \vertex at ($(i8f) + (0.5cm, 0cm)$)  (i10f);
      \vertex at ($(i10f) + (0cm, 0.9cm)$)  (i11f);
      \diagram* {
        (i1f)--[soft](i2f)--[soft](i3f), (i6f)--[soft](i7f)--[soft](i8f)--[soft](i10f), (i2f)--[graviton](i4f)--[graviton](i9f)--[graviton](i7f), (i4f)--[graviton](i11f), (i9f)--[graviton](i8f)
      };
      \draw node[left] at (i1f) {$u_1$};
      \draw node[left] at (i6f) {$u_2$};
        \end{feynman}
        \node at ($(i7f) + (0.5cm, -0.4cm)$) {\MC{6}};
  \begin{feynman}
      \vertex at ($(i3f) + (2.1cm, 0cm)$) (i1g);
      \vertex[blob, shape=circle, minimum size=0.3cm, fill=gray] at ($(i1g) + (1cm, 0cm)$) (i2g) {};
      \vertex[blob, shape=circle, minimum size=0.3cm, fill=gray] at ($(i2g) + (1cm, 0cm)$)  (i3g);
      \vertex[blob, shape=circle, minimum size=0.3cm, fill=gray] at ($(i2g) + (0cm, -0.6cm)$)  (i4g) {};
      \vertex[blob, shape=circle, minimum size=0.3cm, fill=gray] at ($(i4g) + (0.25cm, -0.6cm)$)  (i9g) {};
      \vertex at ($(i1g) + (0cm, -1.8cm)$)  (i6g);
      \vertex[blob, shape=circle, minimum size=0.3cm, fill=gray] at ($(i6g) + (0.5cm, 0cm)$)  (i7g){};
      \vertex[blob, shape=circle, minimum size=0.3cm, fill=gray] at ($(i7g) + (1cm, 0cm)$)  (i8g) {};
      \vertex at ($(i8g) + (0.5cm, 0cm)$)  (i10g);
      \vertex at ($(i10g) + (0cm, 0.9cm)$)  (i11g);
      \diagram* {
        (i1g)--[soft](i2g)--[soft](i3g), (i6g)--[soft](i7g)--[soft](i8g)--[soft](i10g), (i2g)--[graviton](i4g)--[graviton](i9g)--[graviton](i8g), (i4g)--[graviton](i7g), (i9g)--[graviton](i11g)
      };
      \draw node[left] at (i1g) {$u_1$};
      \draw node[left] at (i6g) {$u_2$};
        \end{feynman}
        \node at ($(i7g) + (0.5cm, -0.4cm)$) {\MC{7}};
 \begin{feynman}
      \vertex at ($(i3g) + (2.1cm, -0.4cm)$) (i1h);
      \vertex[blob, shape=circle, minimum size=0.3cm, fill=gray] at ($(i1h) + (1cm, 0cm)$) (i2h) {};
      \vertex[blob, shape=circle, minimum size=0.3cm, fill=gray] at ($(i2h) + (1cm, 0cm)$)  (i3h);
      \vertex[blob, shape=circle, minimum size=0.3cm, fill=gray] at ($(i2h) + (0cm, -0.6cm)$)  (i4h) {};
      \vertex at ($(i1h) + (0cm, -1.2cm)$)  (i6h);
      \vertex[blob, shape=circle, minimum size=0.3cm, fill=gray] at ($(i6h) + (0.5cm, 0cm)$)  (i7h){};
      \vertex[blob, shape=circle, minimum size=0.3cm, fill=gray] at ($(i7h) + (1cm, 0cm)$)  (i8h) {};
      \vertex[blob, shape=circle, minimum size=0.3cm, fill=gray] at ($(i8h) + (1cm, 0cm)$)  (i9h) {};
      \vertex at ($(i9h) + (0.5cm, 0cm)$)  (i10h);
      \vertex at ($(i9h) + (0.5cm, 0.6cm)$)  (i11h);
      \diagram* {
        (i1h)--[soft](i2h)--[soft](i3h), (i6h)--[soft](i7h)--[soft](i8h)--[hard  ](i9h)--[soft](i10h), (i2h)--[graviton](i4h)--[graviton](i7h),  (i4h)--[graviton](i8h), (i9h)--[graviton](i11h)
      };
      \draw node[left] at (i1h) {$u_1$};
      \draw node[left] at (i6h) {$u_2$};
        \end{feynman}
        \node at ($(i8h) + (0cm, -0.6cm)$) {\MC{8}};
 \end{tikzpicture}\\
 \vspace{0.3cm}
 \begin{tikzpicture}[baseline=(current bounding box.center)]
  \begin{feynman}
    \vertex (i1g);
    \vertex[blob, shape=circle, minimum size=0.3cm, fill=gray] at ($(i1g) + (0.5cm, 0cm)$) (i2g) {};
    \vertex at ($(i2g) + (1cm, 0cm)$)  (i3g);
    \vertex at ($(i1g) + (0cm, -1.2cm)$)  (i6g);
    \vertex[blob, shape=circle, minimum size=0.3cm, fill=gray] at ($(i6g) + (0.5cm, 0cm)$)  (i7g){};
    \vertex[blob, shape=circle, minimum size=0.3cm, fill=gray] at ($(i7g) + (01cm, 0cm)$)  (i4g){};
    \vertex[blob, shape=circle, minimum size=0.3cm, fill=gray] at ($(i4g) + (1cm, 0cm)$)  (i8g) {};
    \vertex[blob, shape=circle, minimum size=0.3cm, fill=gray] at ($(i4g) + (0.5cm, 0.6cm)$)  (i9g) {};
    \vertex at ($(i8g) + (0.5cm, 0cm)$)  (i10g);
    \vertex at ($(i10g) + (0cm, 0.6cm)$)  (i11g);
    \diagram* {
      (i1g)--[soft](i2g)--[soft](i3g), (i6g)--[soft](i7g)--[hard  ](i4g)--[soft](i8g)--[soft](i10g), (i2g)--[graviton](i7g), (i4g)--[graviton](i9g)--[graviton](i8g),  (i9g)--[graviton](i11g)
    };
    \draw node[left] at (i1g) {$u_1$};
    \draw node[left] at (i6g) {$u_2$};
      \end{feynman}
      \node at ($(i4g) + (0cm, -0.6cm)$) {\MC{9}};
  \begin{feynman}
      \vertex at ($(i3g) + (3cm, 0cm)$) (i1h);
      \vertex[blob, shape=circle, minimum size=0.3cm, fill=gray] at ($(i1h) + (0.5cm, 0cm)$) (i2h) {};
    \vertex at ($(i2h) + (1cm, 0cm)$)  (i3h);
    \vertex at ($(i1h) + (0cm, -1.2cm)$)  (i6h);
    \vertex[blob, shape=circle, minimum size=0.3cm, fill=gray] at ($(i6h) + (0.5cm, 0cm)$)  (i7h){};
    \vertex[blob, shape=circle, minimum size=0.3cm, fill=gray] at ($(i7h) + (0.5cm, 0cm)$)  (i4h){};
    \vertex[blob, shape=circle, minimum size=0.3cm, fill=gray] at ($(i4h) + (0.7cm, 0cm)$)  (i8h) {};
    \vertex[blob, shape=circle, minimum size=0.3cm, fill=gray] at ($(i8h) + (0.5cm, 0cm)$)  (i9h) {};
    \vertex at ($(i9h) + (0.5cm, 0cm)$)  (i10h);
    \vertex at ($(i10h) + (0cm, 0.6cm)$)  (i11h);
      \diagram* {
        (i1h)--[soft](i2h)--[soft](i3h), (i6h)--[soft](i7h)--[hard  ](i4h)--[soft](i8h)--[hard](i9h)--[soft](i10h), (i2h)--[graviton](i7h), (i4h)--[graviton, bend left=50](i8h),  (i9h)--[graviton](i11h)
      };
      \draw node[left] at (i1h) {$u_1$};
      \draw node[left] at (i6h) {$u_2$};
        \end{feynman}
        \node at ($(i8h) + (-0.25cm, -0.6cm)$) {\MC{10}};
\end{tikzpicture}

\caption{Maximal cut topologies for the waveform. }
\label{fig:2PMwaveformmc}
 \end{figure}\allowdisplaybreaks

 \begin{figure}[t]
  \centering
  \begin{tikzpicture}[baseline=(current bounding box.center)]
     \begin{feynman}
      \vertex (i1a);
      \vertex[blob, shape=circle, minimum size=0.3cm, fill=gray] at ($(i1a) + (1cm, 0cm)$) (i2a) {};
      \vertex[blob, shape=circle, minimum size=0.3cm, fill=gray] at ($(i2a) + (1cm, 0cm)$)  (i4a) {};
      \vertex at ($(i4a) + (0.5cm, 0cm)$)  (i5a);
      \vertex at ($(i1a) + (0cm, -1.2cm)$)  (i6a);
      \vertex[blob, shape=circle, minimum size=0.3cm, fill=gray] at ($(i6a) + (0.5cm, 0cm)$)  (i7a){};
      \vertex[blob, shape=circle, minimum size=0.3cm, fill=gray] at ($(i7a) + (1cm, 0cm)$)  (i8a) {};
      \vertex at ($(i4a) + (0.5cm, -0.6cm)$)  (i9a);
      \vertex at ($(i8a) + (0.5cm, 0cm)$)  (i10a);
      \diagram* {
        (i1a)--[soft](i2a)--[hard  ](i4a)--[soft](i5a), (i6a)--[soft](i7a)--[soft](i8a)--[soft](i10a), (i2a)--[graviton](i7a), (i2a)--[graviton](i8a), (i4a)--[graviton](i9a)
      };
      \draw node[left] at (i1a) {$u_1$};
      \draw node[left] at (i6a) {$u_2$};
        \end{feynman}
        \node at ($(i8a) + (-0.5cm, -0.6cm)$) {\NMC{1}{1}};
  \begin{feynman}
      \vertex at ($(i5a) + (1.6cm, 0cm)$) (i1b);
      \vertex[blob, shape=circle, minimum size=0.3cm, fill=gray] at ($(i1b) + (0.5cm, 0cm)$) (i2b) {};
      \vertex[blob, shape=circle, minimum size=0.3cm, fill=gray] at ($(i2b) + (1cm, 0cm)$)  (i3b) {};
      \vertex at ($(i3b) + (0.5cm, 0cm)$)  (i4b);
      \vertex at ($(i1b) + (0cm, -1.2cm)$)  (i6b);
      \vertex[blob, shape=circle, minimum size=0.3cm, fill=gray] at ($(i6b) + (0.5cm, 0cm)$)  (i7b){};
      \vertex[blob, shape=circle, minimum size=0.3cm, fill=gray] at ($(i7b) + (1cm, 0cm)$)  (i8b) {};
      \vertex at ($(i4b) + (0cm, -0.6cm)$)  (i10b);
      \vertex at ($(i8b) + (0.5cm, 0cm)$)  (i11b);
      \diagram* {
        (i1b)--[soft](i2b)--[hard  ](i3b)--[soft](i4b), (i6b)--[soft](i7b)--[soft](i8b)--[soft](i11b), (i2b)--[graviton](i7b),  (i3b)--[graviton](i8b), (i3b)--[graviton](i10b)
      };
      \draw node[left] at (i1b) {$u_1$};
      \draw node[left] at (i6b) {$u_2$};
        \end{feynman}
        \node at ($(i8b) + (-0.5cm, -0.6cm)$) {\NMC{1}{2}};
 \begin{feynman}
      \vertex at ($(i4b) + (1.6cm, 0cm)$) (i1c);
      \vertex[blob, shape=circle, minimum size=0.3cm, fill=gray] at ($(i1c) + (1cm, 0cm)$) (i2c) {};
      \vertex at ($(i2c) + (1cm, 0cm)$)  (i4c);
      \vertex at ($(i1c) + (0cm, -1.2cm)$)  (i6c);
      \vertex[blob, shape=circle, minimum size=0.3cm, fill=gray] at ($(i6c) + (0.5cm, 0cm)$)  (i7c){};
      \vertex[blob, shape=circle, minimum size=0.3cm, fill=gray] at ($(i7c) + (1cm, 0cm)$)  (i8c) {};
      \vertex[blob, shape=circle, minimum size=0.3cm, fill=gray] at ($(i2c) + (0.25cm, -0.6cm)$)  (i9c) {};
      \vertex at ($(i8c) + (0.5cm, 0cm)$)  (i10c);
      \vertex at ($(i9c) + (0.5cm, 0cm)$)  (i11c);
      \diagram* {
        (i1c)--[soft](i2c)--[soft](i4c), (i6c)--[soft](i7c)--[soft](i8c)--[soft](i10c), (i2c)--[graviton](i7c),  (i2c)--[graviton](i9c)--[graviton](i8c), (i9c)--[graviton](i11c)
      };
      \draw node[left] at (i1c) {$u_1$};
      \draw node[left] at (i6c) {$u_2$};
        \end{feynman}
        \node at ($(i8c) + (-0.5cm, -0.6cm)$) {\NMC{1}{3}};
        \begin{feynman}
      \vertex at ($(i4c) + (1.6cm, 0cm)$) (i1d);
      \vertex[blob, shape=circle, minimum size=0.3cm, fill=gray] at ($(i1d) + (0.5cm, 0cm)$) (i2d) {};
      \vertex[blob, shape=circle, minimum size=0.3cm, fill=gray] at ($(i2d) + (1cm, 0cm)$)  (i3d) {};
      \vertex at ($(i3d) + (0.5cm, 0cm)$)  (i4d);
      \vertex at ($(i1d) + (0cm, -1.2cm)$)  (i6d);
      \vertex[blob, shape=circle, minimum size=0.3cm, fill=gray] at ($(i6d) + (0.5cm, 0cm)$)  (i7d){};
      \vertex[blob, shape=circle, minimum size=0.3cm, fill=gray] at ($(i7d) + (1cm, 0cm)$)  (i8d) {};
      \vertex at ($(i8d) + (0.5cm, 0cm)$)  (i10d);
      \vertex at ($(i8d) + (0.5cm, 0.6cm)$)  (i11d);
      \diagram* {
        (i1d)--[soft](i2d)--[hard  ](i3d)--[soft](i4d), (i6d)--[soft](i7d)--[soft](i8d)--[soft](i10d), (i2d)--[graviton](i7d),  (i3d)--[graviton](i8d), (i8d)--[graviton](i11d)
      };
      \draw node[left] at (i1d) {$u_1$};
      \draw node[left] at (i6d) {$u_2$};
        \end{feynman}
        \node at ($(i8d) + (-0.5cm, -0.6cm)$) {\NMC{1}{4}};
\begin{feynman}
      \vertex at ($(i6a) + (0cm, -1.5cm)$) (i1e);
      \vertex[blob, shape=circle, minimum size=0.3cm, fill=gray] at ($(i1e) + (1cm, 0cm)$) (i2e) {};
      \vertex[blob, shape=circle, minimum size=0.3cm, fill=gray] at ($(i2e) + (1cm, 0cm)$)  (i3e);
      \vertex at ($(i1e) + (0cm, -1.2cm)$)  (i6e);
      \vertex[blob, shape=circle, minimum size=0.3cm, fill=gray] at ($(i6e) + (0.5cm, 0cm)$)  (i7e){};
      \vertex[blob, shape=circle, minimum size=0.3cm, fill=gray] at ($(i7e) + (1cm, 0cm)$)  (i8e) {};
      \vertex[blob, shape=circle, minimum size=0.3cm, fill=gray] at ($(i8e) + (1cm, 0cm)$)  (i9e) {};
      \vertex at ($(i9e) + (0.5cm, 0cm)$)  (i10e);
      \vertex at ($(i9e) + (0.5cm, 0.6cm)$)  (i11e);
      \diagram* {
        (i1e)--[soft](i2e)--[soft](i3e), (i6e)--[soft](i7e)--[soft](i8e)--[hard  ](i9e)--[soft](i10e), (i2e)--[graviton](i7e),  (i2e)--[graviton](i8e), (i9e)--[graviton](i11e)
      };
      \draw node[left] at (i1e) {$u_1$};
      \draw node[left] at (i6e) {$u_2$};
        \end{feynman}
        \node at ($(i8e) + (-0.5cm, -0.6cm)$) {\NMC{1}{5}};
  \begin{feynman}
      \vertex at ($(i6b) + (0cm, -1.5cm)$) (i1f);
      \vertex[blob, shape=circle, minimum size=0.3cm, fill=gray] at ($(i1f) + (1cm, 0cm)$) (i2f) {};
      \vertex[blob, shape=circle, minimum size=0.3cm, fill=gray] at ($(i2f) + (1cm, 0cm)$)  (i3f);
      \vertex[blob, shape=circle, minimum size=0.3cm, fill=gray] at ($(i2f) + (0cm, -0.6cm)$)  (i9f) {};
      \vertex at ($(i1f) + (0cm, -1.2cm)$)  (i6f);
      \vertex[blob, shape=circle, minimum size=0.3cm, fill=gray] at ($(i6f) + (0.5cm, 0cm)$)  (i7f){};
      \vertex[blob, shape=circle, minimum size=0.3cm, fill=gray] at ($(i7f) + (1cm, 0cm)$)  (i8f) {};
      \vertex at ($(i8f) + (0.5cm, 0cm)$)  (i10f);
      \vertex at ($(i10f) + (0cm, 0.6cm)$)  (i11f);
      \diagram* {
        (i1f)--[soft](i2f)--[soft](i3f), (i6f)--[soft](i7f)--[soft](i8f)--[soft](i10f), (i2f)--[graviton](i9f)--[graviton](i7f), (i2f)--[graviton](i11f), (i9f)--[graviton](i8f)
      };
      \draw node[left] at (i1f) {$u_1$};
      \draw node[left] at (i6f) {$u_2$};
        \end{feynman}
        \node at ($(i7f) + (0.5cm, -0.6cm)$) {\NMC{1}{6}};
  \begin{feynman}
      \vertex at ($(i3f) + (1.6cm, 0cm)$) (i1g);
      \vertex[blob, shape=circle, minimum size=0.3cm, fill=gray] at ($(i1g) + (1cm, 0cm)$) (i2g) {};
      \vertex[blob, shape=circle, minimum size=0.3cm, fill=gray] at ($(i2g) + (1cm, 0cm)$)  (i3g);
      \vertex[blob, shape=circle, minimum size=0.3cm, fill=gray] at ($(i2g) + (0cm, -0.6cm)$)  (i4g) {};
      \vertex at ($(i1g) + (0cm, -1.2cm)$)  (i6g);
      \vertex[blob, shape=circle, minimum size=0.3cm, fill=gray] at ($(i6g) + (0.5cm, 0cm)$)  (i7g){};
      \vertex[blob, shape=circle, minimum size=0.3cm, fill=gray] at ($(i7g) + (1cm, 0cm)$)  (i8g) {};
      \vertex at ($(i8g) + (0.5cm, 0cm)$)  (i10g);
      \vertex at ($(i10g) + (0cm, 0.6cm)$)  (i11g);
      \diagram* {
        (i1g)--[soft](i2g)--[soft](i3g), (i6g)--[soft](i7g)--[soft](i8g)--[soft](i10g), (i2g)--[graviton](i4g)--[graviton](i8g), (i4g)--[graviton](i7g), (i4g)--[graviton](i11g)
      };
      \draw node[left] at (i1g) {$u_1$};
      \draw node[left] at (i6g) {$u_2$};
        \end{feynman}
        \node at ($(i7g) + (0.5cm, -0.6cm)$) {\NMC{1}{7}};
 \begin{feynman}
      \vertex at ($(i3g) + (1.6cm, 0cm)$) (i1h);
      \vertex[blob, shape=circle, minimum size=0.3cm, fill=gray] at ($(i1h) + (1cm, 0cm)$) (i2h) {};
      \vertex[blob, shape=circle, minimum size=0.3cm, fill=gray] at ($(i2h) + (1cm, 0cm)$)  (i3h);
      \vertex[blob, shape=circle, minimum size=0.3cm, fill=gray] at ($(i2h) + (0cm, -0.6cm)$)  (i4h) {};
      \vertex at ($(i1h) + (0cm, -1.2cm)$)  (i6h);
      \vertex[blob, shape=circle, minimum size=0.3cm, fill=gray] at ($(i6h) + (0.5cm, 0cm)$)  (i7h){};
      \vertex[blob, shape=circle, minimum size=0.3cm, fill=gray] at ($(i7h) + (1cm, 0cm)$)  (i8h) {};
      \vertex at ($(i8h) + (0.5cm, 0cm)$)  (i10h);
      \vertex at ($(i8h) + (0.5cm, 0.6cm)$)  (i11h);
      \diagram* {
        (i1h)--[soft](i2h)--[soft](i3h), (i6h)--[soft](i7h)--[soft](i8h)--[soft](i10h), (i2h)--[graviton](i4h)--[graviton](i7h),  (i4h)--[graviton](i8h), (i8h)--[graviton](i11h)
      };
      \draw node[left] at (i1h) {$u_1$};
      \draw node[left] at (i6h) {$u_2$};
        \end{feynman}
        \node at ($(i8h) + (-0.5cm, -0.6cm)$) {\NMC{1}{8}};
 \end{tikzpicture}\\
 \vspace{0.3cm}
 \begin{tikzpicture}[baseline=(current bounding box.center)]
  \begin{feynman}
    \vertex (i1g);
    \vertex[blob, shape=circle, minimum size=0.3cm, fill=gray] at ($(i1g) + (1cm, 0cm)$) (i2g) {};
    \vertex[blob, shape=circle, minimum size=0.3cm, fill=gray] at ($(i2g) + (1cm, 0cm)$)  (i3g);
    \vertex[blob, shape=circle, minimum size=0.3cm, fill=gray] at ($(i2g) + (0.5cm, -0.6cm)$)  (i9g) {};
    \vertex at ($(i1g) + (0cm, -1.2cm)$)  (i6g);
    \vertex[blob, shape=circle, minimum size=0.3cm, fill=gray] at ($(i6g) + (0.5cm, 0cm)$)  (i7g){};
    \vertex[blob, shape=circle, minimum size=0.3cm, fill=gray] at ($(i7g) + (1cm, 0cm)$)  (i8g) {};
    \vertex at ($(i8g) + (0.5cm, 0cm)$)  (i10g);
    \vertex at ($(i10g) + (0cm, 0.6cm)$)  (i11g);
    \diagram* {
      (i1g)--[soft](i2g)--[soft](i3g), (i6g)--[soft](i7g)--[soft](i8g)--[soft](i10g), (i2g)--[graviton](i7g)--[graviton](i9g)--[graviton](i8g),  (i9g)--[graviton](i11g)
    };
    \draw node[left] at (i1g) {$u_1$};
    \draw node[left] at (i6g) {$u_2$};
      \end{feynman}
      \node at ($(i8g) + (-0.5cm, -0.6cm)$) {\NMC{1}{9}};
    \begin{feynman}
      \vertex at ($(i3g) + (2cm, 0cm)$) (i1h);
      \vertex[blob, shape=circle, minimum size=0.3cm, fill=gray] at ($(i1h) + (0.5cm, 0cm)$) (i2h) {};
    \vertex at ($(i2h) + (1cm, 0cm)$)  (i3h);
    \vertex at ($(i1h) + (0cm, -1.2cm)$)  (i6h);
    \vertex[blob, shape=circle, minimum size=0.3cm, fill=gray] at ($(i6h) + (0.5cm, 0cm)$)  (i7h){};
    \vertex[blob, shape=circle, minimum size=0.3cm, fill=gray] at ($(i7h) + (1cm, 0cm)$)  (i8h) {};
    \vertex[blob, shape=circle, minimum size=0.3cm, fill=gray] at ($(i8h) + (0.5cm, 0cm)$)  (i9h) {};
    \vertex at ($(i9h) + (0.5cm, 0cm)$)  (i10h);
    \vertex at ($(i10h) + (0cm, 0.6cm)$)  (i11h);
      \diagram* {
        (i1h)--[soft](i2h)--[soft](i3h), (i6h)--[soft](i7h)--[soft](i8h)--[hard](i9h)--[soft](i10h), (i2h)--[graviton](i7h), (i7h)--[graviton, bend left=50](i8h),  (i9h)--[graviton](i11h)
      };
      \draw node[left] at (i1h) {$u_1$};
      \draw node[left] at (i6h) {$u_2$};
        \end{feynman}
        \node at ($(i8h) + (-0.25cm, -0.6cm)$) {\NMC{1}{10}};
  \begin{feynman}
      \vertex at ($(i3h) + (2cm, 0cm)$) (i1i);
      \vertex[blob, shape=circle, minimum size=0.3cm, fill=gray] at ($(i1i) + (0.5cm, 0cm)$) (i2i) {};
    \vertex at ($(i2i) + (1cm, 0cm)$)  (i3i);
    \vertex at ($(i1i) + (0cm, -1.2cm)$)  (i6i);
    \vertex[blob, shape=circle, minimum size=0.3cm, fill=gray] at ($(i6i) + (0.5cm, 0cm)$)  (i7i){};
    \vertex[blob, shape=circle, minimum size=0.3cm, fill=gray] at ($(i7i) + (0.5cm, 0cm)$)  (i4i){};
    \vertex[blob, shape=circle, minimum size=0.3cm, fill=gray] at ($(i4i) + (0.7cm, 0cm)$)  (i8i) {};
    \vertex at ($(i8i) + (0.5cm, 0cm)$)  (i10i);
    \vertex at ($(i10i) + (0cm, 0.6cm)$)  (i11i);
      \diagram* {
        (i1i)--[soft](i2i)--[soft](i3i), (i6i)--[soft](i7i)--[hard  ](i4i)--[soft](i8i)--[soft](i10i), (i2i)--[graviton](i7i), (i4i)--[graviton, bend left=50](i8i),  (i8i)--[graviton](i11i)
      };
      \draw node[left] at (i1i) {$u_1$};
      \draw node[left] at (i6i) {$u_2$};
        \end{feynman}
        \node at ($(i8i) + (-0.25cm, -0.6cm)$) {\NMC{1}{11}};

\end{tikzpicture}
\caption{Next-to-maximal cut topologies for the waveform. }
\label{fig:2PMwaveformnmc}
 \end{figure}\allowdisplaybreaks
\begin{figure} [t]
  \centering
  \begin{tikzpicture}[baseline=(current bounding box.center)]
    \begin{feynman}
      \vertex (i1a);
      \vertex[blob, shape=circle, minimum size=0.3cm, fill=gray] at ($(i1a) + (1cm, 0cm)$) (i2a) {};
      \vertex at ($(i2a) + (1cm, 0cm)$)  (i5a);
      \vertex at ($(i1a) + (0cm, -1.2cm)$)  (i6a);
      \vertex[blob, shape=circle, minimum size=0.3cm, fill=gray] at ($(i6a) + (0.5cm, 0cm)$)  (i7a){};
      \vertex[blob, shape=circle, minimum size=0.3cm, fill=gray] at ($(i7a) + (1cm, 0cm)$)  (i8a) {};
      \vertex at ($(i2a) + (1cm, -0.6cm)$)  (i9a);
      \vertex at ($(i8a) + (0.5cm, 0cm)$)  (i10a);
      \diagram* {
        (i1a)--[soft](i2a)--[soft](i5a), (i6a)--[soft](i7a)--[soft](i8a)--[soft](i10a), (i2a)--[graviton](i7a), (i2a)--[graviton](i8a), (i2a)--[graviton](i9a)
      };
      \draw node[left] at (i1a) {$u_1$};
      \draw node[left] at (i6a) {$u_2$};
        \end{feynman}
        \node at ($(i8a) + (-0.5cm, -0.6cm)$) {\NMC{2}{1}};
  \end{tikzpicture}\qquad  \begin{tikzpicture}[baseline=(current bounding box.center)]
    \begin{feynman}
      \vertex (i1a);
      \vertex[blob, shape=circle, minimum size=0.3cm, fill=gray] at ($(i1a) + (1cm, 0cm)$) (i2a) {};
      \vertex at ($(i2a) + (1cm, 0cm)$)  (i5a);
      \vertex at ($(i1a) + (0cm, -1.2cm)$)  (i6a);
      \vertex[blob, shape=circle, minimum size=0.3cm, fill=gray] at ($(i6a) + (0.5cm, 0cm)$)  (i7a){};
      \vertex[blob, shape=circle, minimum size=0.3cm, fill=gray] at ($(i7a) + (1cm, 0cm)$)  (i8a) {};
      \vertex at ($(i2a) + (1cm, -0.6cm)$)  (i9a);
      \vertex at ($(i8a) + (0.5cm, 0cm)$)  (i10a);
      \diagram* {
        (i1a)--[soft](i2a)--[soft](i5a), (i6a)--[soft](i7a)--[soft](i8a)--[soft](i10a), (i2a)--[graviton](i7a), (i2a)--[graviton](i8a), (i8a)--[graviton](i9a)
      };
      \draw node[left] at (i1a) {$u_1$};
      \draw node[left] at (i6a) {$u_2$};
        \end{feynman}
        \node at ($(i8a) + (-0.5cm, -0.6cm)$) {\NMC{2}{2}};
  \end{tikzpicture}\qquad  \begin{tikzpicture}[baseline=(current bounding box.center)]
    \begin{feynman}
      \vertex (i1a);
      \vertex[blob, shape=circle, minimum size=0.3cm, fill=gray] at ($(i1a) + (1cm, 0cm)$) (i2a) {};
      \vertex at ($(i2a) + (1cm, 0cm)$)  (i5a);
      \vertex at ($(i1a) + (0cm, -1.2cm)$)  (i6a);
      \vertex[blob, shape=circle, minimum size=0.3cm, fill=gray] at ($(i6a) + (0.5cm, 0cm)$)  (i7a){};
      \vertex[blob, shape=circle, minimum size=0.3cm, fill=gray] at ($(i7a) + (1cm, 0cm)$)  (i8a) {};
      \vertex at ($(i2a) + (1cm, -0.6cm)$)  (i9a);
      \vertex at ($(i8a) + (0.5cm, 0cm)$)  (i10a);
      \diagram* {
        (i1a)--[soft](i2a)--[soft](i5a), (i6a)--[soft](i7a)--[soft](i8a)--[soft](i10a), (i2a)--[graviton](i7a), (i7a)--[graviton, bend left=50](i8a), (i8a)--[graviton](i9a)
      };
      \draw node[left] at (i1a) {$u_1$};
      \draw node[left] at (i6a) {$u_2$};
        \end{feynman}
        \node at ($(i8a) + (-0.5cm, -0.6cm)$) {\NMC{2}{3}};
  \end{tikzpicture}
  \caption{Next-to-next-to-maximal cut topologies for the  waveform. }
  \label{fig:2PMwaveformnnmc}
\end{figure}
\end{center}
\newpage
We do not consider any cuts with a loop made of a single graviton propagator on the external leg, such as
\begin{center}
 \begin{tikzpicture}[baseline=(current bounding box.center)]
   \begin{feynman}
     \vertex (i1);
     \vertex[blob, shape=circle, minimum size=0.3cm, fill=gray] at ($(i1) + (1.5cm, 0cm)$) (i2) {};
     \vertex at ($(i2) + (0.5cm, 0cm)$)  (i3);
     \vertex[blob, shape=circle, minimum size=0.3cm, fill=gray] at ($(i2) + (0cm, -0.6cm)$)  (i11) {};
     \vertex at ($(i1) + (-0.8cm, -1.2cm)$)  (i5);
     \vertex[blob, shape=circle, minimum size=0.3cm, fill=gray] at ($(i5) + (0.5cm, 0cm)$)  (i6){};
     \vertex[blob, shape=circle, minimum size=0.3cm, fill=gray] at ($(i6) + (0.8cm, 0cm)$)  (i7) {};
     \vertex[blob, shape=circle, minimum size=0.3cm, fill=gray] at ($(i7) + (1cm, 0cm)$)  (i8) {};
     \vertex at ($(i8) + (0.5cm, 0cm)$)  (i9);
     \vertex at ($(i9) + (0cm, 0.6cm)$)  (i10);
     \diagram* {
       (i1)--[soft](i2)--[soft](i3), (i5)--[soft](i6)--[soft](i7)--[hard](i8)--[soft](i9), (i2)--[graviton](i11)--[graviton](i8), (i6)--[graviton, bend left=50](i7), (i11)--[graviton](i10)
     };
     \end{feynman}
     \draw node[left] at (i1) {$u_1$};
     \draw node[left] at (i5) {$u_2$};
 \end{tikzpicture}
\end{center}
and its parent topologies, as they produce scaleless integrals that vanish in dimensional regularization. Note that MC10 is not a scaleless bubble as one may expect, because worldline energy conservation introduces a scale into the loop integral.

We implement the maximal cut method and fix the integrand in $D$ dimensions up to the N$^2$MC cuts. The ansatze for topologies 1 through 10 in Fig.~\ref{fig:2PMwaveformmc} are included in the ancillary file. The undetermined coefficients correspond to either the freedom of rearranging contact terms or to master integrals that vanish. As a check, we performed tensor and IBP reduction and we determined the coefficients of each master integral.\footnote{In practice, instead of adding all diagrams, it is useful to reduce the N$^2$MC's separately. The overlapping contributions from each cut then serve as a consistent check.}  Our results for these coefficients are given in the ancillary file for this paper and are found to agree with those in \cite{Brandhuber:2023hhy} near $D=4$.\footnote{We are grateful to Stefano de Angelis for kindly providing these coefficients in computer-readable form for comparison.} (see also \cite{Herderschee:2023fxh,Georgoudis:2023eke}).

\section{Conclusion}
In this paper, we have shown that the familiar generalized unitarity method from quantum field theory can be extended to the construction of integrands in worldline field theory and applied to the computation of classical gravitational scattering observables. This method provides a streamlined way to construct complicated integrands for worldline observables, by recycling calculations of lower-perturbative-order and/or lower-point processes, fully bypassing the need to use Feynman diagrams and rules. We illustrated this method by applying it to explicit examples, including Compton-type amplitudes, the conservative on-shell action, as well as the next-to-leading order waveform, and checking agreement with known results.

The tools we have introduced in this paper open the door for further exploration of the structure of worldline observables. Perhaps the most interesting of these would be a systematic investigation of the double copy, going beyond Refs.~\cite{Goldberger:2016iau,Shen:2018ebu,Shi:2021qsb,Comberiati:2022cpm}.

Further improvements of the method seem within reach. For instance one might attempt to dispense altogether with the need for ansatze by choosing a global basis of worldline integrands along the lines of Ref.~\cite{Bern:2024vqs}. Perhaps, one could also extend the method to the background-field amplitudes that resum the metric and geodesic motion as explained in Refs.~\cite{Cheung:2023lnj,Cheung:2024byb} (see also \cite{Kosmopoulos:2023bwc}). A natural future direction is to extend the method to spinning worldlines. Since the propagators for spin degrees of freedom have simple poles, we expect no conceptual difficulty in achieving this goal.

\paragraph{Note added:} We are grateful to Kays Haddad, Gustav U. Jakobsen, Gustav Mogull and Jan Plefka for sharing a draft of their upcoming and complementary work \cite{Haddad:2025cmw}, and for coordinating submission. The realization that the complexified amplitudes defined in this paper satisfy a subleading soft theorem was triggered by comments in their draft.

\subsection*{Acknowledgments}
We thank Stefano de Angelis and Jonah Berean-Dutcher for many helpful discussions and Jan Plefka for comments on the draft. We also thank Nathan Castet for finding several typos in our first draft. VFH acknowledges support from the Simons Foundation.

 \bibliographystyle{utphys-modified}
 \bibliography{worldline_unitarity}

@article{LIGOScientific:2016aoc,
    author = "Abbott, B. P. and others",
    collaboration = "LIGO Scientific, Virgo",
    title = "{Observation of Gravitational Waves from a Binary Black Hole Merger}",
    eprint = "1602.03837",
    archivePrefix = "arXiv",
    primaryClass = "gr-qc",
    reportNumber = "LIGO-P150914",
    doi = "10.1103/PhysRevLett.116.061102",
    journal = "Phys. Rev. Lett.",
    volume = "116",
    number = "6",
    pages = "061102",
    year = "2016"
}

@article{LIGOScientific:2017vwq,
    author = "Abbott, B. P. and others",
    collaboration = "LIGO Scientific, Virgo",
    title = "{GW170817: Observation of Gravitational Waves from a Binary Neutron Star Inspiral}",
    eprint = "1710.05832",
    archivePrefix = "arXiv",
    primaryClass = "gr-qc",
    reportNumber = "LIGO-P170817",
    doi = "10.1103/PhysRevLett.119.161101",
    journal = "Phys. Rev. Lett.",
    volume = "119",
    number = "16",
    pages = "161101",
    year = "2017"
}

@article{Purrer:2019jcp,
    author = {P{\"u}rrer, Michael and Haster, Carl-Johan},
    title = "{Gravitational waveform accuracy requirements for future ground-based detectors}",
    eprint = "1912.10055",
    archivePrefix = "arXiv",
    primaryClass = "gr-qc",
    doi = "10.1103/PhysRevResearch.2.023151",
    journal = "Phys. Rev. Res.",
    volume = "2",
    number = "2",
    pages = "023151",
    year = "2020"
}

@article{Damour:2017zjx,
    author = "Damour, Thibault",
    title = "{High-energy gravitational scattering and the general relativistic two-body problem}",
    eprint = "1710.10599",
    archivePrefix = "arXiv",
    primaryClass = "gr-qc",
    doi = "10.1103/PhysRevD.97.044038",
    journal = "Phys. Rev. D",
    volume = "97",
    number = "4",
    pages = "044038",
    year = "2018"
}

@article{Cheung:2018wkq,
    author = "Cheung, Clifford and Rothstein, Ira Z. and Solon, Mikhail P.",
    title = "{From Scattering Amplitudes to Classical Potentials in the Post-Minkowskian Expansion}",
    eprint = "1808.02489",
    archivePrefix = "arXiv",
    primaryClass = "hep-th",
    reportNumber = "CALT-TH-2018-031",
    doi = "10.1103/PhysRevLett.121.251101",
    journal = "Phys. Rev. Lett.",
    volume = "121",
    number = "25",
    pages = "251101",
    year = "2018"
}

@article{Bern:2019nnu,
    author = "Bern, Zvi and Cheung, Clifford and Roiban, Radu and Shen, Chia-Hsien and Solon, Mikhail P. and Zeng, Mao",
    title = "{Scattering Amplitudes and the Conservative Hamiltonian for Binary Systems at Third Post-Minkowskian Order}",
    eprint = "1901.04424",
    archivePrefix = "arXiv",
    primaryClass = "hep-th",
    reportNumber = "CALT-TH 2019-002, UCLA/TEP/2019/101",
    doi = "10.1103/PhysRevLett.122.201603",
    journal = "Phys. Rev. Lett.",
    volume = "122",
    number = "20",
    pages = "201603",
    year = "2019"
}

@article{KoemansCollado:2019ggb,
    author = "Koemans Collado, Arnau and Di Vecchia, Paolo and Russo, Rodolfo",
    title = "{Revisiting the second post-Minkowskian eikonal and the dynamics of binary black holes}",
    eprint = "1904.02667",
    archivePrefix = "arXiv",
    primaryClass = "hep-th",
    reportNumber = "QMUL-PH-19-08",
    doi = "10.1103/PhysRevD.100.066028",
    journal = "Phys. Rev. D",
    volume = "100",
    number = "6",
    pages = "066028",
    year = "2019"
}

@article{Bern:2019crd,
    author = "Bern, Zvi and Cheung, Clifford and Roiban, Radu and Shen, Chia-Hsien and Solon, Mikhail P. and Zeng, Mao",
    title = "{Black Hole Binary Dynamics from the Double Copy and Effective Theory}",
    eprint = "1908.01493",
    archivePrefix = "arXiv",
    primaryClass = "hep-th",
    reportNumber = "CERN-TH-2019-128, CALT-TH 2019-026, UCLA/TEP/2019/103",
    doi = "10.1007/JHEP10(2019)206",
    journal = "JHEP",
    volume = "10",
    pages = "206",
    year = "2019"
}

@article{Damour:2019lcq,
    author = "Damour, Thibault",
    title = "{Classical and quantum scattering in post-Minkowskian gravity}",
    eprint = "1912.02139",
    archivePrefix = "arXiv",
    primaryClass = "gr-qc",
    doi = "10.1103/PhysRevD.102.024060",
    journal = "Phys. Rev. D",
    volume = "102",
    number = "2",
    pages = "024060",
    year = "2020"
}

@article{Cristofoli:2020uzm,
    author = "Cristofoli, Andrea and Damgaard, Poul H. and Di Vecchia, Paolo and Heissenberg, Carlo",
    title = "{Second-order Post-Minkowskian scattering in arbitrary dimensions}",
    eprint = "2003.10274",
    archivePrefix = "arXiv",
    primaryClass = "hep-th",
    reportNumber = "SAGEX-20-05-E, NORDITA 2020-026",
    doi = "10.1007/JHEP07(2020)122",
    journal = "JHEP",
    volume = "07",
    pages = "122",
    year = "2020"
}

@article{Damour:2020tta,
    author = "Damour, Thibault",
    title = "{Radiative contribution to classical gravitational scattering at the third order in $G$}",
    eprint = "2010.01641",
    archivePrefix = "arXiv",
    primaryClass = "gr-qc",
    doi = "10.1103/PhysRevD.102.124008",
    journal = "Phys. Rev. D",
    volume = "102",
    number = "12",
    pages = "124008",
    year = "2020"
}

@article{Kalin:2020mvi,
    author = {K{\"a}lin, Gregor and Porto, Rafael A.},
    title = "{Post-Minkowskian Effective Field Theory for Conservative Binary Dynamics}",
    eprint = "2006.01184",
    archivePrefix = "arXiv",
    primaryClass = "hep-th",
    reportNumber = "DESY20-077, SLAC-PUB-17529",
    doi = "10.1007/JHEP11(2020)106",
    journal = "JHEP",
    volume = "11",
    pages = "106",
    year = "2020"
}

@article{AccettulliHuber:2020dal,
    author = "Accettulli Huber, Manuel and Brandhuber, Andreas and De Angelis, Stefano and Travaglini, Gabriele",
    title = "{From amplitudes to gravitational radiation with cubic interactions and tidal effects}",
    eprint = "2012.06548",
    archivePrefix = "arXiv",
    primaryClass = "hep-th",
    reportNumber = "QMUL-PH-20-19, SAGEX-20-19-E",
    doi = "10.1103/PhysRevD.103.045015",
    journal = "Phys. Rev. D",
    volume = "103",
    number = "4",
    pages = "045015",
    year = "2021"
}

@article{Bern:2021yeh,
    author = "Bern, Zvi and Parra-Martinez, Julio and Roiban, Radu and Ruf, Michael S. and Shen, Chia-Hsien and Solon, Mikhail P. and Zeng, Mao",
    title = "{Scattering Amplitudes, the Tail Effect, and Conservative Binary Dynamics at O(G4)}",
    eprint = "2112.10750",
    archivePrefix = "arXiv",
    primaryClass = "hep-th",
    doi = "10.1103/PhysRevLett.128.161103",
    journal = "Phys. Rev. Lett.",
    volume = "128",
    number = "16",
    pages = "161103",
    year = "2022"
}

@article{Bern:2021dqo,
    author = "Bern, Zvi and Parra-Martinez, Julio and Roiban, Radu and Ruf, Michael S. and Shen, Chia-Hsien and Solon, Mikhail P. and Zeng, Mao",
    title = "{Scattering Amplitudes and Conservative Binary Dynamics at ${\cal O}(G^4)$}",
    eprint = "2101.07254",
    archivePrefix = "arXiv",
    primaryClass = "hep-th",
    reportNumber = "CALT-TH-2021-004, FR-PHENO-2021-03, OUTP-21-03P",
    doi = "10.1103/PhysRevLett.126.171601",
    journal = "Phys. Rev. Lett.",
    volume = "126",
    number = "17",
    pages = "171601",
    year = "2021"
}

@article{Herrmann:2021lqe,
    author = "Herrmann, Enrico and Parra-Martinez, Julio and Ruf, Michael S. and Zeng, Mao",
    title = "{Gravitational Bremsstrahlung from Reverse Unitarity}",
    eprint = "2101.07255",
    archivePrefix = "arXiv",
    primaryClass = "hep-th",
    reportNumber = "CALT-TH-2021-003, FR-PHENO-2021-02, OUTP-21-02P",
    doi = "10.1103/PhysRevLett.126.201602",
    journal = "Phys. Rev. Lett.",
    volume = "126",
    number = "20",
    pages = "201602",
    year = "2021"
}

@article{DiVecchia:2021ndb,
    author = "Di Vecchia, Paolo and Heissenberg, Carlo and Russo, Rodolfo and Veneziano, Gabriele",
    title = "{Radiation Reaction from Soft Theorems}",
    eprint = "2101.05772",
    archivePrefix = "arXiv",
    primaryClass = "hep-th",
    reportNumber = "CERN-TH-2021-008, NORDITA 2021-001, QMUL-PH-21-03, UUITP-03/21",
    doi = "10.1016/j.physletb.2021.136379",
    journal = "Phys. Lett. B",
    volume = "818",
    pages = "136379",
    year = "2021"
}

@article{DiVecchia:2021bdo,
    author = "Di Vecchia, Paolo and Heissenberg, Carlo and Russo, Rodolfo and Veneziano, Gabriele",
    title = "{The eikonal approach to gravitational scattering and radiation at $ \mathcal{O} $(G$^{3}$)}",
    eprint = "2104.03256",
    archivePrefix = "arXiv",
    primaryClass = "hep-th",
    reportNumber = "CERN-TH-2021-046, NORDITA 2021-028, QMUL-PH-21-17",
    doi = "10.1007/JHEP07(2021)169",
    journal = "JHEP",
    volume = "07",
    pages = "169",
    year = "2021"
}

@article{Herrmann:2021tct,
    author = "Herrmann, Enrico and Parra-Martinez, Julio and Ruf, Michael S. and Zeng, Mao",
    title = "{Radiative classical gravitational observables at $ \mathcal{O} $(G$^{3}$) from scattering amplitudes}",
    eprint = "2104.03957",
    archivePrefix = "arXiv",
    primaryClass = "hep-th",
    doi = "10.1007/JHEP10(2021)148",
    journal = "JHEP",
    volume = "10",
    pages = "148",
    year = "2021"
}

@article{Bjerrum-Bohr:2021vuf,
    author = "Bjerrum-Bohr, N. Emil J. and Damgaard, Poul H. and Plant{\'e}, Ludovic and Vanhove, Pierre",
    title = "{Classical gravity from loop amplitudes}",
    eprint = "2104.04510",
    archivePrefix = "arXiv",
    primaryClass = "hep-th",
    reportNumber = "IPhT-t21/015, CERN-TH-2021-052",
    doi = "10.1103/PhysRevD.104.026009",
    journal = "Phys. Rev. D",
    volume = "104",
    number = "2",
    pages = "026009",
    year = "2021"
}

@article{Bjerrum-Bohr:2021din,
    author = "Bjerrum-Bohr, N. Emil J. and Damgaard, Poul H. and Plant{\'e}, Ludovic and Vanhove, Pierre",
    title = "{The amplitude for classical gravitational scattering at third Post-Minkowskian order}",
    eprint = "2105.05218",
    archivePrefix = "arXiv",
    primaryClass = "hep-th",
    reportNumber = "IPhT-t21/028, CERN-TH-2021-073",
    doi = "10.1007/JHEP08(2021)172",
    journal = "JHEP",
    volume = "08",
    pages = "172",
    year = "2021"
}

@article{Jakobsen:2021smu,
    author = "Jakobsen, Gustav Uhre and Mogull, Gustav and Plefka, Jan and Steinhoff, Jan",
    title = "{Classical Gravitational Bremsstrahlung from a Worldline Quantum Field Theory}",
    eprint = "2101.12688",
    archivePrefix = "arXiv",
    primaryClass = "gr-qc",
    reportNumber = "HU-EP-21/03-RTG",
    doi = "10.1103/PhysRevLett.126.201103",
    journal = "Phys. Rev. Lett.",
    volume = "126",
    number = "20",
    pages = "201103",
    year = "2021"
}

@article{Dlapa:2021npj,
    author = {Dlapa, Christoph and K{\"a}lin, Gregor and Liu, Zhengwen and Porto, Rafael A.},
    title = "{Dynamics of binary systems to fourth Post-Minkowskian order from the effective field theory approach}",
    eprint = "2106.08276",
    archivePrefix = "arXiv",
    primaryClass = "hep-th",
    reportNumber = "DESY 21-093, DESY-21-093, MPP-2021-83",
    doi = "10.1016/j.physletb.2022.137203",
    journal = "Phys. Lett. B",
    volume = "831",
    pages = "137203",
    year = "2022"
}

@article{Brandhuber:2021eyq,
    author = "Brandhuber, Andreas and Chen, Gang and Travaglini, Gabriele and Wen, Congkao",
    title = "{Classical gravitational scattering from a gauge-invariant double copy}",
    eprint = "2108.04216",
    archivePrefix = "arXiv",
    primaryClass = "hep-th",
    reportNumber = "QMUL-PH-21-18, SAGEX-21-07",
    doi = "10.1007/JHEP10(2021)118",
    journal = "JHEP",
    volume = "10",
    pages = "118",
    year = "2021"
}

@article{Dlapa:2021vgp,
    author = {Dlapa, Christoph and K{\"a}lin, Gregor and Liu, Zhengwen and Porto, Rafael A.},
    title = "{Conservative Dynamics of Binary Systems at Fourth Post-Minkowskian Order in the Large-Eccentricity Expansion}",
    eprint = "2112.11296",
    archivePrefix = "arXiv",
    primaryClass = "hep-th",
    reportNumber = "DESY 21-226",
    doi = "10.1103/PhysRevLett.128.161104",
    journal = "Phys. Rev. Lett.",
    volume = "128",
    number = "16",
    pages = "161104",
    year = "2022"
}

@article{Cristofoli:2021vyo,
    author = "Cristofoli, Andrea and Gonzo, Riccardo and Kosower, David A. and O'Connell, Donal",
    title = "{Waveforms from amplitudes}",
    eprint = "2107.10193",
    archivePrefix = "arXiv",
    primaryClass = "hep-th",
    doi = "10.1103/PhysRevD.106.056007",
    journal = "Phys. Rev. D",
    volume = "106",
    number = "5",
    pages = "056007",
    year = "2022"
}

@article{Manohar:2022dea,
    author = "Manohar, Aneesh V. and Ridgway, Alexander K. and Shen, Chia-Hsien",
    title = "{Radiated Angular Momentum and Dissipative Effects in Classical Scattering}",
    eprint = "2203.04283",
    archivePrefix = "arXiv",
    primaryClass = "hep-th",
    doi = "10.1103/PhysRevLett.129.121601",
    journal = "Phys. Rev. Lett.",
    volume = "129",
    number = "12",
    pages = "121601",
    year = "2022"
}

@article{Bern:2022jvn,
    author = "Bern, Zvi and Parra-Martinez, Julio and Roiban, Radu and Ruf, Michael S. and Shen, Chia-Hsien and Solon, Mikhail P. and Zeng, Mao",
    title = "{Scattering amplitudes and conservative dynamics at the fourth post-Minkowskian order}",
    doi = "10.22323/1.416.0051",
    journal = "PoS",
    volume = "LL2022",
    pages = "051",
    year = "2022"
}

@article{Dlapa:2022lmu,
    author = {Dlapa, Christoph and K{\"a}lin, Gregor and Liu, Zhengwen and Neef, Jakob and Porto, Rafael A.},
    title = "{Radiation Reaction and Gravitational Waves at Fourth Post-Minkowskian Order}",
    eprint = "2210.05541",
    archivePrefix = "arXiv",
    primaryClass = "hep-th",
    doi = "10.1103/PhysRevLett.130.101401",
    journal = "Phys. Rev. Lett.",
    volume = "130",
    number = "10",
    pages = "101401",
    year = "2023"
}

@article{Herderschee:2023fxh,
    author = "Herderschee, Aidan and Roiban, Radu and Teng, Fei",
    title = "{The sub-leading scattering waveform from amplitudes}",
    eprint = "2303.06112",
    archivePrefix = "arXiv",
    primaryClass = "hep-th",
    reportNumber = "LCTP-23-04",
    doi = "10.1007/JHEP06(2023)004",
    journal = "JHEP",
    volume = "06",
    pages = "004",
    year = "2023"
}

@article{Georgoudis:2023eke,
    author = "Georgoudis, Alessandro and Heissenberg, Carlo and Russo, Rodolfo",
    title = "{An eikonal-inspired approach to the gravitational scattering waveform}",
    eprint = "2312.07452",
    archivePrefix = "arXiv",
    primaryClass = "hep-th",
    reportNumber = "QMUL-PH-23-34",
    doi = "10.1007/JHEP03(2024)089",
    journal = "JHEP",
    volume = "03",
    pages = "089",
    year = "2024"
}

@article{Brandhuber:2023hhy,
    author = "Brandhuber, Andreas and Brown, Graham R. and Chen, Gang and De Angelis, Stefano and Gowdy, Joshua and Travaglini, Gabriele",
    title = "{One-loop gravitational bremsstrahlung and waveforms from a heavy-mass effective field theory}",
    eprint = "2303.06111",
    archivePrefix = "arXiv",
    primaryClass = "hep-th",
    reportNumber = "QMUL-22-28,SAGEX-22-32-E",
    doi = "10.1007/JHEP06(2023)048",
    journal = "JHEP",
    volume = "06",
    pages = "048",
    year = "2023"
}

@article{Barack:2023oqp,
    author = "Barack, Leor and others",
    title = "{Comparison of post-Minkowskian and self-force expansions: Scattering in a scalar charge toy model}",
    eprint = "2304.09200",
    archivePrefix = "arXiv",
    primaryClass = "hep-th",
    doi = "10.1103/PhysRevD.108.024025",
    journal = "Phys. Rev. D",
    volume = "108",
    number = "2",
    pages = "024025",
    year = "2023"
}

@article{Dlapa:2023hsl,
    author = {Dlapa, Christoph and K{\"a}lin, Gregor and Liu, Zhengwen and Porto, Rafael A.},
    title = "{Bootstrapping the relativistic two-body problem}",
    eprint = "2304.01275",
    archivePrefix = "arXiv",
    primaryClass = "hep-th",
    reportNumber = "DESY 23-041",
    doi = "10.1007/JHEP08(2023)109",
    journal = "JHEP",
    volume = "08",
    pages = "109",
    year = "2023"
}

@article{Damgaard:2023ttc,
    author = "Damgaard, Poul H. and Hansen, Elias Roos and Plant{\'e}, Ludovic and Vanhove, Pierre",
    title = "{Classical observables from the exponential representation of the gravitational S-matrix}",
    eprint = "2307.04746",
    archivePrefix = "arXiv",
    primaryClass = "hep-th",
    reportNumber = "CERN-TH-2023-135, IPhT-T23/041, LAPTh-029/23",
    doi = "10.1007/JHEP09(2023)183",
    journal = "JHEP",
    volume = "09",
    pages = "183",
    year = "2023"
}

@article{Cheung:2023lnj,
    author = "Cheung, Clifford and Parra-Martinez, Julio and Rothstein, Ira Z. and Shah, Nabha and Wilson-Gerow, Jordan",
    title = "{Effective Field Theory for Extreme Mass Ratio Binaries}",
    eprint = "2308.14832",
    archivePrefix = "arXiv",
    primaryClass = "hep-th",
    reportNumber = "CALT-TH 2023-035",
    doi = "10.1103/PhysRevLett.132.091402",
    journal = "Phys. Rev. Lett.",
    volume = "132",
    number = "9",
    pages = "091402",
    year = "2024"
}

@article{Kosmopoulos:2023bwc,
    author = "Kosmopoulos, Dimitrios and Solon, Mikhail P.",
    title = "{Gravitational self force from scattering amplitudes in curved space}",
    eprint = "2308.15304",
    archivePrefix = "arXiv",
    primaryClass = "hep-th",
    doi = "10.1007/JHEP03(2024)125",
    journal = "JHEP",
    volume = "03",
    pages = "125",
    year = "2024"
}

@article{Ivanov:2024sds,
    author = "Ivanov, Mikhail M. and Li, Yue-Zhou and Parra-Martinez, Julio and Zhou, Zihan",
    title = "{Gravitational Raman Scattering in Effective Field Theory: A Scalar Tidal Matching at O(G3)}",
    eprint = "2401.08752",
    archivePrefix = "arXiv",
    primaryClass = "hep-th",
    reportNumber = "MIT-CTP/5664",
    doi = "10.1103/PhysRevLett.132.131401",
    journal = "Phys. Rev. Lett.",
    volume = "132",
    number = "13",
    pages = "131401",
    year = "2024",
    note = "[Erratum: Phys.Rev.Lett. 134, 159901 (2025)]"
}

@article{Correia:2024jgr,
    author = "Correia, Miguel and Isabella, Giulia",
    title = "{The Born regime of gravitational amplitudes}",
    eprint = "2406.13737",
    archivePrefix = "arXiv",
    primaryClass = "hep-th",
    doi = "10.1007/JHEP03(2025)144",
    journal = "JHEP",
    volume = "03",
    pages = "144",
    year = "2025"
}

@article{Driesse:2024xad,
    author = "Driesse, Mathias and Jakobsen, Gustav Uhre and Mogull, Gustav and Plefka, Jan and Sauer, Benjamin and Usovitsch, Johann",
    title = "{Conservative Black Hole Scattering at Fifth Post-Minkowskian and First Self-Force Order}",
    eprint = "2403.07781",
    archivePrefix = "arXiv",
    primaryClass = "hep-th",
    reportNumber = "HU-EP-24/08-RTG, CERN-TH-2024-032",
    doi = "10.1103/PhysRevLett.132.241402",
    journal = "Phys. Rev. Lett.",
    volume = "132",
    number = "24",
    pages = "241402",
    year = "2024"
}

@article{Bern:2024adl,
    author = "Bern, Zvi and Herrmann, Enrico and Roiban, Radu and Ruf, Michael S. and Smirnov, Alexander V. and Smirnov, Vladimir A. and Zeng, Mao",
    title = "{Amplitudes, supersymmetric black hole scattering at $ \mathcal{O}\left({G}^5\right) $, and loop integration}",
    eprint = "2406.01554",
    archivePrefix = "arXiv",
    primaryClass = "hep-th",
    doi = "10.1007/JHEP10(2024)023",
    journal = "JHEP",
    volume = "10",
    pages = "023",
    year = "2024"
}

@article{Cheung:2024byb,
    author = "Cheung, Clifford and Parra-Martinez, Julio and Rothstein, Ira Z. and Shah, Nabha and Wilson-Gerow, Jordan",
    title = "{Gravitational scattering and beyond from extreme mass ratio effective field theory}",
    eprint = "2406.14770",
    archivePrefix = "arXiv",
    primaryClass = "hep-th",
    reportNumber = "CALT-TH 2024-023",
    doi = "10.1007/JHEP10(2024)005",
    journal = "JHEP",
    volume = "10",
    pages = "005",
    year = "2024"
}

@article{Driesse:2024feo,
    author = "Driesse, Mathias and Jakobsen, Gustav Uhre and Klemm, Albrecht and Mogull, Gustav and Nega, Christoph and Plefka, Jan and Sauer, Benjamin and Usovitsch, Johann",
    title = "{Emergence of Calabi{\textendash}Yau manifolds in high-precision black-hole scattering}",
    eprint = "2411.11846",
    archivePrefix = "arXiv",
    primaryClass = "hep-th",
    reportNumber = "HU-EP-24/32-RTG, QMUL-PH-24-26, BONN-TH-2024-15, TUM-HEP-1532/24",
    doi = "10.1038/s41586-025-08984-2",
    journal = "Nature",
    volume = "641",
    number = "8063",
    pages = "603--607",
    year = "2025"
}

@article{Heissenberg:2025ocy,
    author = "Heissenberg, Carlo",
    title = "{Radiation-reaction and angular momentum loss at O(G4)}",
    eprint = "2501.02904",
    archivePrefix = "arXiv",
    primaryClass = "hep-th",
    doi = "10.1103/xz14-87q7",
    journal = "Phys. Rev. D",
    volume = "111",
    number = "12",
    pages = "126012",
    year = "2025"
}

@article{Caron-Huot:2025tlq,
    author = "Caron-Huot, Simon and Correia, Miguel and Isabella, Giulia and Solon, Mikhail",
    title = "{Gravitational Wave Scattering via the Born Series: Scalar Tidal Matching to O(G7) and Beyond}",
    eprint = "2503.13593",
    archivePrefix = "arXiv",
    primaryClass = "hep-th",
    doi = "10.1103/qd3c-nfz6",
    journal = "Phys. Rev. Lett.",
    volume = "135",
    number = "19",
    pages = "191601",
    year = "2025"
}

@article{Ivanov:2025ozg,
    author = "Ivanov, Mikhail M. and Li, Yue-Zhou and Parra-Martinez, Julio and Zhou, Zihan",
    title = "{Resummation of Universal Tails in Gravitational Waveforms}",
    eprint = "2504.07862",
    archivePrefix = "arXiv",
    primaryClass = "hep-th",
    reportNumber = "MIT-CTP/5863",
    doi = "10.1103/jzd1-qzkt",
    journal = "Phys. Rev. Lett.",
    volume = "135",
    number = "14",
    pages = "141401",
    year = "2025"
}

@article{Georgoudis:2025vkk,
    author = "Georgoudis, Alessandro and Goncalves, Vasco and Heissenberg, Carlo and Parra-Martinez, Julio",
    title = "{Nonlinear Gravitational Memory in the Post-Minkowskian Expansion}",
    eprint = "2506.20733",
    archivePrefix = "arXiv",
    primaryClass = "hep-th",
    doi = "10.1103/8m17-s2y8",
    journal = "Phys. Rev. Lett.",
    volume = "136",
    number = "12",
    pages = "121401",
    year = "2026"
}

@article{Mogull:2025cfn,
    author = "Mogull, Gustav and Plefka, Jan and Stoldt, Kathrin",
    title = "{Radiated angular momentum from spinning black hole scattering trajectories}",
    eprint = "2506.20643",
    archivePrefix = "arXiv",
    primaryClass = "hep-th",
    reportNumber = "HU-EP-25/20-RTG, QMUL-PH-25-10",
    doi = "10.1103/my38-14k5",
    journal = "Phys. Rev. D",
    volume = "112",
    number = "12",
    pages = "124076",
    year = "2025"
}

@article{Hoogeveen:2025tew,
    author = "Hoogeveen, Jitze and Jakobsen, Gustav Uhre and Plefka, Jan",
    title = "{Spinning the probe in Kerr with WQFT}",
    eprint = "2506.14626",
    archivePrefix = "arXiv",
    primaryClass = "hep-th",
    reportNumber = "HU-EP-25/22-RTG",
    doi = "10.1007/JHEP10(2025)201",
    journal = "JHEP",
    volume = "10",
    pages = "201",
    year = "2025"
}

@article{Bern:2025zno,
    author = "Bern, Zvi and Herrmann, Enrico and Roiban, Radu and Ruf, Michael S. and Smirnov, Alexander V. and Smirnov, Vladimir A. and Zeng, Mao",
    title = "{Second-Order Self-Force Potential-Region Binary Dynamics at O(G5) in Supergravity}",
    eprint = "2509.17412",
    archivePrefix = "arXiv",
    primaryClass = "hep-th",
    doi = "10.1103/jmby-htz9",
    journal = "Phys. Rev. Lett.",
    volume = "136",
    number = "8",
    pages = "081401",
    year = "2026"
}

@article{Bertotti:1956pxu,
    author = "Bertotti, B.",
    title = "{On gravitational motion}",
    doi = "10.1007/bf02746175",
    journal = "Nuovo Cim.",
    volume = "4",
    number = "4",
    pages = "898--906",
    year = "1956"
}

@article{Kerr:1959zlt,
    author = "Kerr, R. P.",
    title = "{The Lorentz-covariant approximation method in general relativity I}",
    doi = "10.1007/bf02732767",
    journal = "Nuovo Cim.",
    volume = "13",
    number = "3",
    pages = "469--491",
    year = "1959"
}

@article{Bertotti:1960wuq,
    author = "Bertotti, B. and Plebanski, J.",
    title = "{Theory of gravitational perturbations in the fast motion approximation}",
    doi = "10.1016/0003-4916(60)90132-9",
    journal = "Annals Phys.",
    volume = "11",
    number = "2",
    pages = "169--200",
    year = "1960"
}

@article{Portilla:1979xx,
    author = "Portilla, M.",
    title = "{MOMENTUM AND ANGULAR MOMENTUM OF TWO GRAVITATING PARTICLES}",
    doi = "10.1088/0305-4470/12/7/025",
    journal = "J. Phys. A",
    volume = "12",
    pages = "1075--1090",
    year = "1979"
}

@article{Westpfahl:1979gu,
    author = "Westpfahl, K. and Goller, M.",
    title = "{GRAVITATIONAL SCATTERING OF TWO RELATIVISTIC PARTICLES IN POSTLINEAR APPROXIMATION}",
    doi = "10.1007/BF02817047",
    journal = "Lett. Nuovo Cim.",
    volume = "26",
    pages = "573--576",
    year = "1979"
}

@article{Portilla:1980uz,
    author = "Portilla, M.",
    title = "{SCATTERING OF TWO GRAVITATING PARTICLES: CLASSICAL APPROACH}",
    doi = "10.1088/0305-4470/13/12/017",
    journal = "J. Phys. A",
    volume = "13",
    pages = "3677--3683",
    year = "1980"
}

@article{Bel:1981be,
    author = "Bel, LLuis and Damour, T. and Deruelle, N. and Ibanez, J. and Martin, J.",
    title = "{Poincar{\'e}-invariant gravitational field and equations of motion of two pointlike objects: The postlinear approximation of general relativity}",
    reportNumber = "PRINT-81-0534 (MEUDON)",
    doi = "10.1007/BF00756073",
    journal = "Gen. Rel. Grav.",
    volume = "13",
    pages = "963--1004",
    year = "1981"
}

@article{Westpfahl:1985tsl,
    author = "Westpfahl, Konradin",
    title = "{High-Speed Scattering of Charged and Uncharged Particles in General Relativity}",
    doi = "10.1002/prop.2190330802",
    journal = "Fortsch. Phys.",
    volume = "33",
    number = "8",
    pages = "417--493",
    year = "1985"
}

@article{Ledvinka:2008tk,
    author = "Ledvinka, Tomas and Schaefer, Gerhard and Bicak, Jiri",
    title = "{Relativistic Closed-Form Hamiltonian for Many-Body Gravitating Systems in the Post-Minkowskian Approximation}",
    eprint = "0807.0214",
    archivePrefix = "arXiv",
    primaryClass = "gr-qc",
    doi = "10.1103/PhysRevLett.100.251101",
    journal = "Phys. Rev. Lett.",
    volume = "100",
    pages = "251101",
    year = "2008"
}

@article{Goldberger:2004jt,
    author = "Goldberger, Walter D. and Rothstein, Ira Z.",
    title = "{An Effective field theory of gravity for extended objects}",
    eprint = "hep-th/0409156",
    archivePrefix = "arXiv",
    reportNumber = "UCSD-PTH-04-17, CMU-HEP-04-06",
    doi = "10.1103/PhysRevD.73.104029",
    journal = "Phys. Rev. D",
    volume = "73",
    pages = "104029",
    year = "2006"
}

@article{Porto:2016pyg,
    author = "Porto, Rafael A.",
    title = "{The effective field theorist{\textquoteright}s approach to gravitational dynamics}",
    eprint = "1601.04914",
    archivePrefix = "arXiv",
    primaryClass = "hep-th",
    doi = "10.1016/j.physrep.2016.04.003",
    journal = "Phys. Rept.",
    volume = "633",
    pages = "1--104",
    year = "2016"
}

@article{Mogull:2020sak,
    author = "Mogull, Gustav and Plefka, Jan and Steinhoff, Jan",
    title = "{Classical black hole scattering from a worldline quantum field theory}",
    eprint = "2010.02865",
    archivePrefix = "arXiv",
    primaryClass = "hep-th",
    reportNumber = "UUITP-37/20, HU-EP-20/22-RTG",
    doi = "10.1007/JHEP02(2021)048",
    journal = "JHEP",
    volume = "02",
    pages = "048",
    year = "2021"
}

@article{Jakobsen:2022psy,
    author = "Jakobsen, Gustav Uhre and Mogull, Gustav and Plefka, Jan and Sauer, Benjamin",
    title = "{All things retarded: radiation-reaction in worldline quantum field theory}",
    eprint = "2207.00569",
    archivePrefix = "arXiv",
    primaryClass = "hep-th",
    reportNumber = "HU-EP-22/24-RTG",
    doi = "10.1007/JHEP10(2022)128",
    journal = "JHEP",
    volume = "10",
    pages = "128",
    year = "2022"
}

@phdthesis{Jakobsen:2023oow,
    author = "Jakobsen, Gustav Uhre",
    title = "{Gravitational Scattering of Compact Bodies from Worldline Quantum Field Theory}",
    eprint = "2308.04388",
    archivePrefix = "arXiv",
    primaryClass = "hep-th",
    reportNumber = "HU-EP-23/45-RTG",
    doi = "10.18452/27075",
    school = "Humboldt U., Berlin, Humboldt U., Berlin (main)",
    year = "2023"
}

@article{Jakobsen:2021zvh,
    author = "Jakobsen, Gustav Uhre and Mogull, Gustav and Plefka, Jan and Steinhoff, Jan",
    title = "{SUSY in the sky with gravitons}",
    eprint = "2109.04465",
    archivePrefix = "arXiv",
    primaryClass = "hep-th",
    reportNumber = "HU-EP-21/28-RTG",
    doi = "10.1007/JHEP01(2022)027",
    journal = "JHEP",
    volume = "01",
    pages = "027",
    year = "2022"
}

@article{Tkachov:1981wb,
    author = "Tkachov, F. V.",
    title = "{A theorem on analytical calculability of 4-loop renormalization group functions}",
    doi = "10.1016/0370-2693(81)90288-4",
    journal = "Phys. Lett. B",
    volume = "100",
    pages = "65--68",
    year = "1981"
}

@article{Chetyrkin:1981qh,
    author = "Chetyrkin, K. G. and Tkachov, F. V.",
    title = "{Integration by parts: The algorithm to calculate $\beta$-functions in 4 loops}",
    doi = "10.1016/0550-3213(81)90199-1",
    journal = "Nucl. Phys. B",
    volume = "192",
    pages = "159--204",
    year = "1981"
}

@article{Kotikov:1990kg,
    author = "Kotikov, A. V.",
    title = "{Differential equations method: New technique for massive Feynman diagrams calculation}",
    reportNumber = "ITF-90-31E",
    doi = "10.1016/0370-2693(91)90413-K",
    journal = "Phys. Lett. B",
    volume = "254",
    pages = "158--164",
    year = "1991"
}

@article{Bern:1992em,
    author = "Bern, Zvi and Dixon, Lance J. and Kosower, David A.",
    title = "{Dimensionally regulated one loop integrals}",
    eprint = "hep-ph/9212308",
    archivePrefix = "arXiv",
    reportNumber = "SLAC-PUB-6001, CERN-TH-6756-92, UCLA-92-42",
    doi = "10.1016/0370-2693(93)90400-C",
    journal = "Phys. Lett. B",
    volume = "302",
    pages = "299--308",
    year = "1993",
    note = "[Erratum: Phys.Lett.B 318, 649 (1993)]"
}

@article{Gehrmann:1999as,
    author = "Gehrmann, T. and Remiddi, E.",
    title = "{Differential equations for two-loop four-point functions}",
    eprint = "hep-ph/9912329",
    archivePrefix = "arXiv",
    reportNumber = "TTP-99-49",
    doi = "10.1016/S0550-3213(00)00223-6",
    journal = "Nucl. Phys. B",
    volume = "580",
    pages = "485--518",
    year = "2000"
}

@article{Henn:2013pwa,
    author = "Henn, Johannes M.",
    title = "{Multiloop integrals in dimensional regularization made simple}",
    eprint = "1304.1806",
    archivePrefix = "arXiv",
    primaryClass = "hep-th",
    doi = "10.1103/PhysRevLett.110.251601",
    journal = "Phys. Rev. Lett.",
    volume = "110",
    pages = "251601",
    year = "2013"
}

@article{Henn:2014qga,
    author = "Henn, Johannes M.",
    title = "{Lectures on differential equations for Feynman integrals}",
    eprint = "1412.2296",
    archivePrefix = "arXiv",
    primaryClass = "hep-ph",
    doi = "10.1088/1751-8113/48/15/153001",
    journal = "J. Phys. A",
    volume = "48",
    pages = "153001",
    year = "2015"
}

@article{Anastasiou:2002yz,
    author = "Anastasiou, Charalampos and Melnikov, Kirill",
    title = "{Higgs boson production at hadron colliders in NNLO QCD}",
    eprint = "hep-ph/0207004",
    archivePrefix = "arXiv",
    reportNumber = "SLAC-PUB-9273",
    doi = "10.1016/S0550-3213(02)00837-4",
    journal = "Nucl. Phys. B",
    volume = "646",
    pages = "220--256",
    year = "2002"
}

@article{Anastasiou:2002qz,
    author = "Anastasiou, Charalampos and Dixon, Lance J. and Melnikov, Kirill",
    editor = "Blumlein, J. and Jegerlehner, F. and Riemann, T. and Hollik, W. and Kuhn, Johann H.",
    title = "{NLO Higgs boson rapidity distributions at hadron colliders}",
    eprint = "hep-ph/0211141",
    archivePrefix = "arXiv",
    reportNumber = "SLAC-PUB-9571",
    doi = "10.1016/S0920-5632(03)80168-8",
    journal = "Nucl. Phys. B Proc. Suppl.",
    volume = "116",
    pages = "193--197",
    year = "2003"
}

@article{Anastasiou:2003yy,
    author = "Anastasiou, Charalampos and Dixon, Lance J. and Melnikov, Kirill and Petriello, Frank",
    title = "{Dilepton rapidity distribution in the Drell-Yan process at NNLO in QCD}",
    eprint = "hep-ph/0306192",
    archivePrefix = "arXiv",
    reportNumber = "SLAC-PUB-10000, UH-511-1029-03",
    doi = "10.1103/PhysRevLett.91.182002",
    journal = "Phys. Rev. Lett.",
    volume = "91",
    pages = "182002",
    year = "2003"
}

@article{Anastasiou:2015yha,
    author = "Anastasiou, Charalampos and Duhr, Claude and Dulat, Falko and Furlan, Elisabetta and Herzog, Franz and Mistlberger, Bernhard",
    title = "{Soft expansion of double-real-virtual corrections to Higgs production at N$^{3}$LO}",
    eprint = "1505.04110",
    archivePrefix = "arXiv",
    primaryClass = "hep-ph",
    reportNumber = "CERN-PH-TH-2015-092, CP3-15-11, FERMILAB-PUB-15-089-T, NIKHEF-2015-016",
    doi = "10.1007/JHEP08(2015)051",
    journal = "JHEP",
    volume = "08",
    pages = "051",
    year = "2015"
}

@article{Parra-Martinez:2020dzs,
    author = "Parra-Martinez, Julio and Ruf, Michael S. and Zeng, Mao",
    title = "{Extremal black hole scattering at $\mathcal{O}(G^3)$: graviton dominance, eikonal exponentiation, and differential equations}",
    eprint = "2005.04236",
    archivePrefix = "arXiv",
    primaryClass = "hep-th",
    reportNumber = "FR-PHENO-2020-007, UCLA/TEP/2020/103",
    doi = "10.1007/JHEP11(2020)023",
    journal = "JHEP",
    volume = "11",
    pages = "023",
    year = "2020"
}

@article{Bern:1994zx,
    author = "Bern, Zvi and Dixon, Lance J. and Dunbar, David C. and Kosower, David A.",
    title = "{One loop n point gauge theory amplitudes, unitarity and collinear limits}",
    eprint = "hep-ph/9403226",
    archivePrefix = "arXiv",
    reportNumber = "SLAC-PUB-6415, SACLAY-SPH-T-94-20, UCLA-TEP-94-4, SWAT-94-17",
    doi = "10.1016/0550-3213(94)90179-1",
    journal = "Nucl. Phys. B",
    volume = "425",
    pages = "217--260",
    year = "1994"
}

@article{Bern:1994cg,
    author = "Bern, Zvi and Dixon, Lance J. and Dunbar, David C. and Kosower, David A.",
    title = "{Fusing gauge theory tree amplitudes into loop amplitudes}",
    eprint = "hep-ph/9409265",
    archivePrefix = "arXiv",
    reportNumber = "SLAC-PUB-6563, SACLAY-SPH-T-94-95, UCLA-TEP-94-29, SWAT-94-36",
    doi = "10.1016/0550-3213(94)00488-Z",
    journal = "Nucl. Phys. B",
    volume = "435",
    pages = "59--101",
    year = "1995"
}

@article{Britto:2004nc,
    author = "Britto, Ruth and Cachazo, Freddy and Feng, Bo",
    title = "{Generalized unitarity and one-loop amplitudes in N=4 super-Yang-Mills}",
    eprint = "hep-th/0412103",
    archivePrefix = "arXiv",
    doi = "10.1016/j.nuclphysb.2005.07.014",
    journal = "Nucl. Phys. B",
    volume = "725",
    pages = "275--305",
    year = "2005"
}

@article{Bern:2007ct,
    author = "Bern, Z. and Carrasco, J. J. M. and Johansson, Henrik and Kosower, D. A.",
    title = "{Maximally supersymmetric planar Yang-Mills amplitudes at five loops}",
    eprint = "0705.1864",
    archivePrefix = "arXiv",
    primaryClass = "hep-th",
    reportNumber = "UCLA-07-TEP-04, ZU-TH-12-07, SACLAY-SPHT-T07-050",
    doi = "10.1103/PhysRevD.76.125020",
    journal = "Phys. Rev. D",
    volume = "76",
    pages = "125020",
    year = "2007"
}

@article{Bern:2008pv,
    author = "Bern, Z. and Carrasco, J. J. M. and Dixon, Lance J. and Johansson, Henrik and Roiban, R.",
    title = "{Manifest Ultraviolet Behavior for the Three-Loop Four-Point Amplitude of N=8 Supergravity}",
    eprint = "0808.4112",
    archivePrefix = "arXiv",
    primaryClass = "hep-th",
    reportNumber = "SLAC-PUB-13361, UCLA-08-TEP-24",
    doi = "10.1103/PhysRevD.78.105019",
    journal = "Phys. Rev. D",
    volume = "78",
    pages = "105019",
    year = "2008"
}

@article{Bern:2010tq,
    author = "Bern, Z. and Carrasco, J. J. M. and Dixon, Lance J. and Johansson, H. and Roiban, R.",
    title = "{The Complete Four-Loop Four-Point Amplitude in N=4 Super-Yang-Mills Theory}",
    eprint = "1008.3327",
    archivePrefix = "arXiv",
    primaryClass = "hep-th",
    reportNumber = "UCLA-10-TEP-105, SACLAY-IPHT-T10-075, SLAC-PUB-14137, CERN-TH-2010-186",
    doi = "10.1103/PhysRevD.82.125040",
    journal = "Phys. Rev. D",
    volume = "82",
    pages = "125040",
    year = "2010"
}

@article{Carrasco:2011hw,
    author = "Carrasco, John Joseph M. and Johansson, Henrik",
    title = "{Generic multiloop methods and application to N=4 super-Yang-Mills}",
    eprint = "1103.3298",
    archivePrefix = "arXiv",
    primaryClass = "hep-th",
    reportNumber = "SU-ITP-11-07, SACLAY-IPHT-T11-029",
    doi = "10.1088/1751-8113/44/45/454004",
    journal = "J. Phys. A",
    volume = "44",
    pages = "454004",
    year = "2011"
}

@article{Bern:2024vqs,
    author = "Bern, Zvi and Herrmann, Enrico and Roiban, Radu and Ruf, Michael S. and Zeng, Mao",
    title = "{Global bases for nonplanar loop integrands, generalized unitarity, and the double copy to all loop orders}",
    eprint = "2408.06686",
    archivePrefix = "arXiv",
    primaryClass = "hep-th",
    doi = "10.1007/JHEP06(2025)115",
    journal = "JHEP",
    volume = "06",
    pages = "115",
    year = "2025"
}

@article{Bern:2008qj,
    author = "Bern, Z. and Carrasco, J. J. M. and Johansson, Henrik",
    title = "{New Relations for Gauge-Theory Amplitudes}",
    eprint = "0805.3993",
    archivePrefix = "arXiv",
    primaryClass = "hep-ph",
    reportNumber = "UCLA-07-TEP-15",
    doi = "10.1103/PhysRevD.78.085011",
    journal = "Phys. Rev. D",
    volume = "78",
    pages = "085011",
    year = "2008"
}

@article{Bern:2010ue,
    author = "Bern, Zvi and Carrasco, John Joseph M. and Johansson, Henrik",
    title = "{Perturbative Quantum Gravity as a Double Copy of Gauge Theory}",
    eprint = "1004.0476",
    archivePrefix = "arXiv",
    primaryClass = "hep-th",
    reportNumber = "UCLA-10-TEP-102, SACLAY-IPHT-T10-044",
    doi = "10.1103/PhysRevLett.105.061602",
    journal = "Phys. Rev. Lett.",
    volume = "105",
    pages = "061602",
    year = "2010"
}

@article{Bern:2019prr,
    author = "Bern, Zvi and Carrasco, John Joseph and Chiodaroli, Marco and Johansson, Henrik and Roiban, Radu",
    title = "{The duality between color and kinematics and its applications}",
    eprint = "1909.01358",
    archivePrefix = "arXiv",
    primaryClass = "hep-th",
    reportNumber = "CERN-TH-2019-135, UCLA/TEP/2019/104, NUHEP-TH/19-11, UUITP-35/19,
  NORDITA 2019-079",
    doi = "10.1088/1751-8121/ad5fd0",
    journal = "J. Phys. A",
    volume = "57",
    number = "33",
    pages = "333002",
    year = "2024"
}

@article{Goldberger:2016iau,
    author = "Goldberger, Walter D. and Ridgway, Alexander K.",
    title = "{Radiation and the classical double copy for color charges}",
    eprint = "1611.03493",
    archivePrefix = "arXiv",
    primaryClass = "hep-th",
    doi = "10.1103/PhysRevD.95.125010",
    journal = "Phys. Rev. D",
    volume = "95",
    number = "12",
    pages = "125010",
    year = "2017"
}

@article{Shen:2018ebu,
    author = "Shen, Chia-Hsien",
    title = "{Gravitational Radiation from Color-Kinematics Duality}",
    eprint = "1806.07388",
    archivePrefix = "arXiv",
    primaryClass = "hep-th",
    doi = "10.1007/JHEP11(2018)162",
    journal = "JHEP",
    volume = "11",
    pages = "162",
    year = "2018"
}

@article{Shi:2021qsb,
    author = "Shi, Canxin and Plefka, Jan",
    title = "{Classical double copy of worldline quantum field theory}",
    eprint = "2109.10345",
    archivePrefix = "arXiv",
    primaryClass = "hep-th",
    doi = "10.1103/PhysRevD.105.026007",
    journal = "Phys. Rev. D",
    volume = "105",
    number = "2",
    pages = "026007",
    year = "2022"
}

@article{Comberiati:2022cpm,
    author = "Comberiati, Francesco and Shi, Canxin",
    title = "{Classical Double Copy of Spinning Worldline Quantum Field Theory}",
    eprint = "2212.13855",
    archivePrefix = "arXiv",
    primaryClass = "hep-th",
    doi = "10.1007/JHEP04(2023)008",
    journal = "JHEP",
    volume = "04",
    pages = "008",
    year = "2023"
}

@article{Edison:2022cdu,
    author = "Edison, Alex and Levi, Mich{\`e}le",
    title = "{A tale of tails through generalized unitarity}",
    eprint = "2202.04674",
    archivePrefix = "arXiv",
    primaryClass = "hep-th",
    doi = "10.1016/j.physletb.2022.137634",
    journal = "Phys. Lett. B",
    volume = "837",
    pages = "137634",
    year = "2023"
}

@article{Edison:2023qvg,
    author = "Edison, Alex and Levi, Mich{\`e}le",
    title = "{Higher-order tails and RG flows due to scattering of gravitational radiation from binary inspirals}",
    eprint = "2310.20066",
    archivePrefix = "arXiv",
    primaryClass = "hep-th",
    doi = "10.1007/JHEP08(2024)161",
    journal = "JHEP",
    volume = "08",
    pages = "161",
    year = "2024"
}

@article{Edison:2024owb,
    author = "Edison, Alex",
    title = "{Parting gravity{\textquoteright}s tail: quadrupole tails at fifth order and beyond via integer partitions}",
    eprint = "2409.17222",
    archivePrefix = "arXiv",
    primaryClass = "hep-th",
    doi = "10.1007/JHEP02(2025)016",
    journal = "JHEP",
    volume = "02",
    pages = "016",
    year = "2025"
}

@article{Benincasa:2007xk,
    author = "Benincasa, Paolo and Cachazo, Freddy",
    title = "{Consistency Conditions on the S-Matrix of Massless Particles}",
    eprint = "0705.4305",
    archivePrefix = "arXiv",
    primaryClass = "hep-th",
    reportNumber = "UWO-TH-07-09",
    month = "5",
    year = "2007"
}

@article{Kalin:2022hph,
    author = {K{\"a}lin, Gregor and Neef, Jakob and Porto, Rafael A.},
    title = "{Radiation-reaction in the Effective Field Theory approach to Post-Minkowskian dynamics}",
    eprint = "2207.00580",
    archivePrefix = "arXiv",
    primaryClass = "hep-th",
    reportNumber = "DESY-22-109, DESY 22-109",
    doi = "10.1007/JHEP01(2023)140",
    journal = "JHEP",
    volume = "01",
    pages = "140",
    year = "2023"
}

@article{Keldysh:1964ud,
    author = "Keldysh, L. V.",
    title = "{Diagram Technique for Nonequilibrium Processes}",
    doi = "10.1142/9789811279461_0007",
    journal = "Sov. Phys. JETP",
    volume = "20",
    pages = "1018--1026",
    year = "1965"
}

@article{Schwinger:1960qe,
    author = "Schwinger, Julian S.",
    title = "{Brownian motion of a quantum oscillator}",
    doi = "10.1063/1.1703727",
    journal = "J. Math. Phys.",
    volume = "2",
    pages = "407--432",
    year = "1961"
}

@article{Caron-Huot:2010fvq,
    author = "Caron-Huot, Simon",
    title = "{Loops and trees}",
    eprint = "1007.3224",
    archivePrefix = "arXiv",
    primaryClass = "hep-ph",
    doi = "10.1007/JHEP05(2011)080",
    journal = "JHEP",
    volume = "05",
    pages = "080",
    year = "2011"
}

@article{Biswas:2024ept,
    author = "Biswas, Shovon and Parra-Martinez, Julio",
    title = "{Classical observables from causal response functions}",
    eprint = "2411.09016",
    archivePrefix = "arXiv",
    primaryClass = "hep-th",
    doi = "10.1007/JHEP07(2025)037",
    journal = "JHEP",
    volume = "07",
    pages = "037",
    year = "2025"
}

@article{Bern:2011qt,
    author = "Bern, Zvi and Huang, Yu-tin",
    title = "{Basics of Generalized Unitarity}",
    eprint = "1103.1869",
    archivePrefix = "arXiv",
    primaryClass = "hep-th",
    reportNumber = "UCLA-11-TEP-103",
    doi = "10.1088/1751-8113/44/45/454003",
    journal = "J. Phys. A",
    volume = "44",
    pages = "454003",
    year = "2011"
}

@article{Cheung:2021yog,
    author = "Cheung, Clifford and Helset, Andreas and Parra-Martinez, Julio",
    title = "{Geometric soft theorems}",
    eprint = "2111.03045",
    archivePrefix = "arXiv",
    primaryClass = "hep-th",
    reportNumber = "CALT-TH-2021-038",
    doi = "10.1007/JHEP04(2022)011",
    journal = "JHEP",
    volume = "04",
    pages = "011",
    year = "2022"
}

@article{Derda:2024jvo,
    author = "Derda, Maria and Helset, Andreas and Parra-Martinez, Julio",
    title = "{Soft scalars in effective field theory}",
    eprint = "2403.12142",
    archivePrefix = "arXiv",
    primaryClass = "hep-th",
    reportNumber = "CALT-TH-2024-011, CERN-TH-2024-035",
    doi = "10.1007/JHEP06(2024)133",
    journal = "JHEP",
    volume = "06",
    pages = "133",
    year = "2024"
}

@article{Haddad:2025cmw,
    author = "Haddad, Kays and Jakobsen, Gustav Uhre and Mogull, Gustav and Plefka, Jan",
    title = "{Unitarity and the On-Shell Action of Worldline Quantum Field Theory}",
    eprint = "2510.00988",
    archivePrefix = "arXiv",
    primaryClass = "hep-th",
    reportNumber = "HU-EP-25/33, QMUL-PH-25-29",
    doi = "10.1007/JHEP02(2026)008",
    journal = "JHEP",
    volume = "02",
    pages = "008",
    year = "2026"
}

@article{Arkani-Hamed:2016rak,
    author = "Arkani-Hamed, Nima and Rodina, Laurentiu and Trnka, Jaroslav",
    title = "{Locality and Unitarity of Scattering Amplitudes from Singularities and Gauge Invariance}",
    eprint = "1612.02797",
    archivePrefix = "arXiv",
    primaryClass = "hep-th",
    doi = "10.1103/PhysRevLett.120.231602",
    journal = "Phys. Rev. Lett.",
    volume = "120",
    number = "23",
    pages = "231602",
    year = "2018"
}

@article{Rodina:2016jyz,
    author = "Rodina, Laurentiu",
    title = "{Uniqueness from gauge invariance and the Adler zero}",
    eprint = "1612.06342",
    archivePrefix = "arXiv",
    primaryClass = "hep-th",
    doi = "10.1007/JHEP09(2019)084",
    journal = "JHEP",
    volume = "09",
    pages = "084",
    year = "2019"
}

@article{Cachazo:2005ca,
    author = "Cachazo, Freddy and Svrcek, Peter",
    title = "{Tree level recursion relations in general relativity}",
    eprint = "hep-th/0502160",
    archivePrefix = "arXiv",
    month = "2",
    year = "2005"
}

@article{Ivanov:2026icp,
    author = "Ivanov, Mikhail M. and Li, Yue-Zhou and Parra-Martinez, Julio and Zhou, Zihan",
    title = "{Gravitational Raman Scattering: a Systematic Toolkit for Tidal Effects in General Relativity}",
    eprint = "2602.06951",
    archivePrefix = "arXiv",
    primaryClass = "hep-th",
    reportNumber = "MIT-CTP/6001",
    month = "2",
    year = "2026"
}

@article{Bjerrum-Bohr:2025bqg,
    author = "Bjerrum-Bohr, N. Emil J. and Chen, Gang and Eriksen, Carl Jordan and Shah, Nabha",
    title = "{The gravitational Compton amplitude from flat and curved spacetimes at second post-Minkowskian order}",
    eprint = "2506.19705",
    archivePrefix = "arXiv",
    primaryClass = "hep-th",
    doi = "10.1007/JHEP10(2025)235",
    journal = "JHEP",
    volume = "10",
    pages = "235",
    year = "2025"
}

@article{Damgaard:2021ipf,
    author = "Damgaard, Poul H. and Plante, Ludovic and Vanhove, Pierre",
    title = "{On an exponential representation of the gravitational S-matrix}",
    eprint = "2107.12891",
    archivePrefix = "arXiv",
    primaryClass = "hep-th",
    reportNumber = "IPhT-t21/037, CERN-TH-2021-111",
    doi = "10.1007/JHEP11(2021)213",
    journal = "JHEP",
    volume = "11",
    pages = "213",
    year = "2021"
}

\end{document}